\documentclass[apj]{emulateapj}
\usepackage{graphics}
\usepackage{natbib}
\usepackage{rotating}
\usepackage{color,soul}

\newcommand{\as}{$''$}
\newcommand{\am}{$'$}
\newcommand{\nh}{$N_{\rm{H}}$}
\newcommand{\nustar}{\textit{NuSTAR}}
\newcommand{\xmm}{\textit{XMM-Newton}}
\newcommand{\chandra}{\textit{Chandra}}
\newcommand{\suzaku}{\textit{Suzaku}}

\newcommand{\swift}{\textit{Swift}}
\newcommand{\pksb}{PKS\,2155-304}
\newcommand{\qcs}{3C\,273}

\begin{document}

\title{IACHEC$^1$ Cross-Calibration of \chandra, \nustar, \swift, \suzaku, and \xmm\ with \qcs\ and \pksb.}
\author{Kristin K. Madsen$^1$, Andrew P. Beardmore$^2$, Karl Forster$^1$, Matteo Guainazzi$^{3,5}$, Herman L. Marshall$^4$, Eric D. Miller$^4$, Kim L. Page$^2$, Martin Stuhlinger$^5$}
\begin{abstract}
On behalf of the International Astronomical Consortium for High Energy Calibration (IACHEC), we present results from the cross-calibration campaigns in 2012 on \qcs\ and in 2013 on \pksb\ between the then active X-ray observatories \chandra, \nustar, \suzaku, \swift\ and \xmm. We compare measured fluxes between instrument pairs in two energy bands, 1--5\,keV and 3--7\,keV and calculate an average cross-normalization constant for each energy range. We review known cross-calibration features and provide a series of tables and figures to be used for evaluating cross-normalization constants obtained from other observations with the above mentioned observatories.
\end{abstract}

\affiliation{
$^1$ Cahill Center for Astronomy and Astrophysics, California Institute of Technology, Pasadena, CA 91125, USA\\
$^2$ X-ray and Observational Astronomy Group, Department of Physics and Astronomy, University of Leicester, Leicester LE1 7RH, UK\\
$^3$ Japan Aerospace Exploration Agency, Institute of Space and Astronautical Science, 3-1-1, Yoshinodai, Sagamihara, Kanagawa, Japan, 252-5201\\
$^4$ Kavli Institute for Astrophysics and Space Research, Massachusetts Institute of Technology, 77 Massachusetts Ave., Cambridge, MA 02139, USA\\
$^5$ European Space Astronomy Centre (ESAC), P.O. Box 78, 28691 Villanueva de la Caada, Madrid, Spain\\
}

\section{Introduction}
It is common to have simultaneous or near simultaneous X-ray coverage of astrophysical sources with multiple observations. The community is often faced with the question on how to properly fit joint data sets spanning multiple observatories. It is the mission of the International Astronomical Consortium for High Energy Calibration \citep[IACHEC,][]{Sembay2010} to provide the proper guidance and cross-calibration information to help the community avoid pitfalls and approach cross-observatory fitting in the correct manner. 

Several papers have been published as a result of IACHEC effort, using a variety of methods and targets. \citet{Nevalainen2010,Kettula2013, Schellenberger2015} used galaxy clusters to measure the differences in measured temperatures between instruments, and \citet{Tsujimoto2011} used the Pulsar Wind Nebula G21.5-0.9 to measure power-law slopes and fluxes. These studies were focused on extended sources and the differences in measured spectral parameters. For the soft X-ray CCD instruments \citet{Plucinsky2012, Plucinsky2016} used the supernova remnant 1E\,0102.2-7219 to compare the fluxes of the line spectrum, developing an empirical model as a reference spectrum for future cross-calibrations campaigns. For point source flux comparison \citet{Ishida2011} used \pksb. Furthermore, recent cross-calibration studies have been made by \citet{Guver2016} using simultaneous observations of thermonuclear X-ray bursts from GS\,1826-238.

Observatory calibration is under continuing development as instruments age and change. Most calibration updates deal with time dependent changes such as the evolution of instrument gain or contamination, but occasionally errors are discovered and corrected that affect data from the entire mission lifetime. It is therefore necessary to repeat and update cross-calibration campaigns on a regular basis using the newest instrument calibrations available. As instruments get decommissioned and new ones launched, it is also beneficial to the community to relate the newest member to the rest of the group.

We examine in this paper two cross-calibration campaigns conducted in 2012 and 2013 on two sources, the quasar \qcs, and the BL Lac object \pksb. Both sources have been used as calibration targets in the past, and they are well suited for several reasons: they are not too bright to cause severe pileup issues for the CCD instruments, their absorbing Galactic column is very low, and the spectra can be well fit by a power-law between 1--20 keV for \qcs\ and 1--10 keV for \pksb. Both targets are variable so that calibration observations must be simultaneous. In addition, \qcs\ can enter states where it has a curving spectrum above 20 keV, so caution is required for comparing slopes across a wide broadband \citep{madsen2015a}, and \pksb\ is very soft with $\Gamma \sim 2.7$, so that very long integration times would be required for sufficient statistics above 10\,keV.

In this paper we perform two different analyses. First, we find the best fit for each instrument and compare the differences between them, focusing primarily on the flux from which we will derive the cross-calibration constant. Second, we explore the change in flux and cross-normalization constant when we require the same spectral parameters for all data sets, rather than allowing each to take on its best fit. This second situation is what most astronomers do when fitting data from multiple instruments. It is therefore important to understand what systematics one might encounter due to systematic calibration errors that are different among instruments.

The participating instruments were: \chandra\ with HETGS for \qcs\ and LETGS for \pksb, \nustar, \suzaku\ with XIS for both targets and HXD-PIN for \qcs\ (there was no detection of the source in GSO), \swift\ with XRT only (no BAT), and \xmm.

This paper is a summary of the activity of a working group aiming at calibrating the effective areas of different X-ray missions within the framework of the IACHEC. The consortium aims to provide standards for high-energy calibration and to supervise cross calibration between different missions. We refer the readers to the website\footnote{International Astronomical Consortium for High Energy Calibration, http://web.mit.edu/iachec/} for more details on IACHEC activity and meetings.

\section{The Calibration Targets}
\subsection{\qcs}
At a redshift of $z = 0.158$ \citep{Schmidt1963}, \qcs\ is the nearest high luminosity quasar and has been extensively studied at all wavelengths since its discovery in 1963 \citep[for a review, see][]{courvoisier1998}. It is characterized as a radio-loud quasar with a jet showing apparent superluminal motion, and \qcs\ exhibits large flux variations across nearly all energies \citep{Soldi2008}.

As observed in many other AGN, there is sometimes a soft excess at low-energy X-rays ($<2$\,keV), possibly due to Comptonized UV photons \citep{Page2004}. Above  2\,keV and up to $\sim1$\,MeV, previous observations report that the spectrum of \qcs\ is a power-law, as is common for jet-dominated Active Galactic Nuclei (AGN). Over the 30 years that the source has been reliably monitored, there appears to be a long term spectral evolution underlying the short term variations. \qcs\ was in its softest observed state in June 2003 ($\Gamma \sim 1.82 \pm 0.01$), a value of $\Delta\Gamma \sim$ 0.3--0.4 above what was measured in the 1980's ($\Gamma \sim 1.5$). Since then the source has hardened again to a value of $\Gamma \sim 1.6 - 1.7$ \citep{Chernyakova2007}.

There is evidence of an intermittently weak iron line, which appears to be broad ($\sigma \sim 0.6$\,keV, Equivalent Width, EW, $\sim$ 20--60\,eV), occasionally neutral \citep{Turner1990,Page2004,Grandi2004} and sometimes ionized \citep{Yaqoob2000,Kataoka2002}. Using the full 244\,ks of the \nustar\ observation, of which we here only use the subsection overlapping the other observatories, \citet{madsen2015a} finds the very weak presence of an iron-line (6.4\,keV fixed) with a width and Equivalent Width of $\sigma=0.65\pm 0.03$\,keV and $EW=23\pm 11$\,eV. Because of the shorter exposure times, it was not detected by the other observatories. The spectrum of \qcs\ during this particular epoch clearly deviated from a power-law above 20\,keV and could be fit with a cutoff power-law of $\Gamma=1.627\pm0.006$ and $E_\mathrm{cutoff}=291^{+90}_{-55}$\,keV. The interpretation of the turnover is that it is due both to the direct coronal signature of the AGN and reflection off the accretion disk. These components typically are not visible since the jet generally dominates this source above $\sim 20$~keV.

The soft excess was not measurable above 1\,keV during the cross-calibration campaign and the turnover was not detectable below 20\,keV, so the spectral shape between 1--20\,keV is best represented by a power-law.

\subsection{\pksb}
\pksb\ is a high-frequency peaked BL Lac (HBL) object at $z=0.116$ and one of the most luminous of its kind. It has been frequently observed at all wavelengths and has consistently shown a soft, $\Gamma > 2.5$, X-ray spectrum in the 2--10\,keV band \citep{Sembay1993,Brinkmann1994,Edelson1995,Urry1997,Zhang1999,Kataoka2000,Tanihata2001}. Rapid variability can be found on hour time scales in the X-ray and optical bands \citep{Zhang1999,Edelson2001,Tanihata2001,Ishida2011}.

The broadband spectrum of \pksb\ is mostly featureless, but displays two prominent peaks located respectively in the far UV/soft X-ray band and in the GeV band. This double-peaked spectral energy distribution (SED) is believed to originate from synchrotron self-Comptonization (SSC) in which photons are scattered by the energetic electrons in the jet \citep{Band1985,Ghisellini1985}. The lower energy peak is presumably due to synchrotron emission, and the higher energy, secondary peak in the GeV band from inverse Compton scattering. The turnover into the second peak is believed to occur somewhere between 20\,keV and 100\,MeV. The broadband SED from X-rays to GeV is typically empirically fitted with a log-parabolic model, but in the narrow energy band we consider (1--10keV), the source has been well described as a power-law \citep{Ishida2011}. 

\section{Observations and Data Reduction}
\begin{figure}
\includegraphics[width=0.45\textwidth]{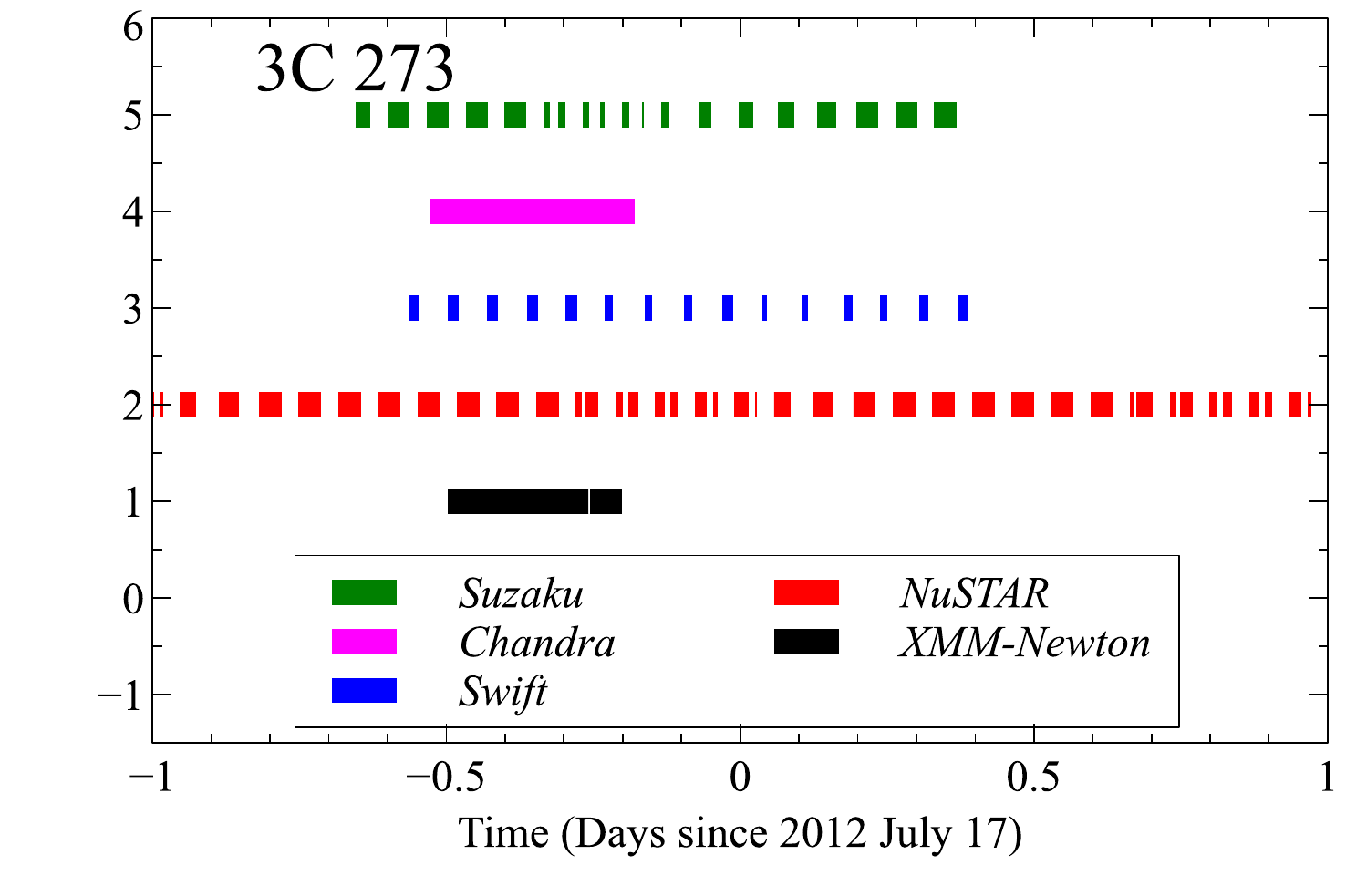}
\includegraphics[width=0.45\textwidth]{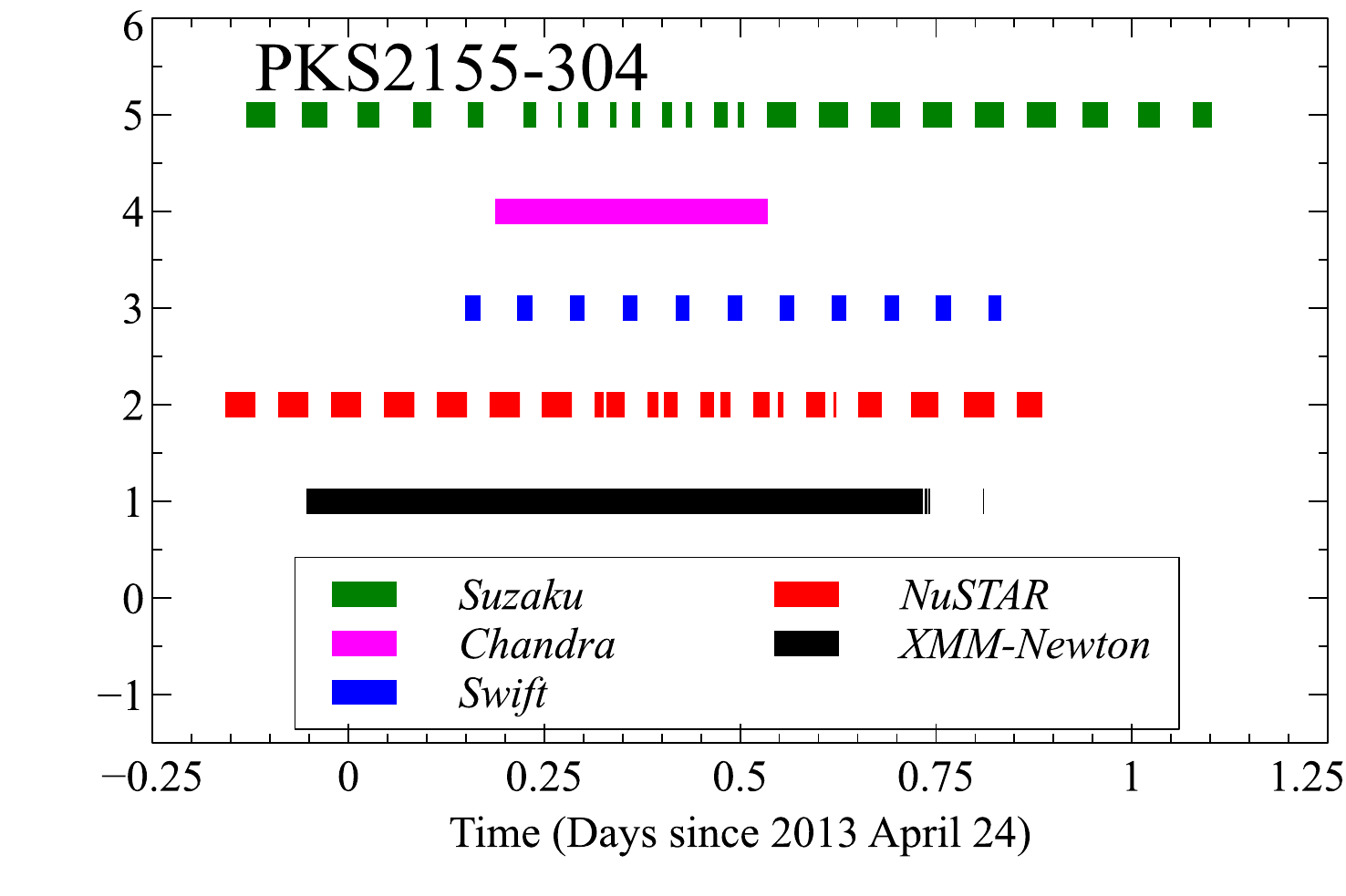}
\caption{Observation duration for each observatory for the two cross-calibration campaigns.}
\label{gtifigure}
\end{figure}

On 2012 July 17, a cross-calibration campaign was conducted between \textit{Chandra, NuSTAR, Suzaku, Swift} and \textit{XMM-Newton} on \qcs, and for \pksb\ on 2013 April 24. Table \ref{crossobsid} lists the observation identification numbers (obsID) and exposure times for the respective observatories. The duration and coverage of each observation is shown schematically in Figure \ref{gtifigure}. Due to the unfortunate relative beating of South Atlantic Anomaly (SAA) passages and occultation periods between the low Earth orbit observatories (\textit{NuSTAR}, \textit{Suzaku},  and \textit{Swift}), we decided to forego strict simultaneity between all observatories. Instead, we analyze observatory pairs truncated to the overlapping exposures, by only matching the good time intervals (GTIs) of the beginning and end of the overlap. 

In this scheme we have 11 observatory pairs (10 for \pksb\ due to the \suzaku/HXD flux being too low) as listed in Table \ref{gti}. The observation START and STOP times are applied as user GTIs to each pipeline, and the resulting total exposure times for each instrument in a pair are recorded in the last column of the table. Because the occultation/SAA periods are not being excluded, the exposure times are quite different between pairs. However, both sources were relatively stable during the observations as shown in the detailed light curves of Figures \ref{lightcurve3c273} and \ref{lightcurvepks2155}, and since the missing exposure times are evenly distributed across the overlapping periods, we determined that it did not impact fitting, since the error of the fit parameters are dominated by the total number of counts, rather than possible short term variations in flux and/or slope. We will discuss the light curves in more detail in Section \ref{sectionlightcurves}.

\subsection{Detailed Reduction and Extraction for each observatory}

\subsubsection{Chandra}
\chandra\ \citep{Weisskopf2002} is a single telescope observatory. Its main instrument, the Advanced CCD Imaging Spectrometer (ACIS) is an X-ray imaging-spectrometer consisting of the ACIS-I and ACIS-S CCD arrays. The instrument can be used together with either the High Energy Transmission Gratings \citep[HETGS,][]{Canizares2005}, which has a medium-energy grating (MEG) and a high-energy grating (HEG) arm, or with the Low Energy Transmission Grating (LETG). For these observations the back-side illuminated ACIS-S chip, sensitive in the 0.2--10\,keV band, was used.

We reduced the grating spectra using CIAO 4.6.1 and calibration database, \texttt{CALDB} 4.6.1.1 and reprocessed using the CIAO \texttt{chandra\_repro} reprocessing script. For \qcs\ the data were taken in grating configuration ACIS+HETG. We computed MEG and HEG spectra, combining the $-1$ and $+1$ orders in both cases. The HEG and MEG spectra were then fit simultaneously in the analysis. For \pksb\ the data were taken with grating configuration ACIS+LETG and we combined orders $-1$ and $+1$. 

We used \texttt{dmextract} to create light curves from the 0 order image in 1--3\,keV and 3--7\,keV bands, binned at a cadence of 5 minutes. In the following, we drop the ``ACIS+" abbreviation and simply refer to the instruments as LETGS and HETGS.

\subsubsection{NuSTAR}
\nustar\ \citep{Harrison2013} flies two co-aligned conical approximation Wolter-I telescoped coated with Pt/C and W/Si multilayers to provide a broad-band coverage in the hard X-rays between 3--79\,keV. The telescopes are focused on the Focal Plane Modules (FPMs) that are CdZnTe crystal hybrid pixel detectors \citep{Kitaguchi2014} aligned in a 2$\times$2 array. The two modules, FPMA and FPMB, are of identical design.

We reduced the data using \texttt{nustardas 09Jun15\_v1.5.1} and \texttt{CALDB} version 20150904. We used an extraction region of radius 50\as\ for \qcs\ and 40\as\ for \pksb. Backgrounds were taken on the same detector from a 75\as\ radius circular region, as close to the source as possible without risking contamination. We used \texttt{nuproducts} to extract spectra and all default point source parameters were applied during the pipeline run and in the extraction and generation of response files. 

We extracted the light curves using \texttt{nuproducts} and applied PSF and vignetting corrections at 7\,keV, since this is where the effective area peaks in the 3--7\,keV band. We binned the light curve in 5 minute intervals.

\subsubsection{Suzaku}
As of the time of writing, \suzaku\ \citep{Mitsuda2007} is no longer operational. The observatory consisted of four co-aligned Wolter-I telescopes, aimed at the X-ray Imaging Spectrometer \citep[XIS,][]{Koyama2007} numbered 0 through 3 and sensitive in the 0.2--12 \,keV band. XIS2 was not operational during the observations due to micro-meteorite hits. In addition to the XIS, \suzaku\ also carried the Hard X-ray Detector (HXD) of which the PIN was a component covering the 10--70\,keV band. It was a non-imaging detector composed of 64 Si PIN diodes at the bottom of well-type collimators surrounded by GSO anti-coincidence scintillators.

For these observations only \qcs\ was detected by HXD and only in PIN. We will in the following only be discussing HXD-PIN and abbreviate it with HXD.

We reduced the \suzaku\ data using HEASoft 6.16 and \texttt{CALDB} version 20150105, and reprocessed using the FTOOL \texttt{aepipeline} v1.1.0 reprocessing script.  Since the data were taken in 1/4 window mode, extraction regions were chosen by eye to include a 4$\times$8\am\ box centered on the sources.  Background regions were selected from the remaining exposed area, avoiding point sources identified by eye.  Response files were produced using the default parameters using a full ray-tracing simulation from a model point source with the FTOOL \texttt{sxisimarfgen} to create the ARF and properly account for photons lost ($\sim$ 15\%) outside of the narrow exposed window.

Light curves were extracted from the same regions with the FTOOL \texttt{lcurve} and binned at a cadence of 5 minutes.

\subsubsection{Swift-XRT}
\swift\ \citep{Gehrels2004} flies a single Wolter-I telescope (XRT) which focuses onto an X-ray CCD device similar to those used by the
\xmm/MOS cameras \citep{Burrows2005}. The instrument operates from 0.3--10\,keV, and since the primary science goal of \swift\ is to respond to gamma-ray bursts (GRBs) it operates autonomously to measure GRB light curves and spectra over seven orders of magnitude in flux. To mitigate pile-up, the instrument switches between CCD readout modes depending on the source brightness. Two frequently-used modes are: Windowed Timing (WT) mode, which provides 1D spectral information, and Photon Counting (PC) mode, which allows full 2D imaging-spectroscopy. \qcs\ was observed in PC mode and \pksb\ in WT mode.

We reduced the data using the \swift\ XRT pipeline \texttt{swxrtdas 17Jul15\_v3.1.1} and \texttt{CALDB} version 20150721. \qcs\ is slightly piled up and we extracted spectra using \texttt{XSELECT} from an annulus of inner radius 16.5\as and outer radius 71\as. \pksb\ was not piled up and we extracted a spectrum from a circle of radius 47\as.

Light curves were extracted from the same regions and binned at a cadence of 5 minutes.

\subsubsection{XMM-Newton}
\xmm\ \citep{Jansen2001} flies three Wolter-I grazing incidence telescopes. Two of the telescopes are each paired with the European Photon Imaging Camera (EPIC) Metal Oxide Semi-conductor CCD arrays (MOS1 and MOS2), and the third with EPIC-pn. All three are sensitive in the 0.15--15\,keV band. MOS1 and MOS2 cameras are front-illuminated, and the pn is back-illuminated. 

We reduced the data using \xmm\ Science Analysis System (SAS) version 14.0.0 and Current Calibration Files (CCFs) as of 2015 July 1 starting from ODF level. For event selection we used default values for best spectrum quality. For spectral extractions of both PKS~2155-304 and 3C~273 we excised the inner core with radius of 10\as\ from the EPIC PSFs to avoid possible pile-up effects. For EPIC-pn, we used local backgrounds taken from the corners of the EPIC-pn small window modes, whereas for EPIC-MOS empty sky fields of the peripheral CCDs were used.

Light curves were extracted using the same regions and event selections as in the spectral extraction, binned in 5 minutes intervals, and corrected for effective area, PSF, quantum efficiency, GTI, dead times as well as for background counts using the SAS task \emph{epiclccorr}.

\begin{table}
\centering
\caption{Cross calibration campaign}
\begin{tabular}{l||c|c||c|c}
\hline
Instrument & obsID & Exposure & obsID & Exposure \\
 &  & (ks) &  &(ks) \\
\hline
&  \multicolumn{2}{c}{\qcs} & \multicolumn{2}{c}{\pksb}\\
\hline
\chandra & 14455 & 30 & 15475 & 30 \\
\nustar  & 10002020001 & 244 & 60002022002 & 45\\
\swift & 00050900019	& 13 & 00030795108 & 17.7\\
\swift & 00050900020	& 6.9 &  & \\
\suzaku & 107013010 & 39.8 & 108010010 & 53.3 \\
\xmm  & 0414191001 & 38.9 & 0411782101 & 76 \\
\hline
\end{tabular}
\label{crossobsid}
\end{table}

\begin{table}
\caption{Observatory pairs}
\centering
\begin{tabular}{l|l|l|l|l}
\hline
GTI Start & GTI Stop & Concatenated & Limiting & Exposure\\
(MJD) & (MJD) & observation & observation &  (ks) \\
\hline
\multicolumn{5}{c}{3C\,273}\\
\hline
56124.346 & 56125.369 & \nustar & \suzaku & 42.3/40.2 \\
56124.475 & 56124.822 & \nustar & \chandra & 15.2/29.5 \\
56124.438 & 56125.389 & \nustar & \swift & 36.3/19.9 \\
56124.504 & 56124.801 & \nustar & \xmm & 12.8/26.9 \\
56124.475 & 56124.822 & \suzaku & \chandra & 13.7/29.5 \\
56124.438 & 56125.389 & \suzaku & \swift & 32.7/19.9  \\
56124.504 & 56124.801 & \suzaku & \xmm & 10.1/26.9 \\
56124.475 & 56124.822 & \swift & \chandra & 7.6/29.5\\
56124.504 & 56124.801 & \swift & \xmm & 7.6/25.3 \\
56124.504 & 56124.801 & \chandra & \xmm & 17.8/26.9\\
\hline
\multicolumn{5}{c}{PKS\,2155-304}\\
\hline
56406.184 & 56406.532 &\suzaku & \chandra & 8.3/30.1 \\
56405.840 & 56406.883 &\suzaku & \nustar & 38.4/45.0 \\
56406.146 & 56406.831 &\suzaku & \swift & 25.8/17.8 \\
56405.944 & 56406.808 &\suzaku & \xmm & 30.7/68.6 \\
56406.184 & 56406.532 &\nustar & \chandra & 16.0/30.1 \\
56405.840 & 56406.883 &\nustar & \swift & 29.0/17.8 \\
56405.944 & 56406.808 &\nustar & \xmm & 31.9/68.6 \\
56406.184 & 56406.532 &\xmm & \chandra & 20.9/30.1 \\
56406.146 & 56406.808 &\xmm & \swift & 28.9/17.7 \\
56406.184 & 56406.532 &\swift & \chandra & 8.1/30.1\\

\hline
\end{tabular}
\label{gti}
\end{table}

\begin{figure*}
\includegraphics[width=0.5\textwidth]{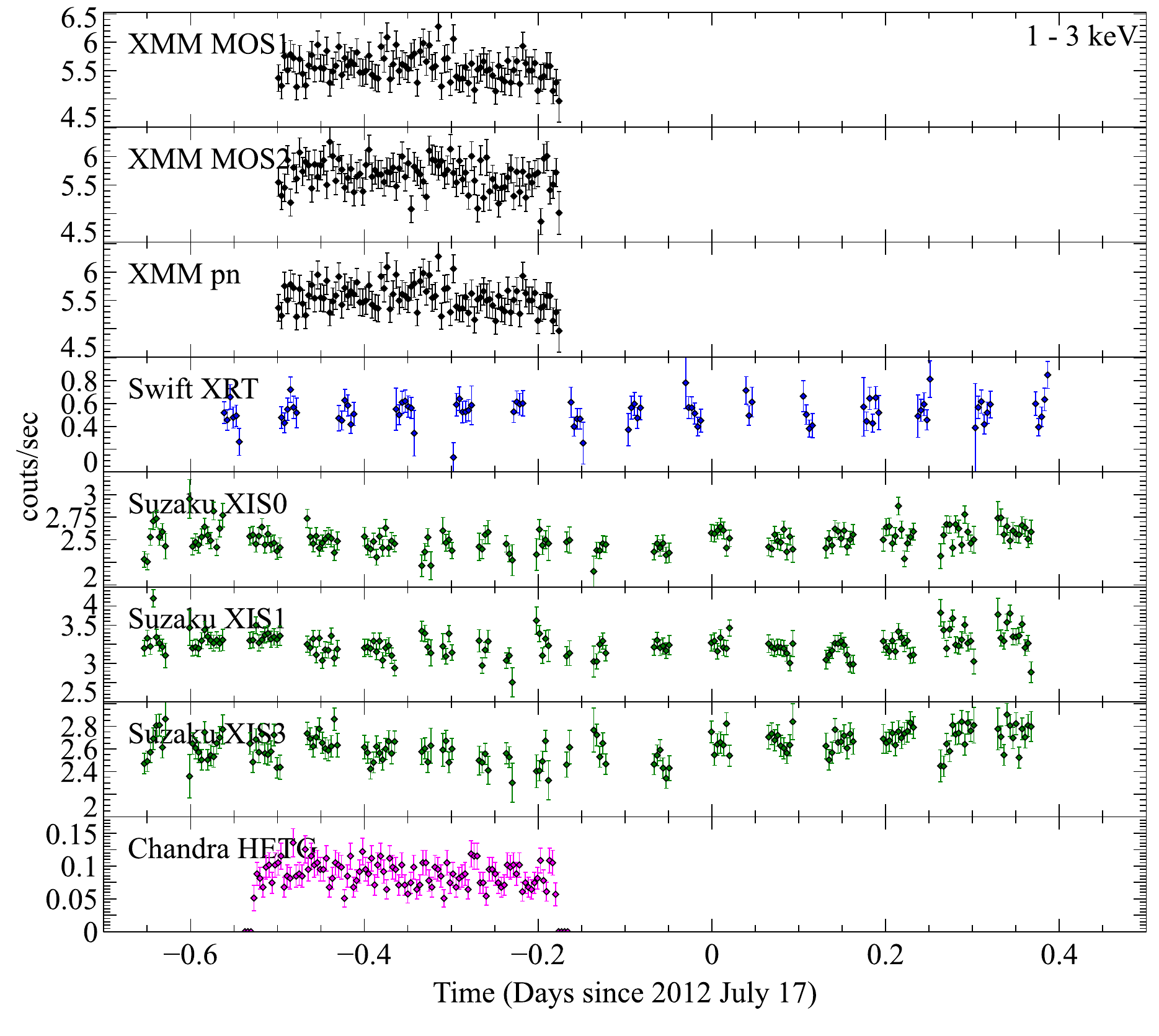}
\includegraphics[width=0.5\textwidth]{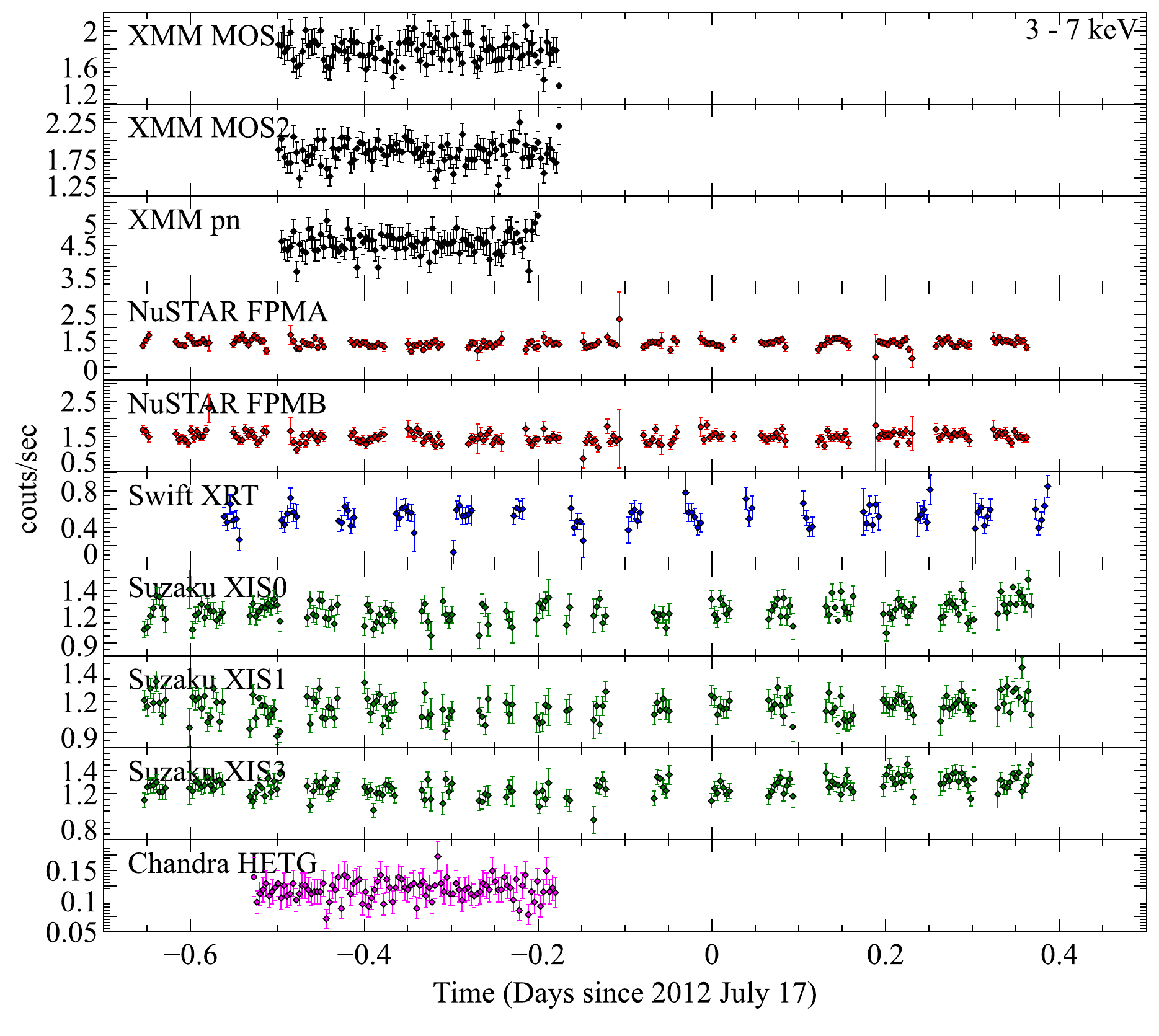} \\
\caption{Light curve of 3C\,273 in instrument counts s$^{-1}$. See text for instrument specific corrections made to each curve. Bin width is 5 minutes. Large error bars (1$\sigma$) at the edge of the occultation/SAA gaps of the low-Earth orbit instruments are typically due to very small fractional exposures within the bin, or an SAA passage that was not entirely filtered out.  The light curves show that there is no rapid variability present.}
\label{lightcurve3c273}
\end{figure*}

\begin{figure*}
\includegraphics[width=0.5\textwidth]{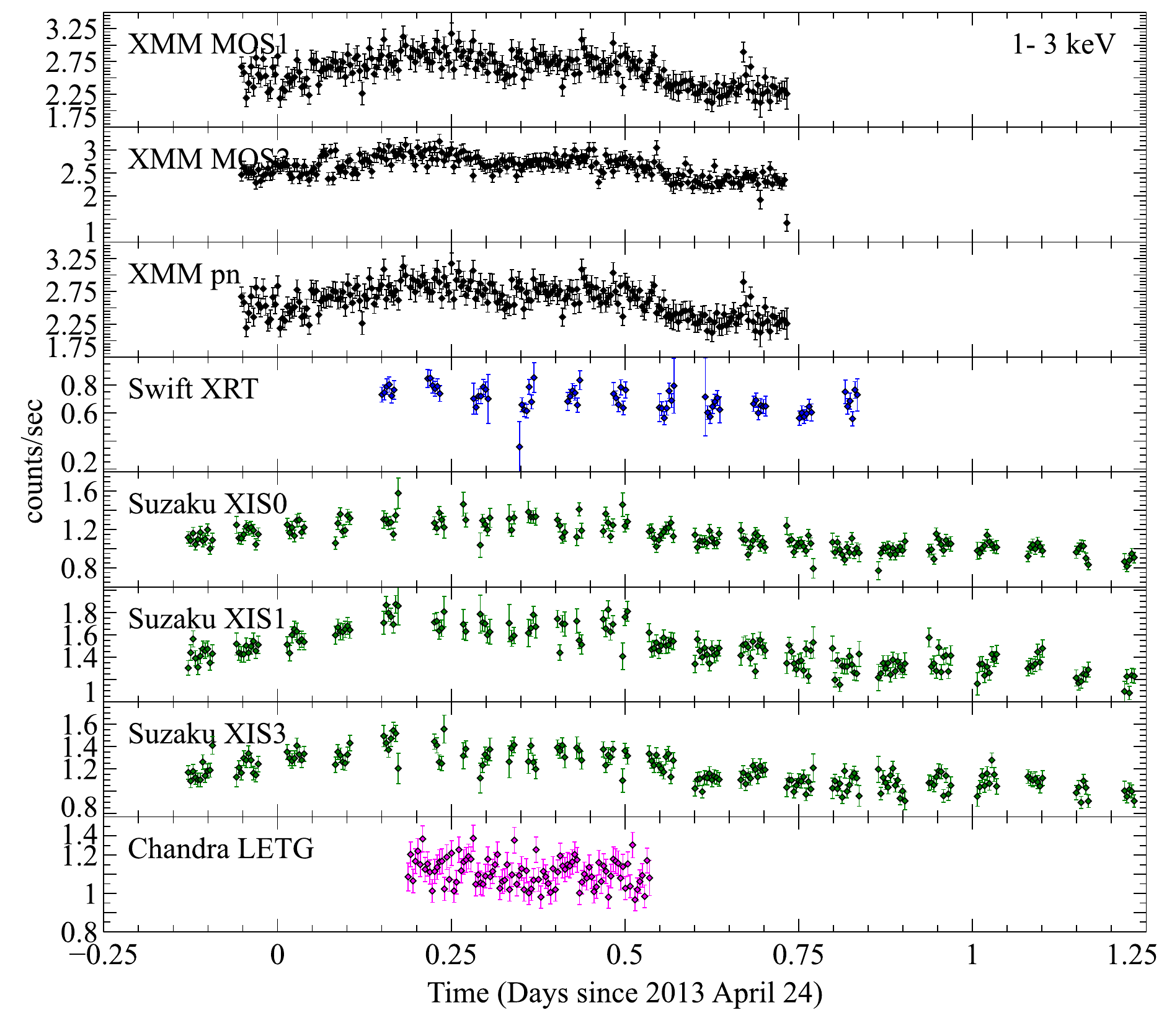}
\includegraphics[width=0.5\textwidth]{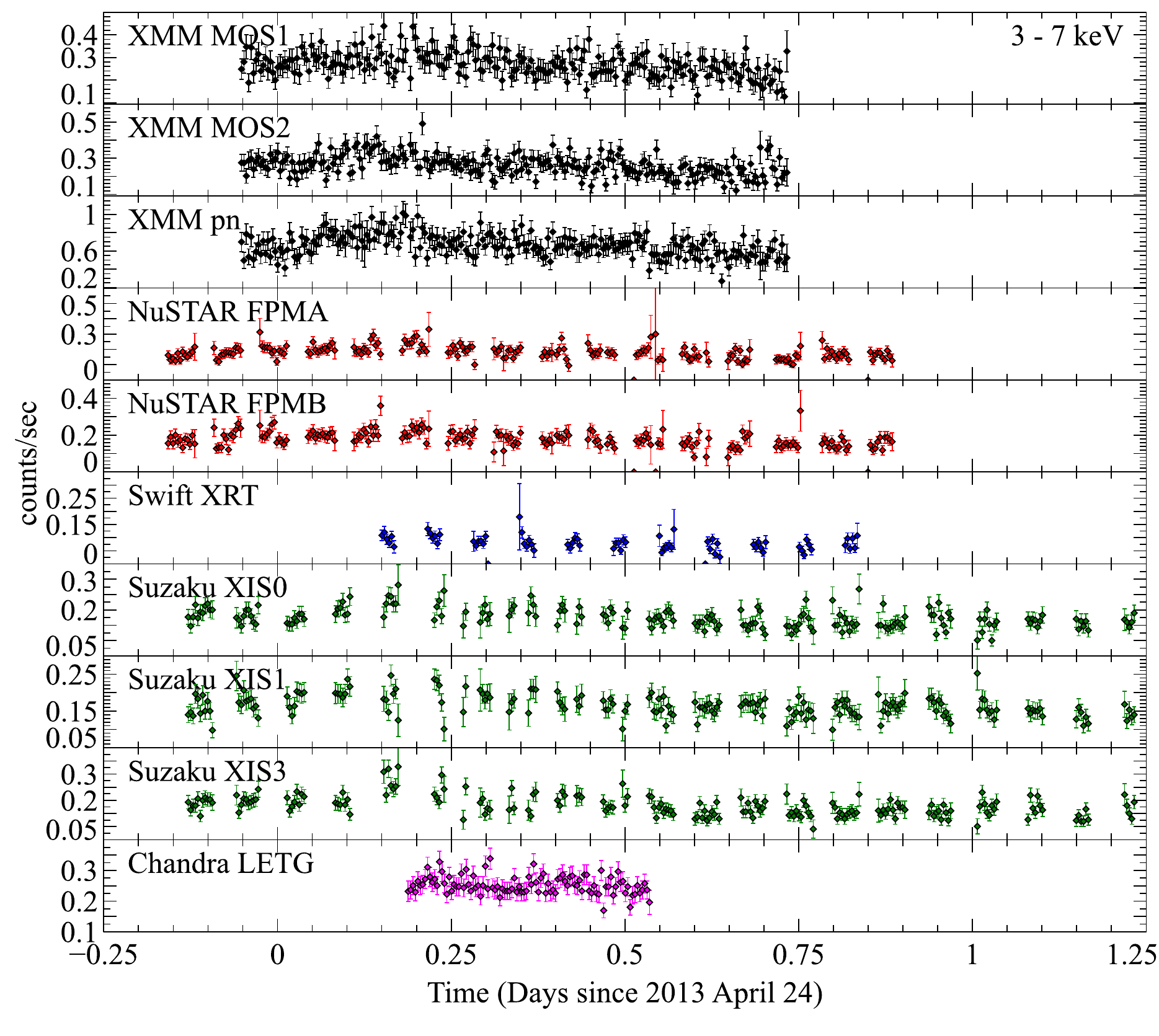} \\
\caption{Same as for Figure \ref{lightcurve3c273} but for \pksb.}
\label{lightcurvepks2155}
\end{figure*}

\subsection{Fitting Procedure}
We performed all fits with XSPEC \citep{Arnaud1996} version 12.8.2 and use Cash C-statistics \citep{Cash1979} (\texttt{cstat} in XSPEC) as implemented in XSPEC, because of the bias associated with $\chi^2$-statistics \citep{Nousek1989,Humphrey2009}. A discussion on the benefits of Cash C-statistics and the bias can be found in the XSPEC manual\footnote{https://heasarc.gsfc.nasa.gov/xanadu/xspec/manual/XSappendixStatistics.html}. We ensure that every bin has at least one count, since bins with zero counts can lead to erroneous results in XSPEC. To prove the goodness of fit we binned the spectra after fitting to a minimum of 30 counts and calculated the $\chi_\mathrm{red}^2$ on the fit obtained with Cash statistics. All errors are 90\% confidence unless otherwise stated.

Each observatory, instrument, and source was first fitted independently using an absorbed power-law model, \texttt{cflux $\times$ tbabs $\times$ pow} (in XSPEC). We used Wilms abundances \citep{Wilms2000} and Verner cross-sections \citep{Verner1996} and fixed the column to the Galactic values \nh=1.79 $\times 10^{20}$ cm$^{-2}$ for 3C\,273 and \nh=1.42 $\times 10^{20}$ cm$^{-2}$ for PKS2155-304 \citep{Dickey1990}. Tests letting \nh\ float resulted in inconsistent and unlikely column values (\nh\ tended towards 0 for both sources). The value of \nh\ is degenerate with slope for the soft instruments and sensitive to their low energy calibration, but allowing the \nh\ to vary within previously measured realistic values, revealed that the changes this caused to the slope were smaller than the measurement error on the slope itself. Fixing \nh\ to the same value for all observatories therefore introduces minimal biases on slope and flux, mainly because at such low \nh\ values the absorption is negligible above $\sim$2\,keV, and we do not fit below 1\,keV.

We also tested the dependence of the flux and slope for different choices of abundance and/or cross-section, as well as choice of photoelectric absorption model. For abundances we tested \citet{Anders1989} and \citet{Lodders2009}, for cross-sections \citet{Balucinska1992}, and for photoelectric model \texttt{tbabs} and \texttt{tbnew}\footnote{http://pulsar.sternwarte.uni-erlangen.de/wilms/research/tbabs/}. Using the EPIC and XIS spectra as reference, and fitting them in their nominal calibrated energy bandpass, the \qcs\ results are essentially indistinguishable, while the fluxes/slope/column densities from the \pksb\ spectral analysis are minimally affected by less than a fraction of percent, 0.01, and $10^{19}$\,cm$^{-2}$, respectively. 

We applied the XSPEC convolution model \texttt{cflux} to fit the flux instead of the power-law normalization, as this directly gives us the absorbed flux, which is the value we are interested in comparing between instruments. We therefore do not report on the normalization or the intrinsic flux of the power-law. 

We use two energy bands for calculating the fitted flux. For the low-energy observatories, \chandra, \suzaku, \swift, and \xmm\, we used 1--5\,keV when paired with each other, and 3--7\,keV when paired with \nustar. These energy ranges were chosen for several reasons. First, \nustar\ is not well calibrated below 3\,keV, which restricted the lower energy range. Second, to ensure that all spectra were of good quality and well above background, 7\,keV was set at the upper limit. Finally, the lower range of 1\,keV was picked to avoid complications with detector contamination layers on the CCD instruments. For \suzaku\ HXD and \nustar\ we used 20--40\,keV. 

We carefully examined the highest signal to noise observations from \suzaku\ and \xmm\ to determine if there is measurable spectral curvature between 1--10 keV in either source. Instead of a simple power-law, we tried fitting with a log-parabolic power-law (\texttt{logpar}) used for blazars, a broken power-law (\texttt{bknpowlw}), and a cutoff power-law (\texttt{cutpowerlw}) used for accretion-dominated sources. The improvements in the fits were not significant for either source. There are changes in the spectral index with different choice in upper and lower energy bound within a single instrument, but the changes are not systematically the same for different observatories and are therefore most likely due to statistical fluctuations or instrument specific calibration.

\section{Light Curves}\label{sectionlightcurves}
We extracted light curves for all instruments in 5 minute bins for the two non-overlapping energy bands: 1--3\,keV (\nustar\ excluded) and 3--7\,keV. The light curves are shown in Figures \ref{lightcurve3c273} and \ref{lightcurvepks2155} for the respective bands and two sources. Neither show rapid variability, and the few large outliers, which can be observed adjacent to the occultation/SAA gaps for the low-Earth orbit observatories, are either small fractional flux bins or SAA periods that were not entirely filtered out, and they do not influence the cross-calibration results.

For both sources, the flux follows the same shape in the low and high band with no indication of rapid variability between gaps. We are therefore confident that including the occulted periods for \chandra\ and \xmm, or between \nustar\, \suzaku\, and \swift\, for non-overlapping periods, should not affect the flux measurements.

\section{Cross-calibration Results}

\begin{figure*}
\includegraphics[width=0.30\textwidth]{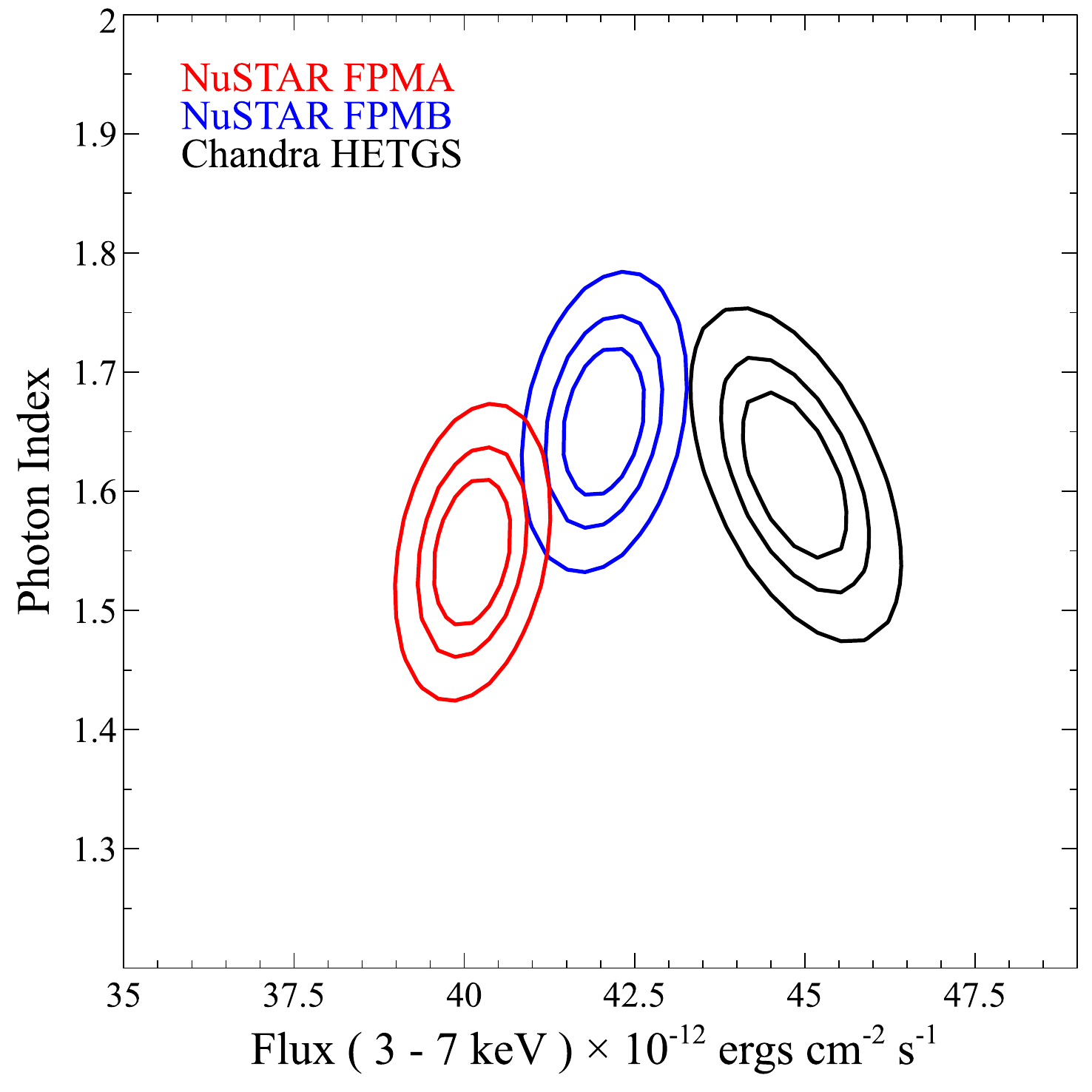}
\includegraphics[width=0.30\textwidth]{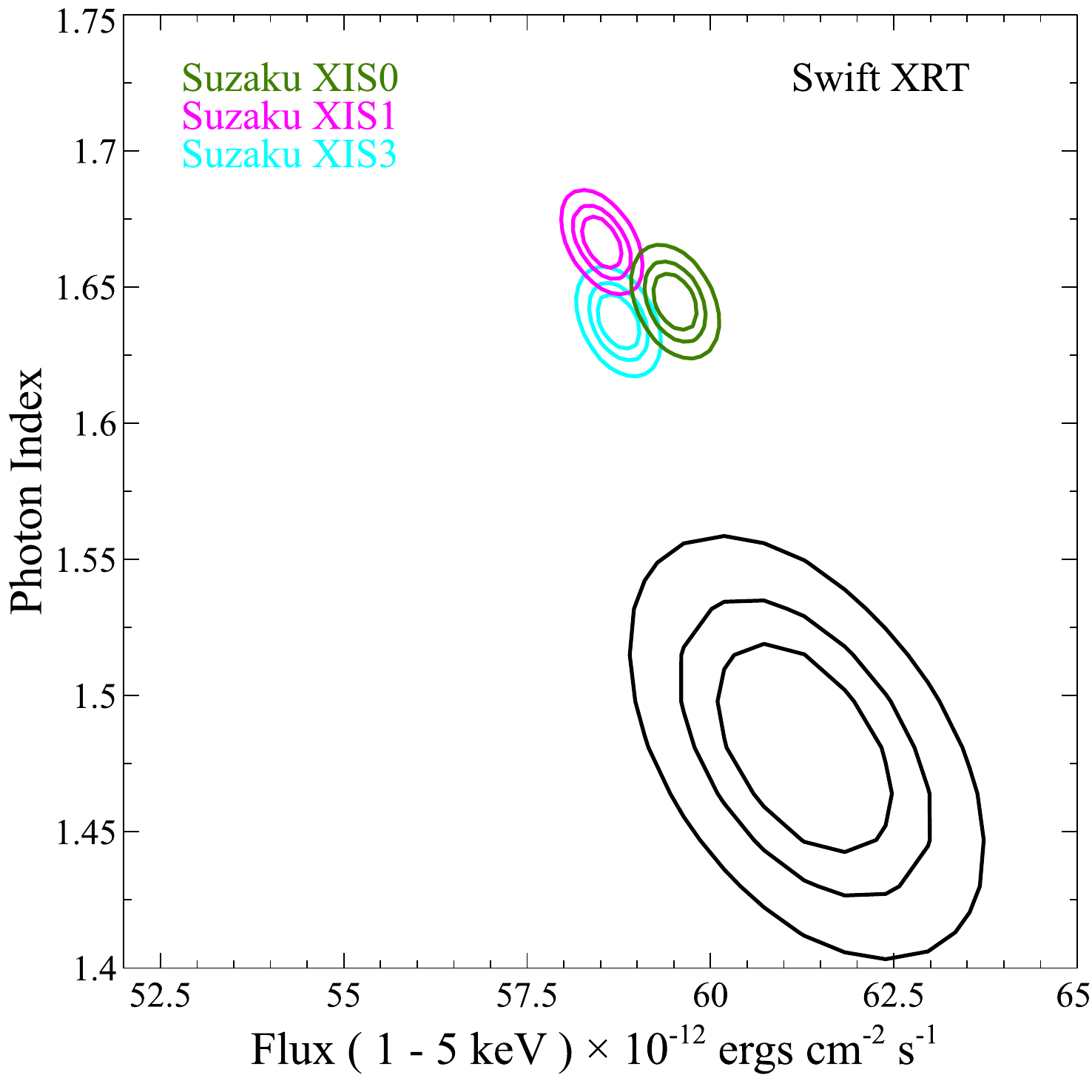}
\includegraphics[width=0.30\textwidth]{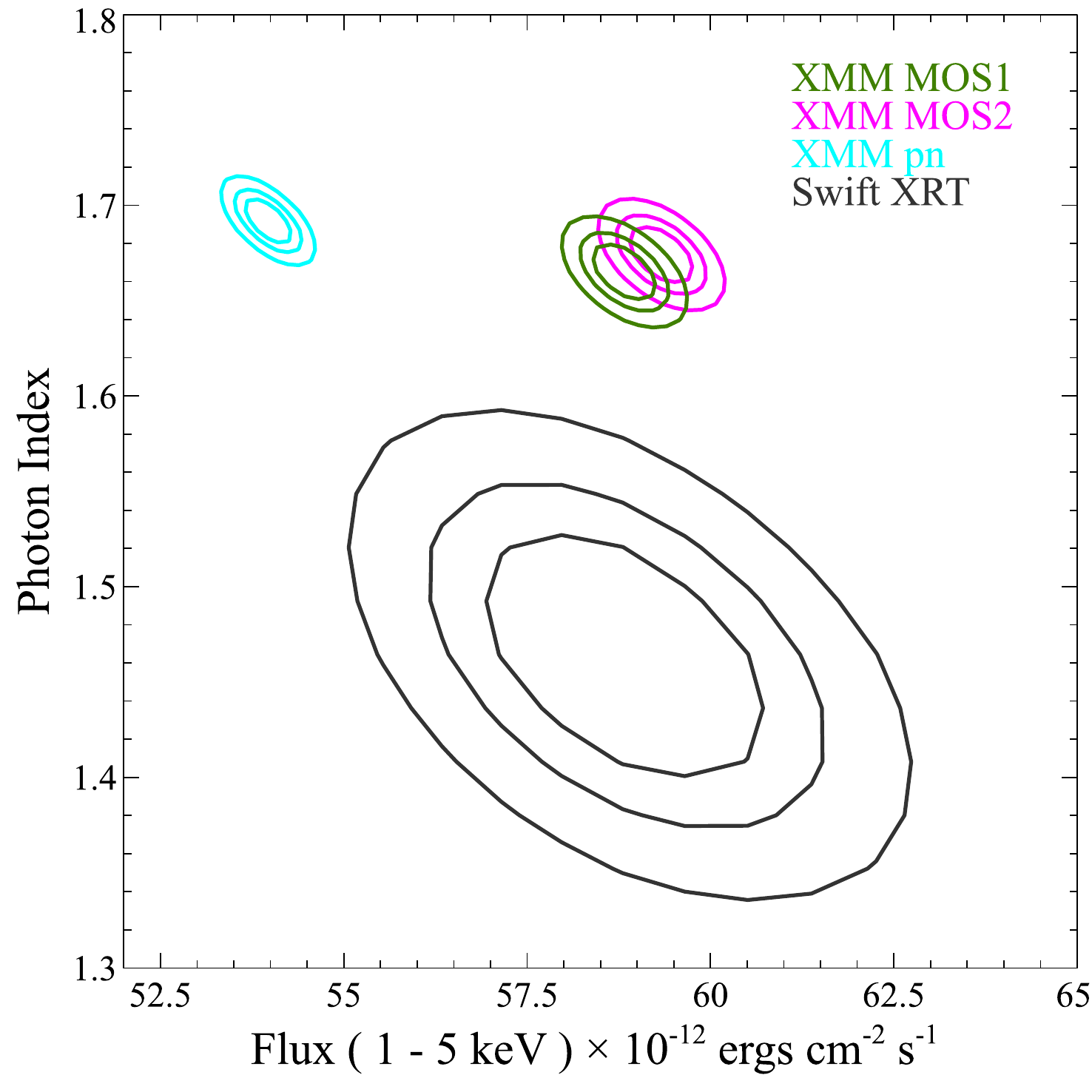}\\
\includegraphics[width=0.30\textwidth]{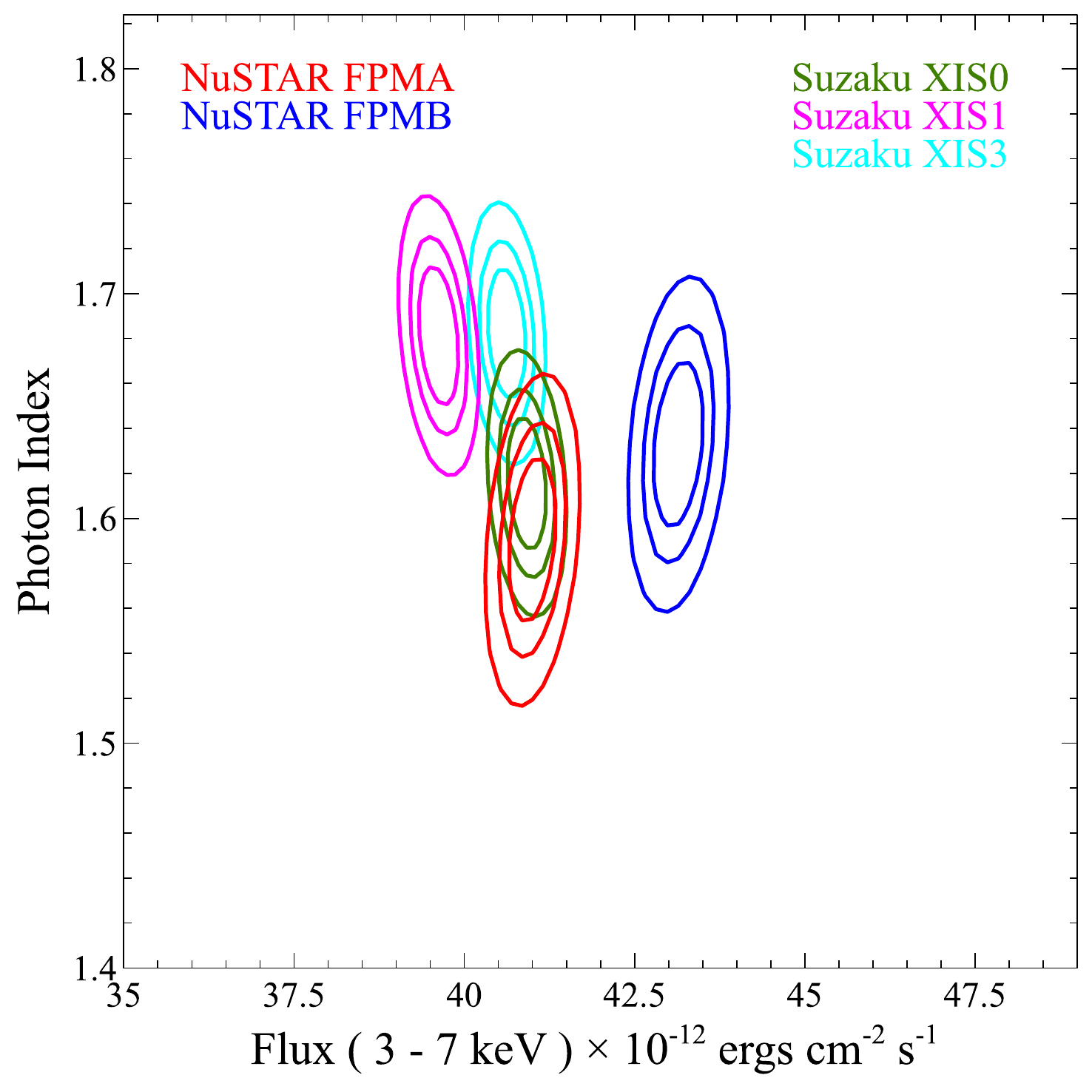}
\includegraphics[width=0.30\textwidth]{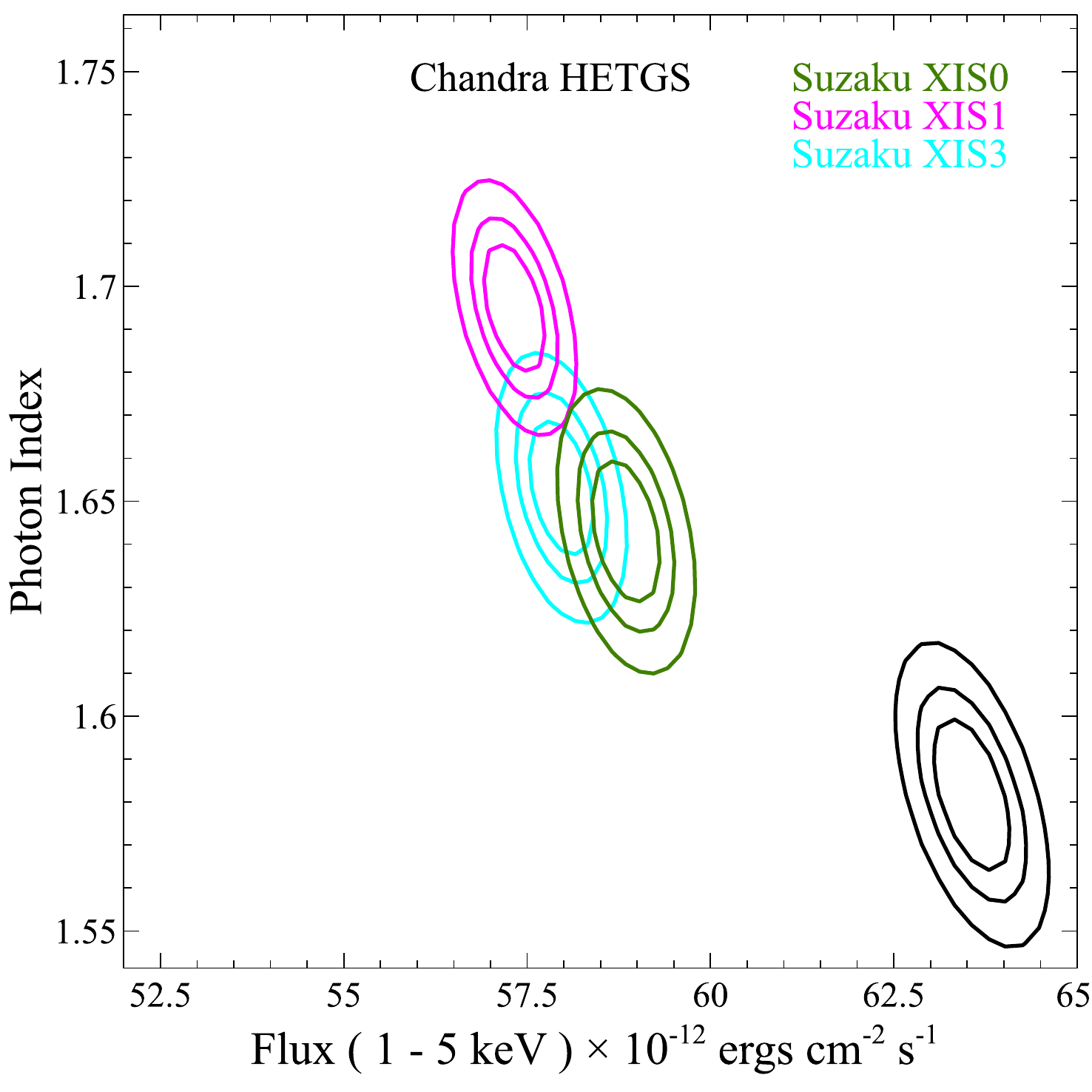}
\includegraphics[width=0.30\textwidth]{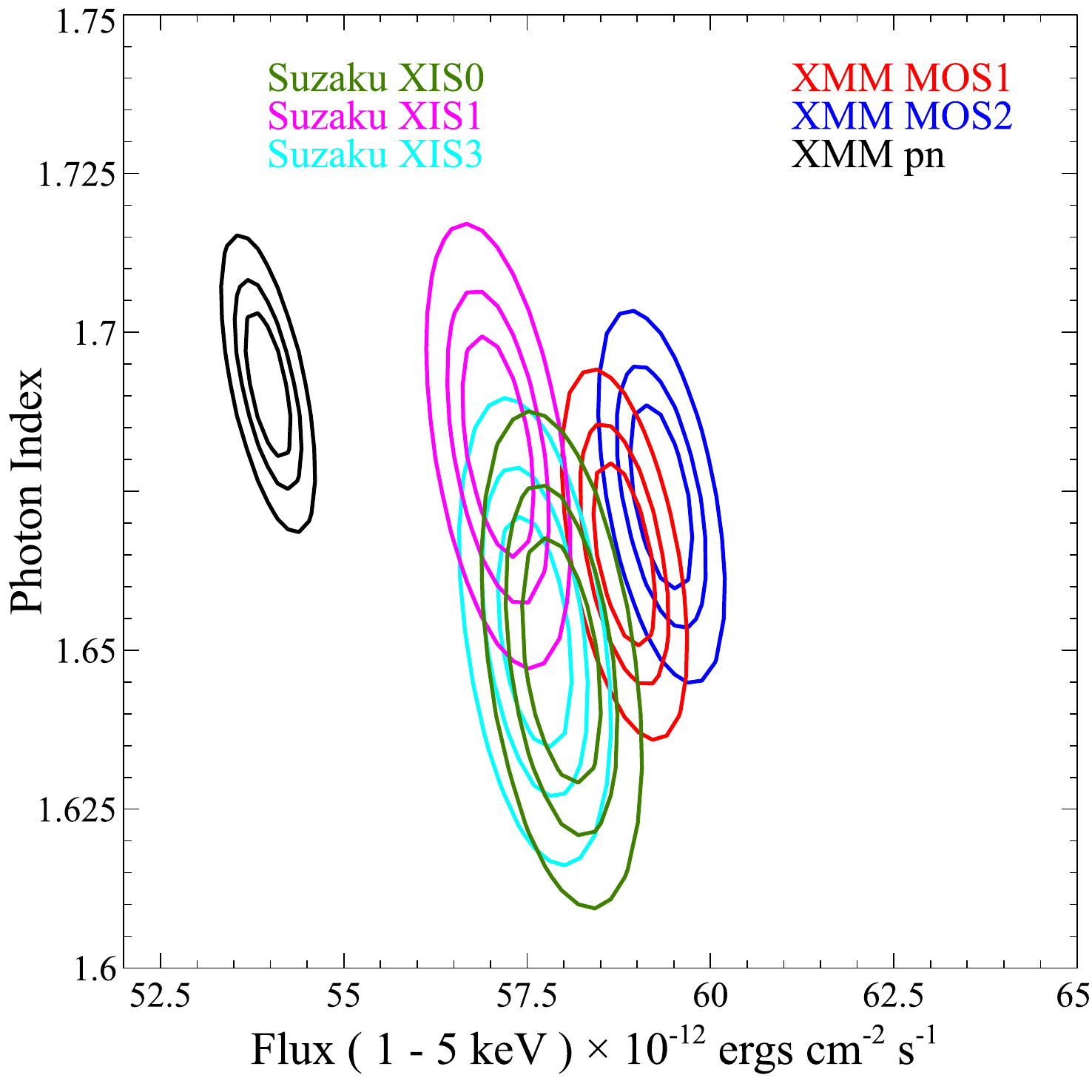}\\
\includegraphics[width=0.30\textwidth]{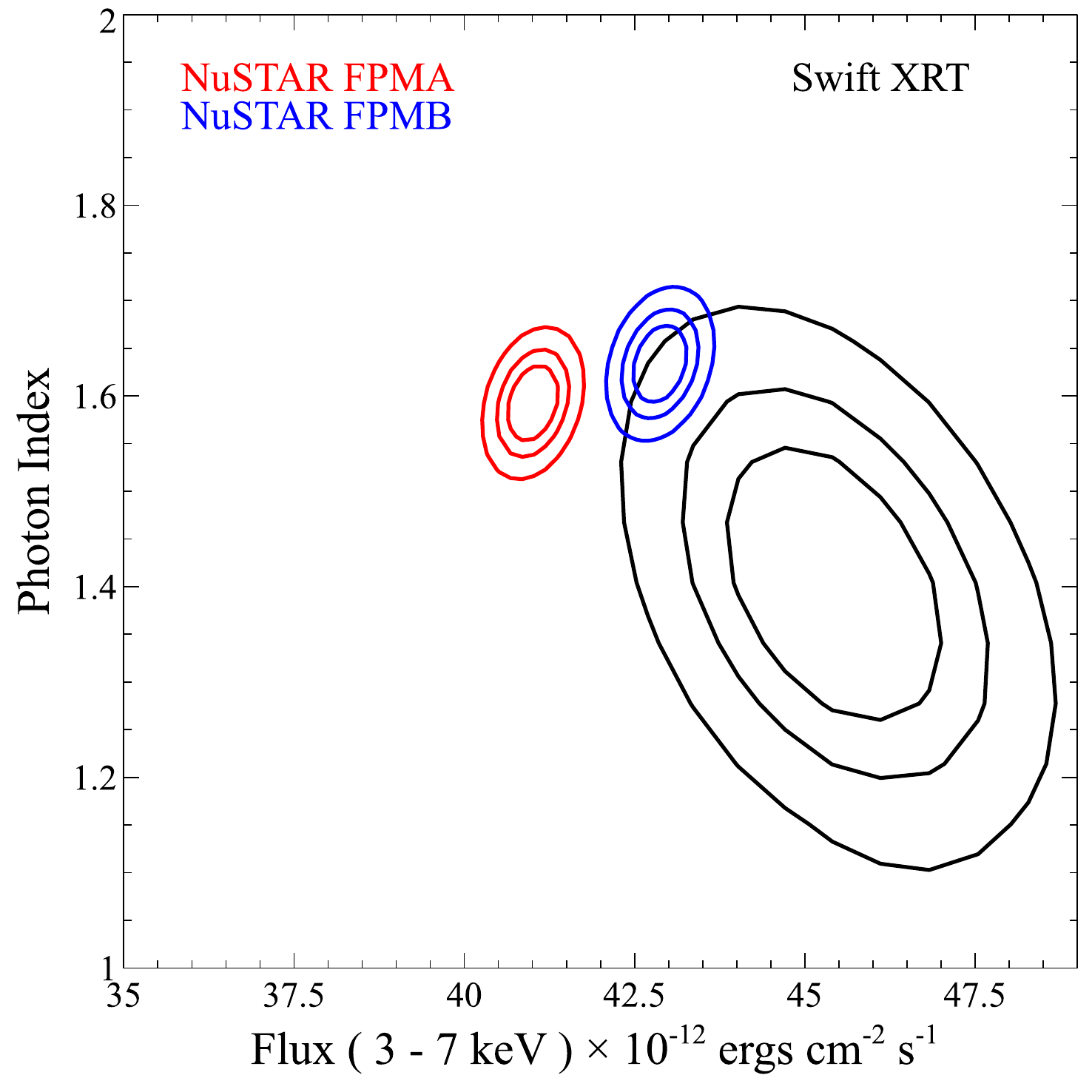}
\includegraphics[width=0.30\textwidth]{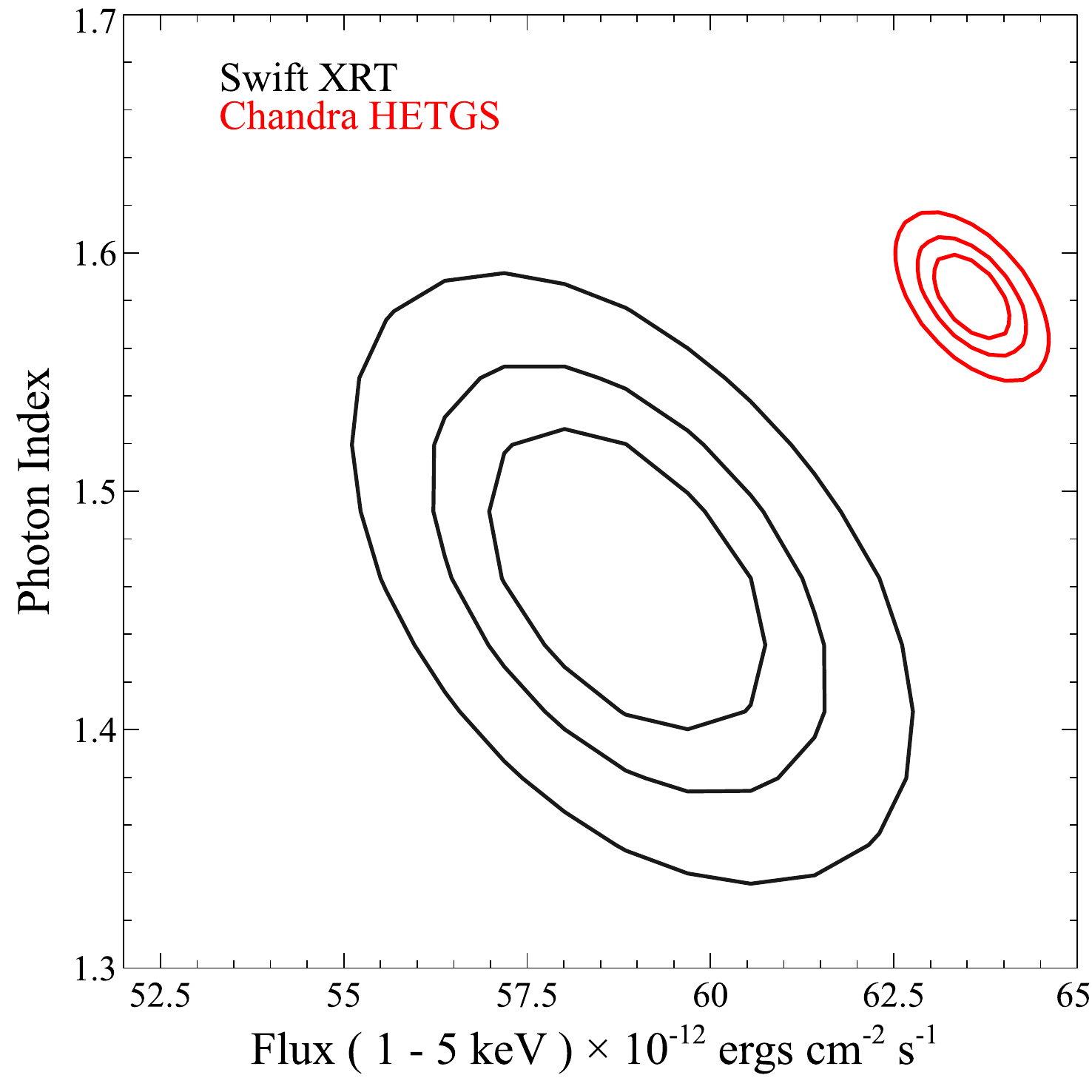}
\includegraphics[width=0.30\textwidth]{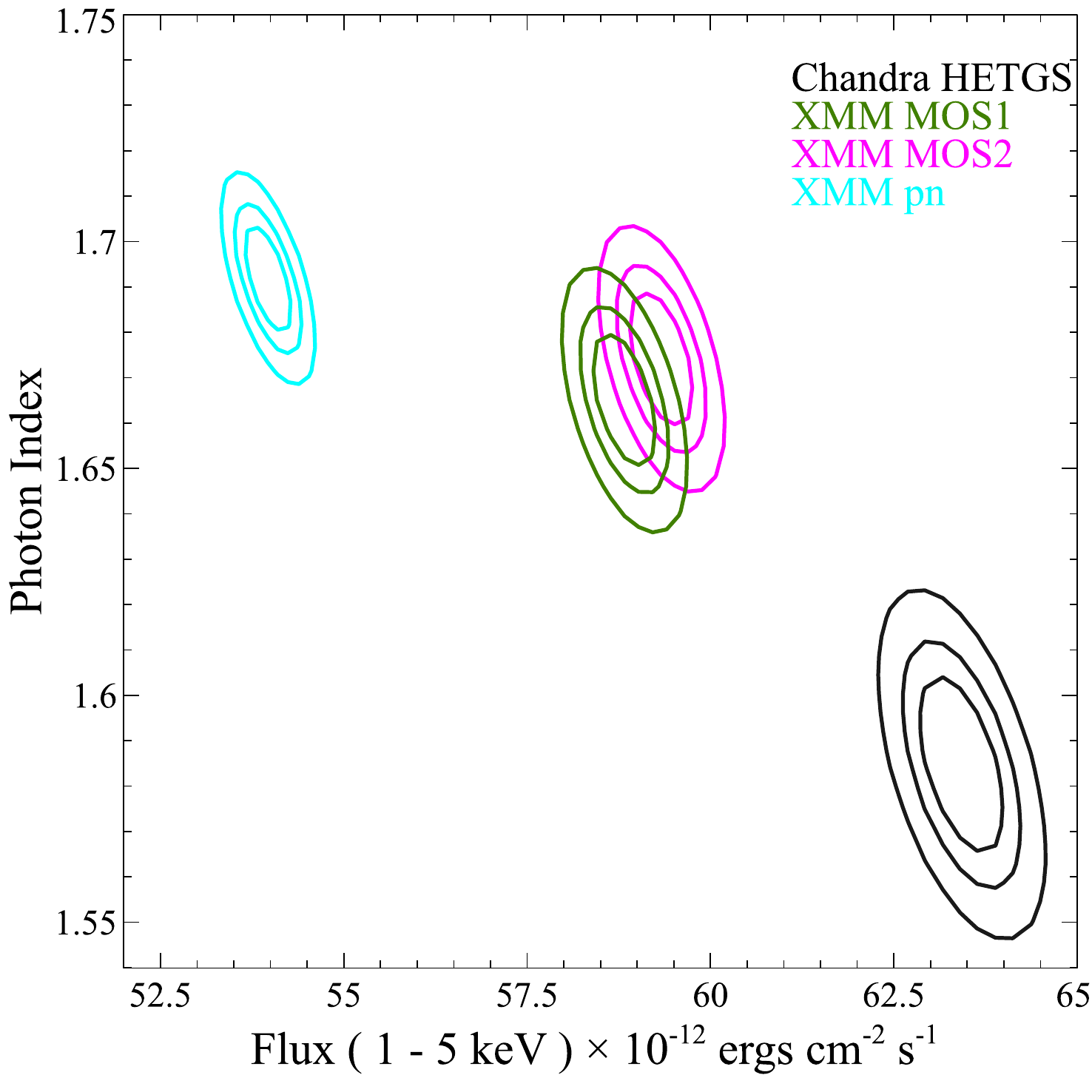}\\
\includegraphics[width=0.30\textwidth]{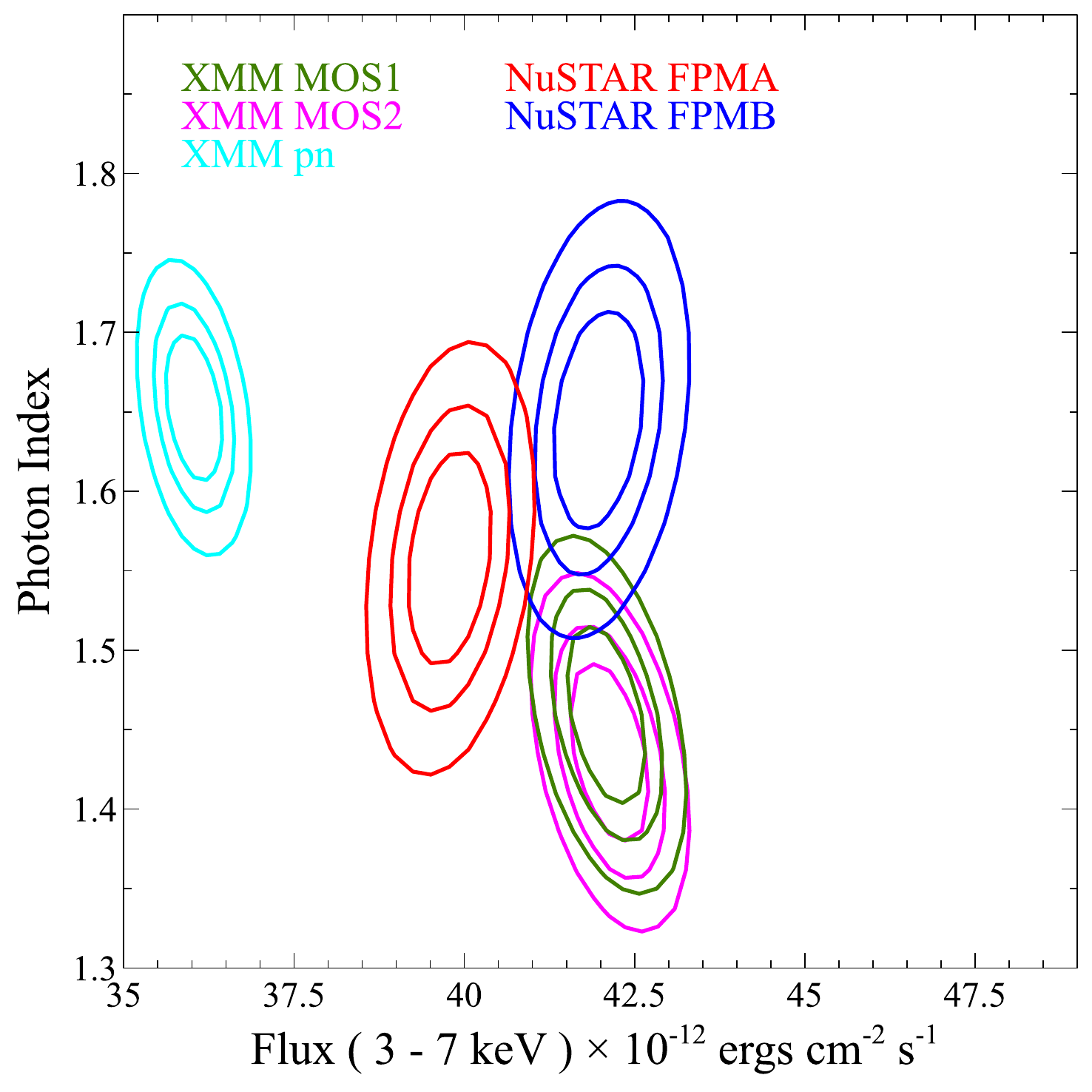}
\includegraphics[width=0.30\textwidth]{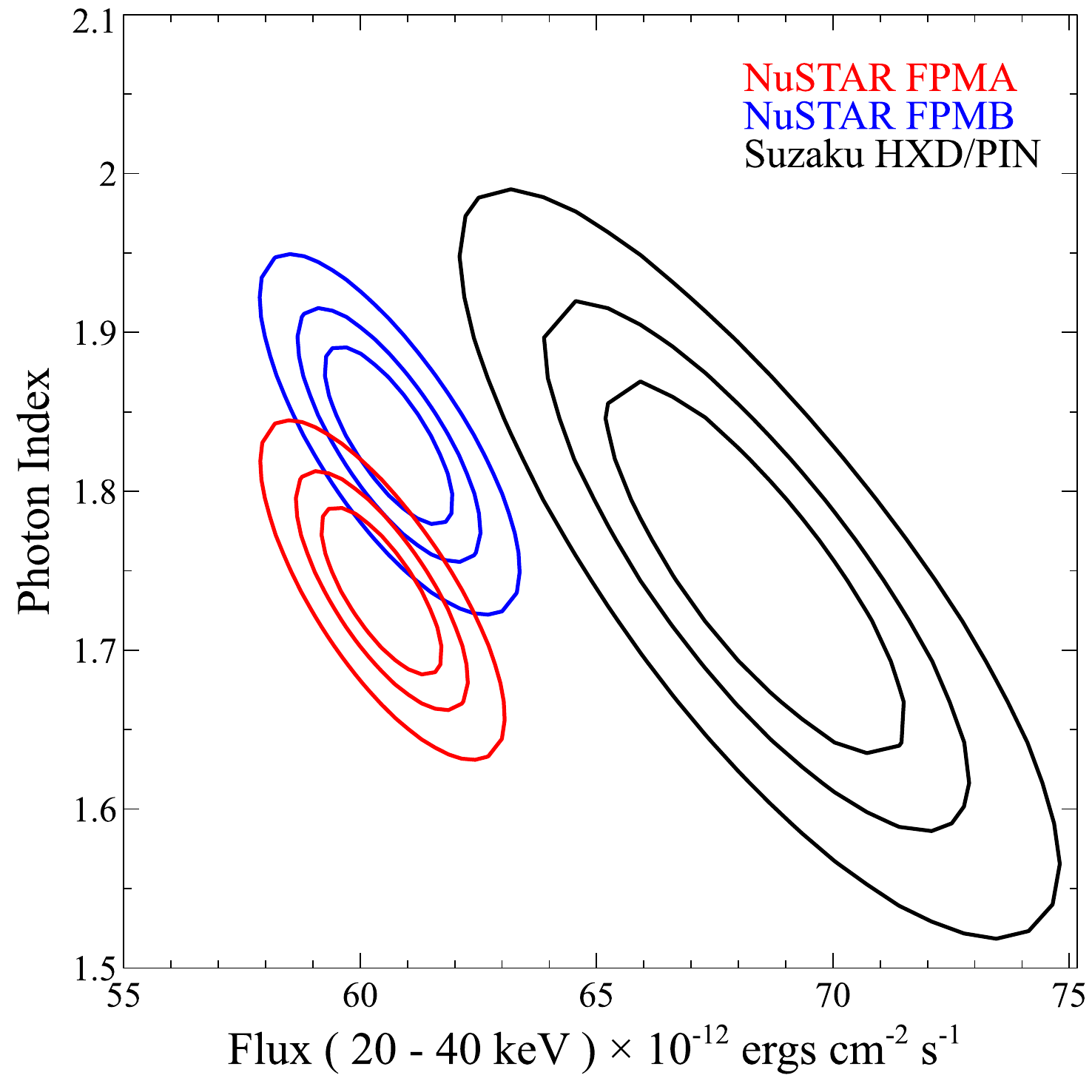}\\
\caption{Confidence contours for 3C\,273. The flux axes for the two energy bands 1--5\,keV and 3--7\,keV have been aligned, but not for the slope axis, $\Gamma$.}
\label{qccontours}
\end{figure*}

\begin{figure*}
\includegraphics[width=0.30\textwidth]{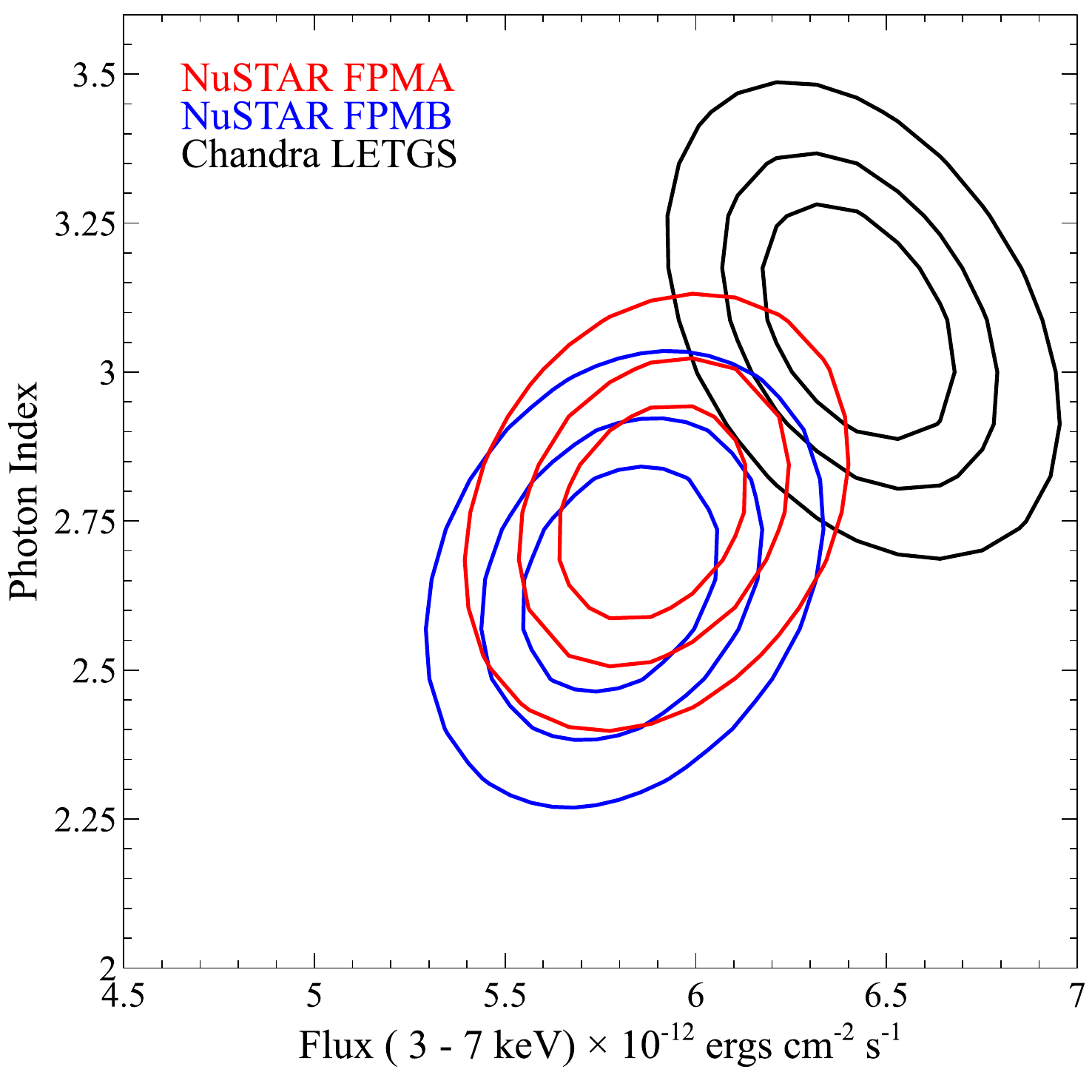}
\includegraphics[width=0.30\textwidth]{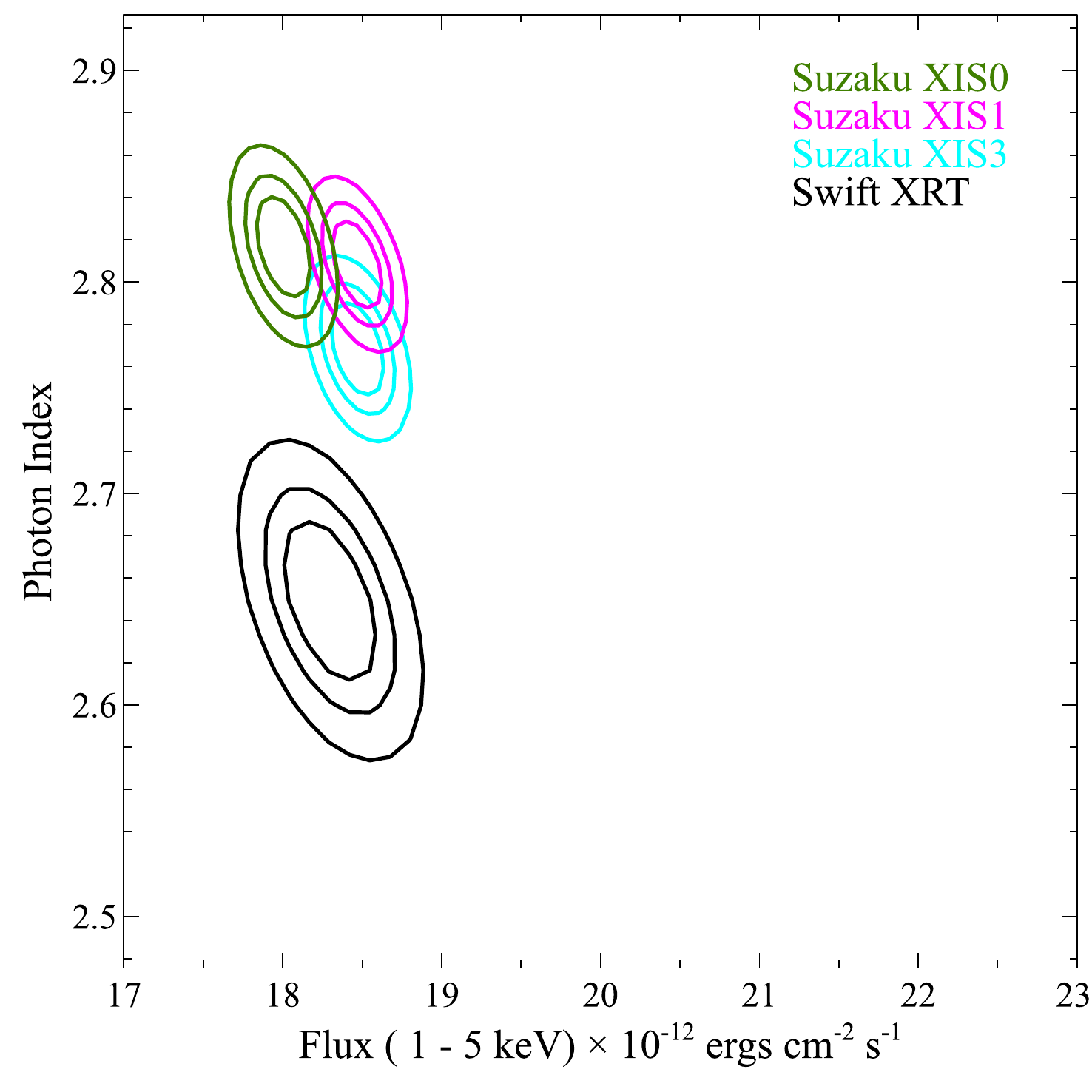}
\includegraphics[width=0.30\textwidth]{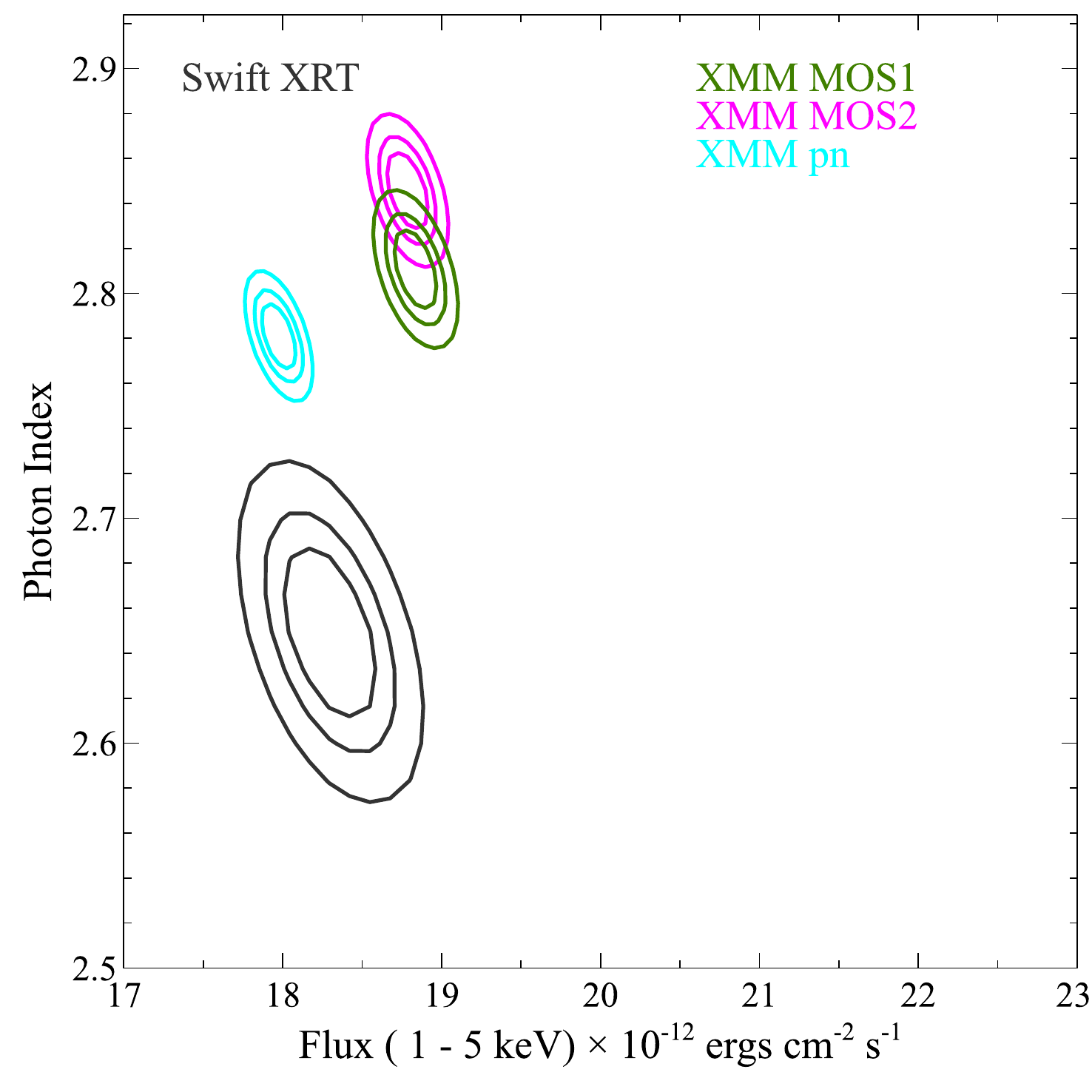}\\
\includegraphics[width=0.30\textwidth]{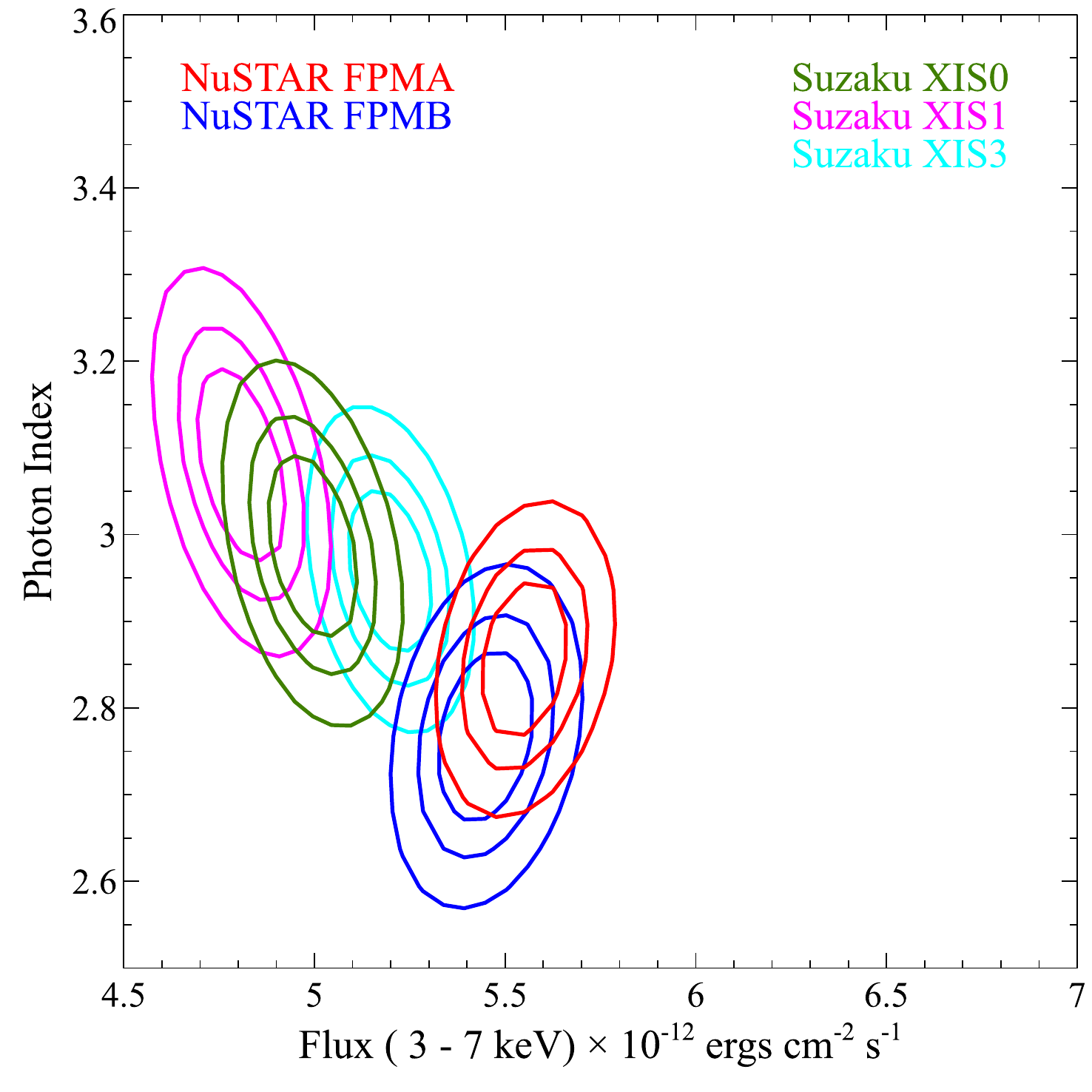}
\includegraphics[width=0.30\textwidth]{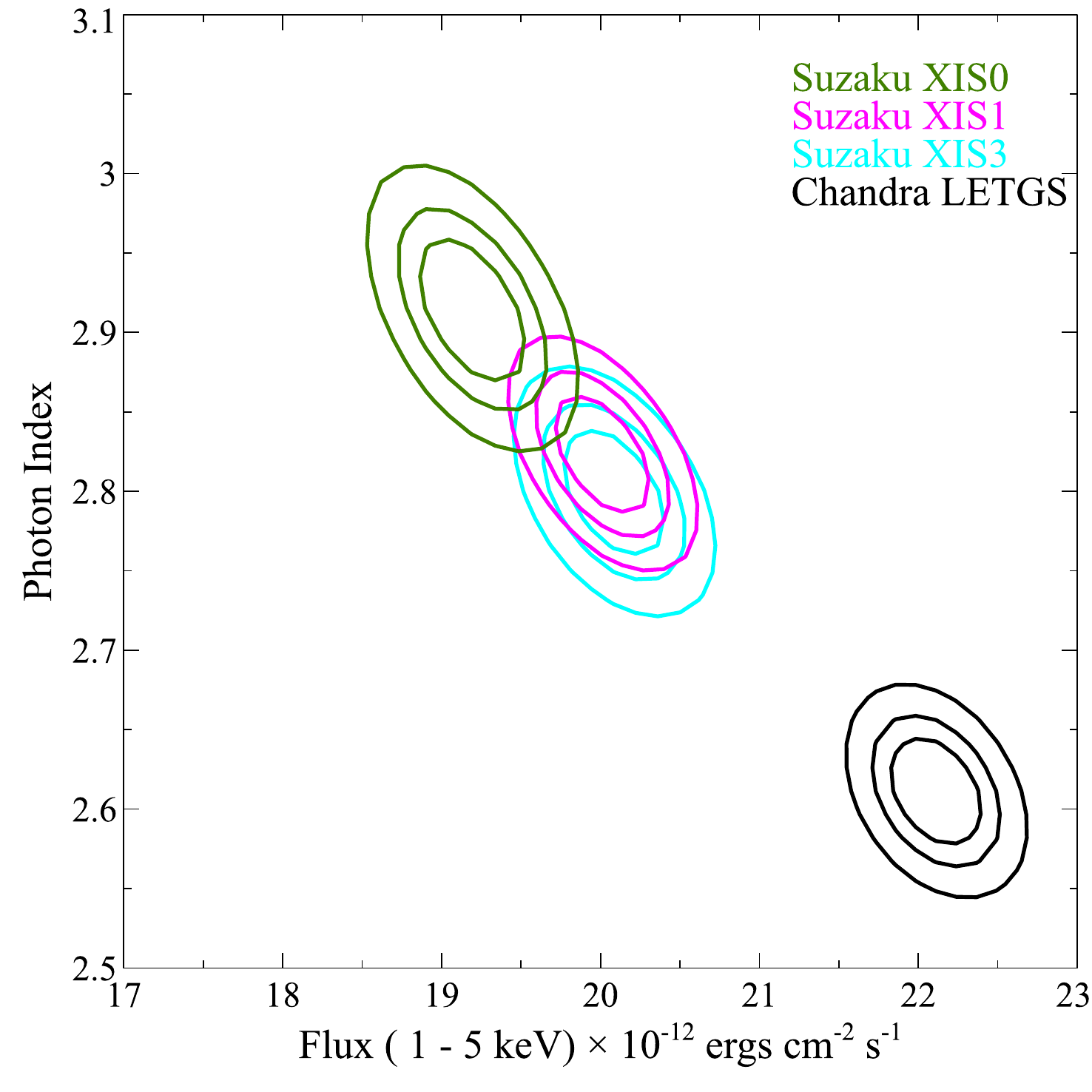}
\includegraphics[width=0.30\textwidth]{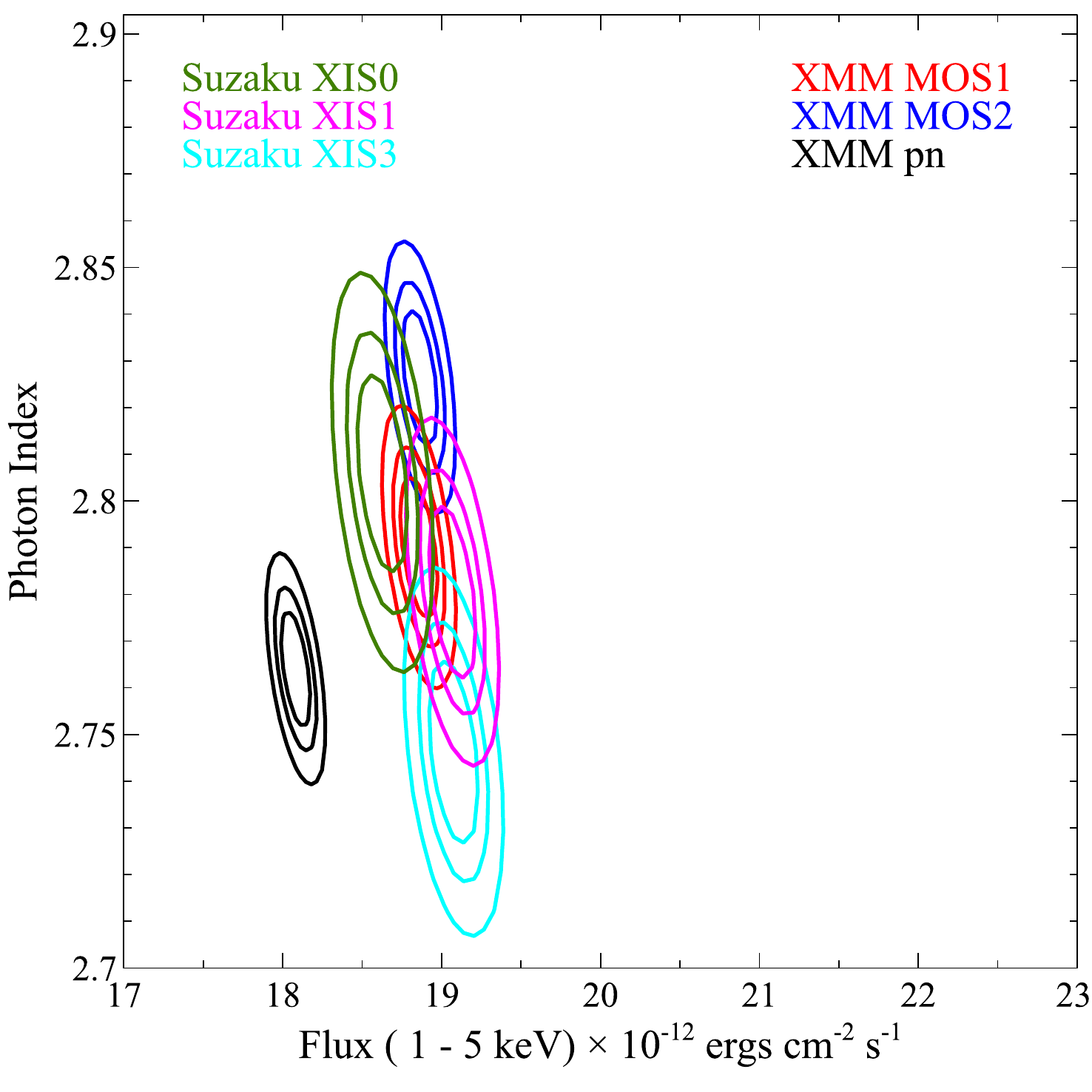}\\
\includegraphics[width=0.30\textwidth]{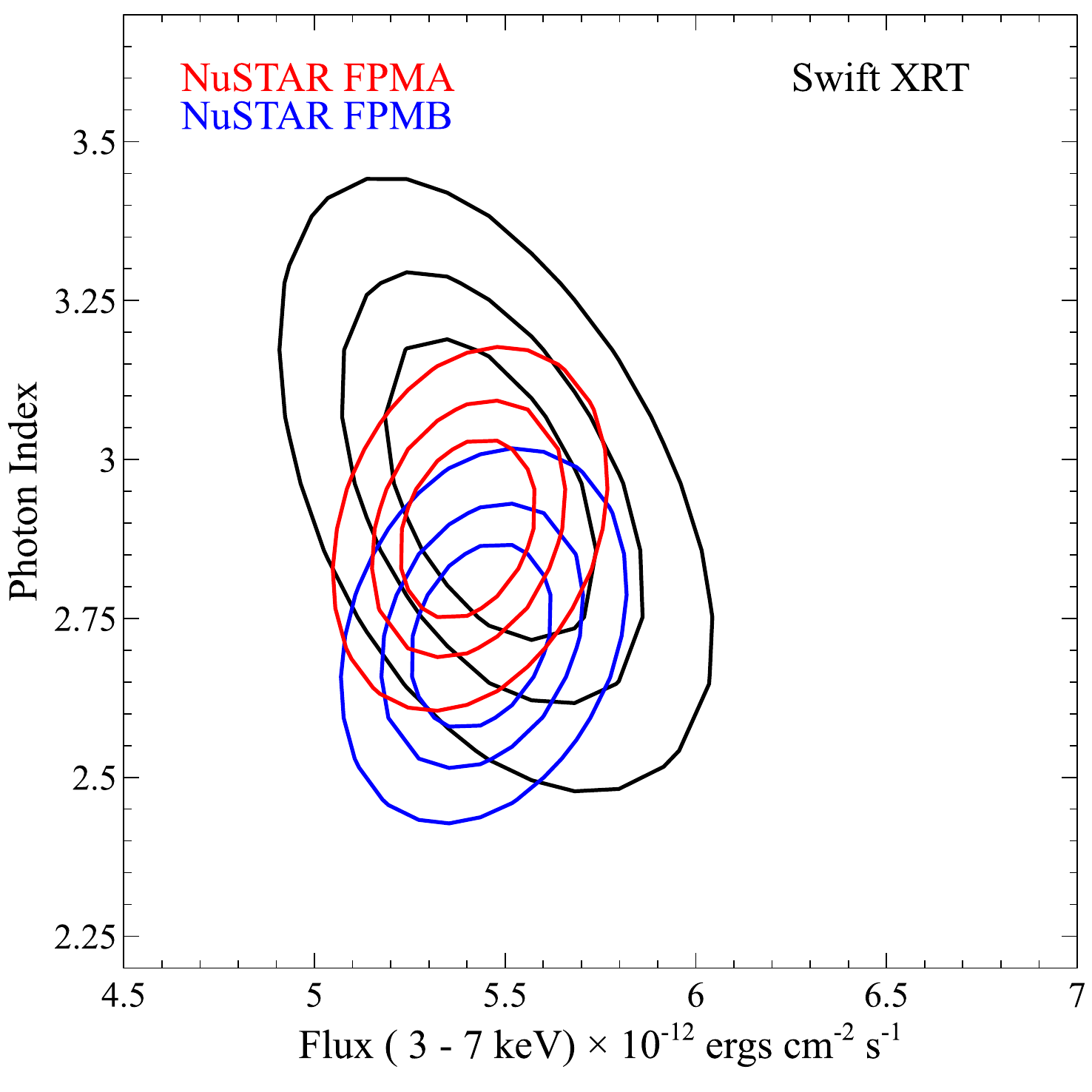}
\includegraphics[width=0.30\textwidth]{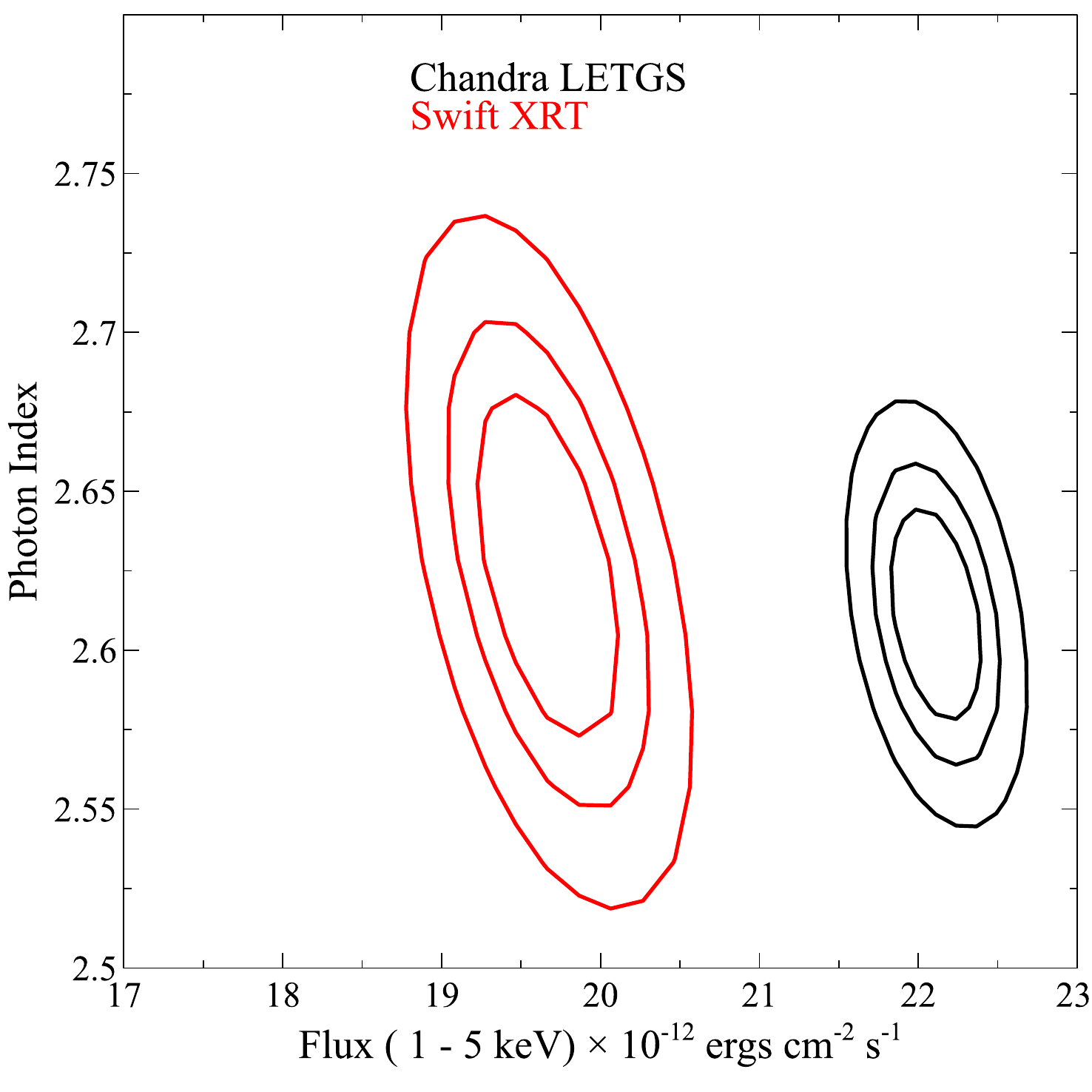}
\includegraphics[width=0.30\textwidth]{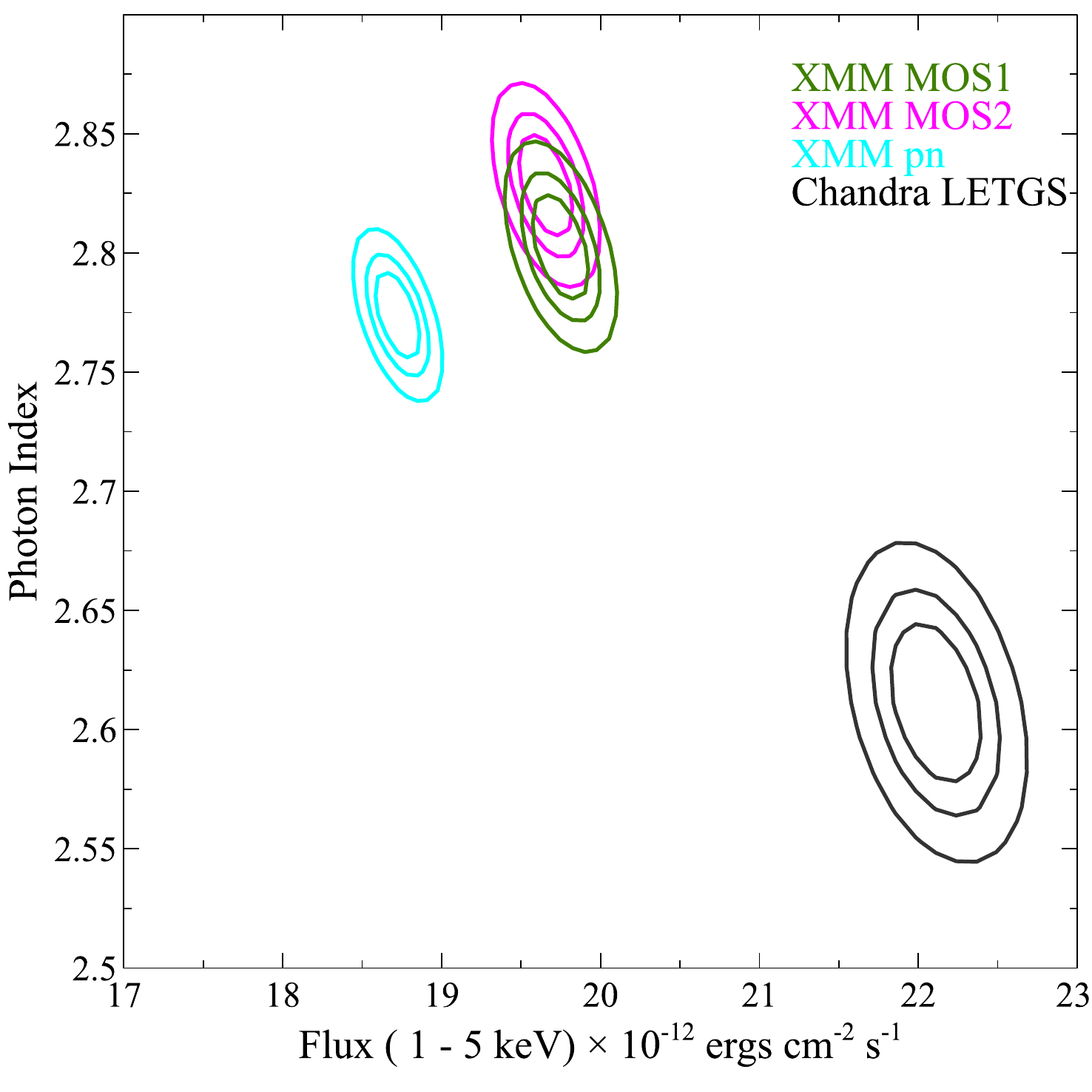}\\
\includegraphics[width=0.30\textwidth]{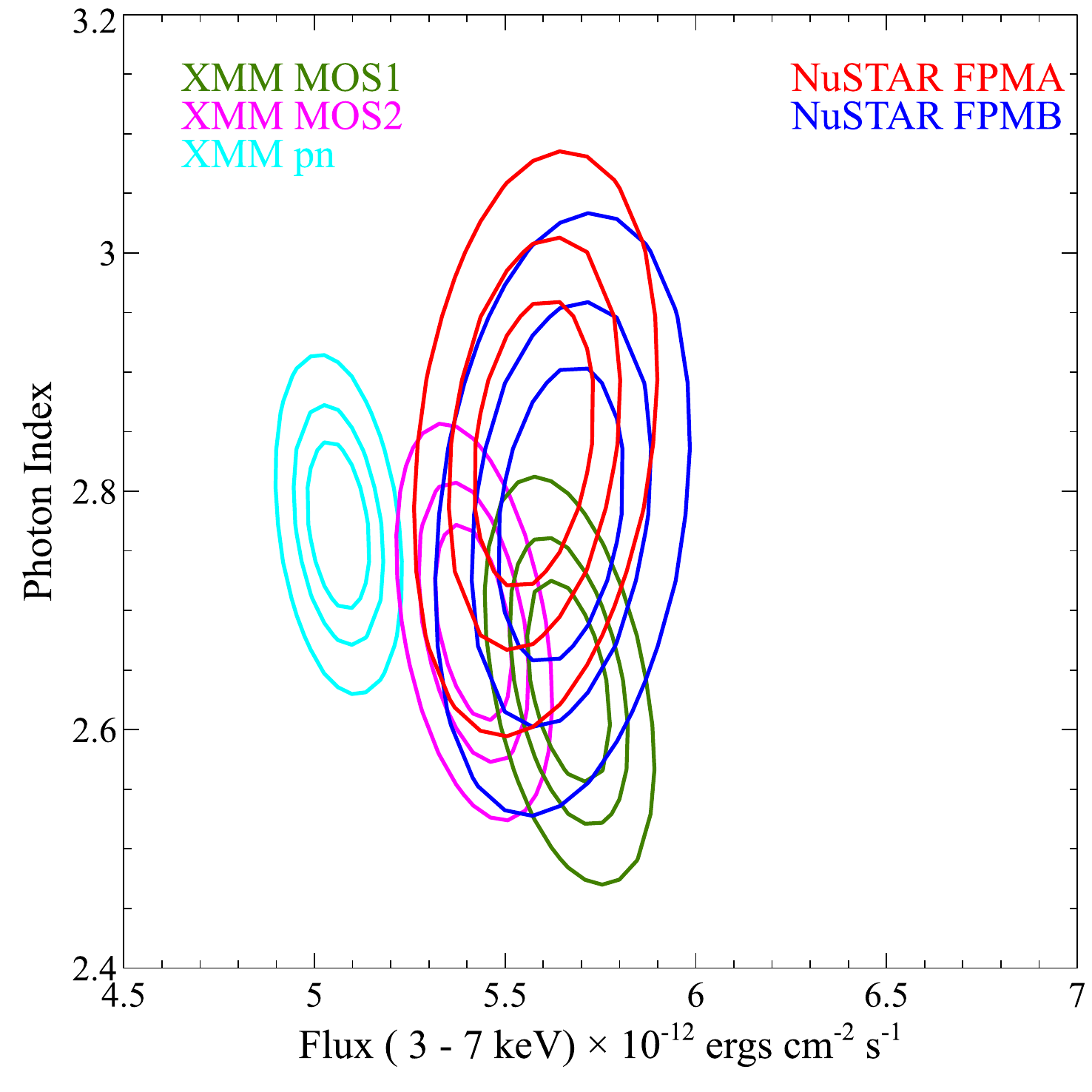}\\
\caption{Confidence contours for PKS2155-304. The flux axes for the two energy bands 1--5\,keV and 3--7\,keV have been aligned, but not for the slope axis, $\Gamma$.}
\label{pkscontours}
\end{figure*}

Tables \ref{qcstable} and \ref{pkstable} summarize the results of the individual fits, and Figures \ref{qccontours} and \ref{pkscontours} show the confidence contours (1, 2, and 3$\sigma$) of flux and power-law index of the individual fits to each observatory pair.

The individual fits aim to quantify the question of how different the observatories are from each other. There is some degeneracy between flux and slope and assuming a fixed cross-normalization factor in multi-instrument spectral fitting is therefore not recommended. Rather cross-normalization constants should be allowed to float and their values subsequently evaluated against Tables \ref{fluxtable} (1--5\,keV) and \ref{fluxtable_2} (3--7\,keV and 20--40\,keV), which summarize the inter-instrumental and observatory relative flux ratios, and/or in conjunction with Figures \ref{qccontours} and \ref{pkscontours}. We stress that the ratios are derived from the fluxes in the specific bands as given in Tables \ref{qcstable} and \ref{pkstable}, and that they only ensure the flux calculated in this restricted band should agree within reasonable uncertainties between observatories. They do not correct for differences in slope, and deviations may occur if applied outside the specified energy range. 

\begin{figure*}
\begin{center}
\includegraphics[width=1.0\textwidth]{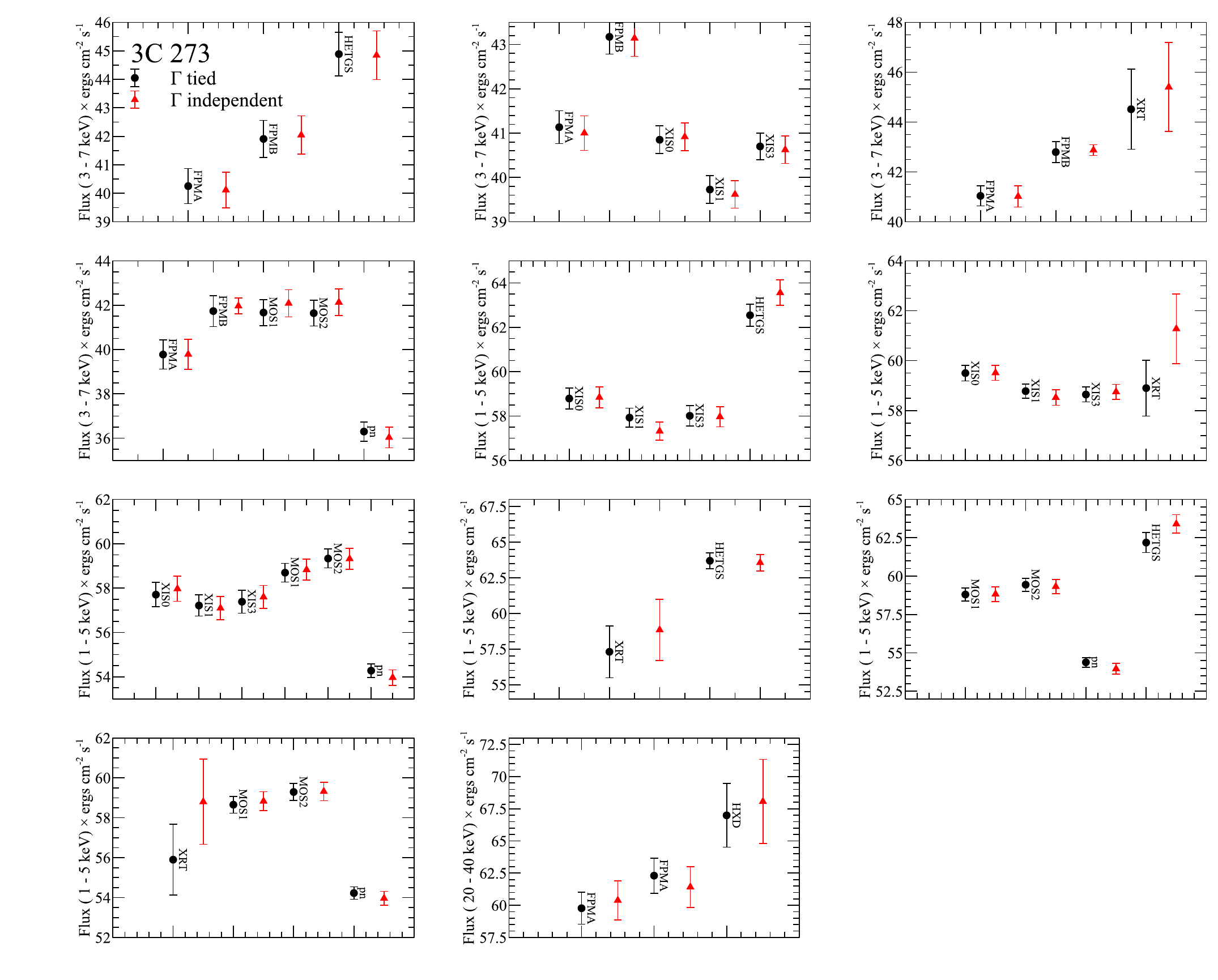}
\end{center}
\caption{Flux comparison between instruments when the slope, $\Gamma$, is tied (black circles) and left free to fit independently (red triangles) for \qcs. In most cases the measured flux is the same within errors, but the spectrum with the lowest statistics does run a risk of being modified by the high statistic spectrum.}
\label{fluxcompareqcs}
\end{figure*}

\begin{figure*}
\begin{center}
\includegraphics[width=1.0\textwidth]{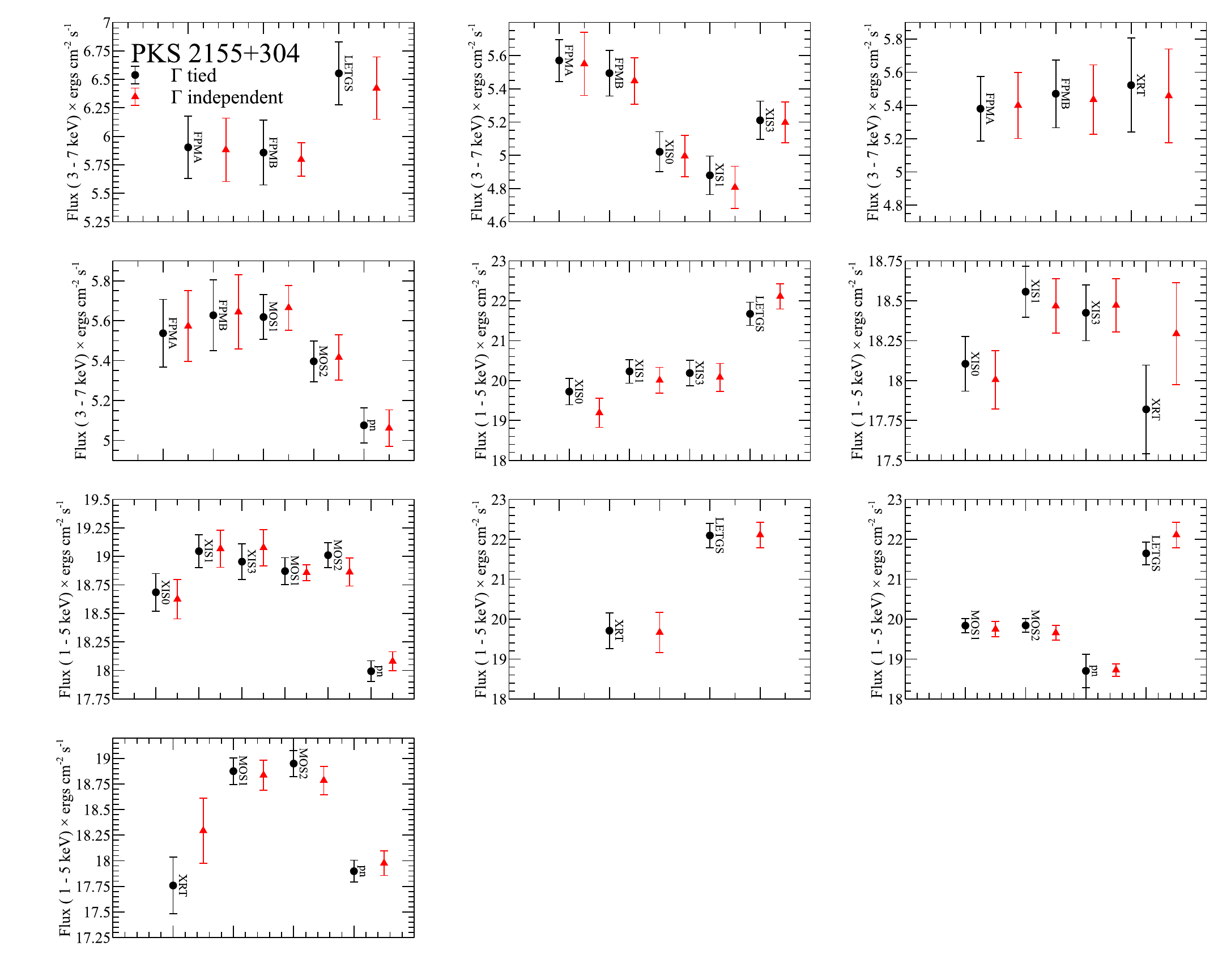}
\end{center}
\caption{Flux comparison between instruments when the slope, $\Gamma$, is tied (black circles) and left free to fit independently (red triangles) for \pksb. In most cases the measured flux is the same within errors, but the spectrum with the lowest statistics does run a risk of being modified by the high statistic spectrum.}
\label{fluxcomparepks}
\end{figure*}

In the individual fits, presented in Tables \ref{fluxtable} and \ref{fluxtable_2}, the source parameters, $\Gamma$ and flux, between instruments take on different values because of calibration. In many cases, however, a source spectrum can not be constrained by one instrument alone and it is necessary to assume that source parameters are the same for all involved instruments and adjust for the differences between observatories with just one constant (\texttt{Constant*Model}). We show in Figure \ref{fluxcompareqcs} and \ref{fluxcomparepks} the difference in calculated fluxes when tying the model slopes, $\Gamma$, together and when allowing them to take on their optimal value from the individual fits in Tables \ref{qcstable} and \ref{pkstable}. Fortunately, there is not much difference in the derived flux, but in the cases where the measured slopes are significantly different, as for \swift/XRT compared to the other instruments, the measured flux of the lowest count rate spectrum gets modified. The exact cross-normalization constant (\texttt{Constant}) between two instruments is therefore sensitive to the relative quality of the spectra, ascribing more influence on the parameters by the higher quality spectrum.

To gain more insight into the observed fluxes and spectral differences between observatories, we calculated fluxes in smaller energy bands tailored to each instrument. For each observatory pair, we used the full valid fitting range for each instrument, but did not go below 1 keV to avoid issues with detector contamination. We then found the best fit to a power-law for the pair leaving all parameters tied and used the model to unfold the spectra and calculate fluxes of the instruments relative to the model. The relative flux ratios for each instrument pair as a function of energy are shown in Figures \ref{qcratios} and \ref{pksratios}. There is some uncertainty in the shapes simply due to statistics, but there are a couple of recognizable overall trends that we discuss in detail below.

\subsection{Internal Observatory Calibration}
In this section we discuss the known features and limitations of the current instrumental calibrations for each observatory respectively.

\subsubsection{Chandra}
The \chandra\ HETGS and LETGS effective area calibrations depend on the effective area of the mirror system, the efficiencies of the gratings, the transmission of the ACIS optical blocking filter, the quantum efficiencies of the various ACIS CCDs, and the opacity of the contaminant on the ACIS filter.  The contaminant contributes significant uncertainty in the effective area only below 1 keV.  The mirror effective area was corrected for an overlayer of static contaminant as of CALDB version 4.1.1 and is estimated to be good to better than 5\% above 1 keV.  Using in-flight observations of blazars, the HEG and MEG efficiencies were reconciled to better than 3\% over the 0.8-5.0 keV range \citep{Marshall2012}, in CALDB version 4.4.6.1. The LETG $\pm 1$ order efficiencies have not been adjusted over the course of the mission but are cross-checked with regular HETGS observations of blazars, agreeing to better than 3\%.  Absolute calibration is estimated for both the HETGS and LETGS to be at about the 10\% level.

\subsubsection{NuSTAR}
The details on the \nustar\ calibration can be found in \citet{madsen2015b} and the section below should be regarded as a brief summary. 

For \pksb\ FPMA and FPMB fluxes are in agreement and the slopes overlap within errors. For \qcs\ there is a global $\sim 3\%$ offset in flux between the two modules with FPMB being higher than FPMA. This is a known offset, which can be between $\pm 5\%$  and is because the optical axis location on the detector is unique for every observation, but only known absolutely to $\sim30$\as. The uncertainty in its location means the vignetting functions could be off by $\sim30$\as\ and manifests itself as flux offsets and slight slope differences between FPMA/FPMB.

\nustar\ is absolutely calibrated against the Crab assuming a spectrum with $\Gamma = 2.1$ and power-law normalization of $N = 8.7$ ph $\mathrm{keV^{-1}cm^{-2}s^{-1}}$ at 1\,keV. The uncertainty in the absolute value of the slope with respect to chosen $\Gamma$ of the Crab is $\pm 0.01$. This is smaller than the slope uncertainty in either source and the \nustar\ calibration uncertainty therefore does not play into the cross-calibration. The dominant effect is statistical and we note that for \pksb\ the slope for FPMA is systematically harder than FPMB, while for \qcs\ the converse is true and FPMB is harder than FPMA, though still within the uncertainty on the measured values.

\subsubsection{Suzaku}
The details of the \suzaku\ XIS and HXD-PIN calibration can be found in a series of memos on the ISAS \suzaku\ web
page\footnote{http://www.astro.isas.ac.jp/suzaku/doc/suzakumemo} and in the \suzaku\ Technical Description available from the US \suzaku\ GOF web page\footnote{http://heasarc.gsfc.nasa.gov/docs/astroe}. Previous cross-calibration studies have shown that the three sensors of the XIS agree with each other at a 5\% level at energies between 1 and 10 keV \citep{Ishida2011,Tsujimoto2011}.  A change in inertial
reference units in 2010 produced attitude instability that could cause a source to partially move out of the XIS frame, especially in 1/4 window mode\footnote{http://www.astro.isas.ac.jp/suzaku/doc/suzakumemo/suzakumemo-2010-06.pdf}. However, this is properly accounted for in the ray-tracing code that generates the response files \citep{Mitsuda2007}, and the effect on 1--10 keV photometry is well below 5\%
\footnote{http://www.astro.isas.ac.jp/suzaku/doc/suzakumemo/suzakumemo-2010-05.pdf}. Below 1 keV, molecular contamination has greatly reduced the effective area in a way that varies with time, off-axis angle, and XIS \citep{Koyama2007}, however this has no effect in the energy band of the current study. Joint fits between the XIS detectors and the HXD-PIN show that the PIN normalization is about 10\% higher than that of the lower energy CCDs\footnote{http://www.astro.isas.ac.jp/suzaku/doc/suzakumemo/suzakumemo-2007-11.pdf}.

\subsubsection{\swift}
The most up-to-date \swift/XRT calibration information can be found on the XRT digest page hosted at the UK Swift Science Data
Centre, which summarizes the latest XRT calibration file releases (along with relevant release notes\footnote{http://www.swift.ac.uk/analysis/xrt/rmfarf.php}) and highlights any known calibration issues.

Previous comparisons of XRT spectra with data obtained from other instruments (such as \xmm/EPIC, \chandra/ACIS and
{\it RXTE}/PCA) in general revealed agreements in the power law photon indices to better than $\sim 6$\% (i.e. $\Delta \Gamma \la 0.1$) and broadband fluxes to within $\sim 10$\%.

We note when analyzing XRT spectra from sources which are piled-up or positioned near the CCD bad-columns, that poorly centered source
extraction regions can result in incorrect point spread function correction factors, causing flux inaccuracies at the $\sim$ 5--10\% level. Furthermore, as XRT star-tracker derived positions have an associated systematic uncertainty of 3\as\ then extraction
regions are best centered for each observation, rather than be based on the expected source coordinates.

High signal-to-noise spectra from sources with featureless continua typically show residuals of about 3\%, for example, near the
Au-M$_{\rm V}$ edge (at 2.205~keV), the Si-K edge (at 1.839~keV), or the O-K edge (at 0.545~keV).  Occasionally, however, residuals nearer
the 10\% level are seen, especially near the O-K edge, which are caused by small energy scale offsets (caused by inaccurate bias
and/or gain corrections). Such residuals can often be improved through careful use of the {\em gain} command in {\sc xspec} (by allowing the
gain offset term to vary by $\sim \pm $10--50~eV).

\subsubsection{XMM-Newton}
The details on \xmm\ EPIC calibration can be found in the EPIC calibration status document on the \xmm\ SOC web page\footnote{http://www.cosmos.esa.int/web/xmm-newton/calibration-documentation\#EPIC}. It is known from previous cross-calibration efforts that EPIC-pn and EPIC-MOS slightly diverge in normalization. We statistically investigated the EPIC cross-calibration using all 23 PKS2155-304 and 21 3C273 observations with all EPIC using small window modes. On average, EPIC-MOS1 (MOS2) return consistently about 5\% (7\%) higher flux compared to EPIC-pn in the energy range of about 0.5--5~keV. At lower energies, the average flux differences are smaller. At higher energies, the average differences increase to 10-15\% in the 5--10~keV band. The average EPIC spectral slopes agree except for the extremes of the sensitivity range. Towards the high end of the band pass (above $\sim 5$\,kev), the average EPIC-pn spectral slopes become softer compared to EPIC-MOS, as can be observed in Figures \ref{qcratios} and \ref{pksratios}.

\subsection{Flux cross-calibration}
We have calculated and tabulated the cross-normalization constant between each instrument as measured from the two sources. The constant is the ratio between the derived flux from Tables \ref{qcstable} and \ref{pkstable} for an instrument pair. We stress that the calculated fluxes between instrument pairs that include \nustar\ are for 3--7\,keV, all others are for 1--5\,keV, except \nustar/\suzaku(HXD), which is for 20--40\,keV. Table \ref{fluxtable} is for the soft instruments only (1--5\,keV) while Table \ref{fluxtable_2} is for instrument pairs including \nustar\ (3--7\,keV) and \suzaku/HXD (20--40\,keV). If a range is shown that means the instrument pair had data for both \qcs\ and \pksb. If there is only one number, as is the case for the \chandra\ instruments and \suzaku/HXD, the instrument pair was only active for one of the observations. 

There is excellent agreement between \suzaku, \swift/XRT, and \xmm/MOS with only a few percent dispersion. \xmm/pn systematically measures a lower flux with respect to \xmm/MOS, which is a well-known offset and already discussed. Both \chandra/LETGS and HETGS measure a systematically higher flux than all other observatories and is a trend that has been noted before in other studies \citep{Nevalainen2010,Tsujimoto2011}. In the harder band, 3-7\,keV, which is together with \nustar, there is excellent agreement between \swift/XRT, \xmm/MOS and \suzaku/XIS0 and XIS3, while XIS1 measures a few percent lower values than the other instruments. \suzaku/HXD has about 10\% higher flux than \nustar\ in the 20--40\,keV band, which was expected from the XIS/PIN internal relative normalizations.

\subsection{Cross-observatory spectral discrepancies}
\begin{figure*}
\begin{center}
\includegraphics[width=0.30\textwidth]{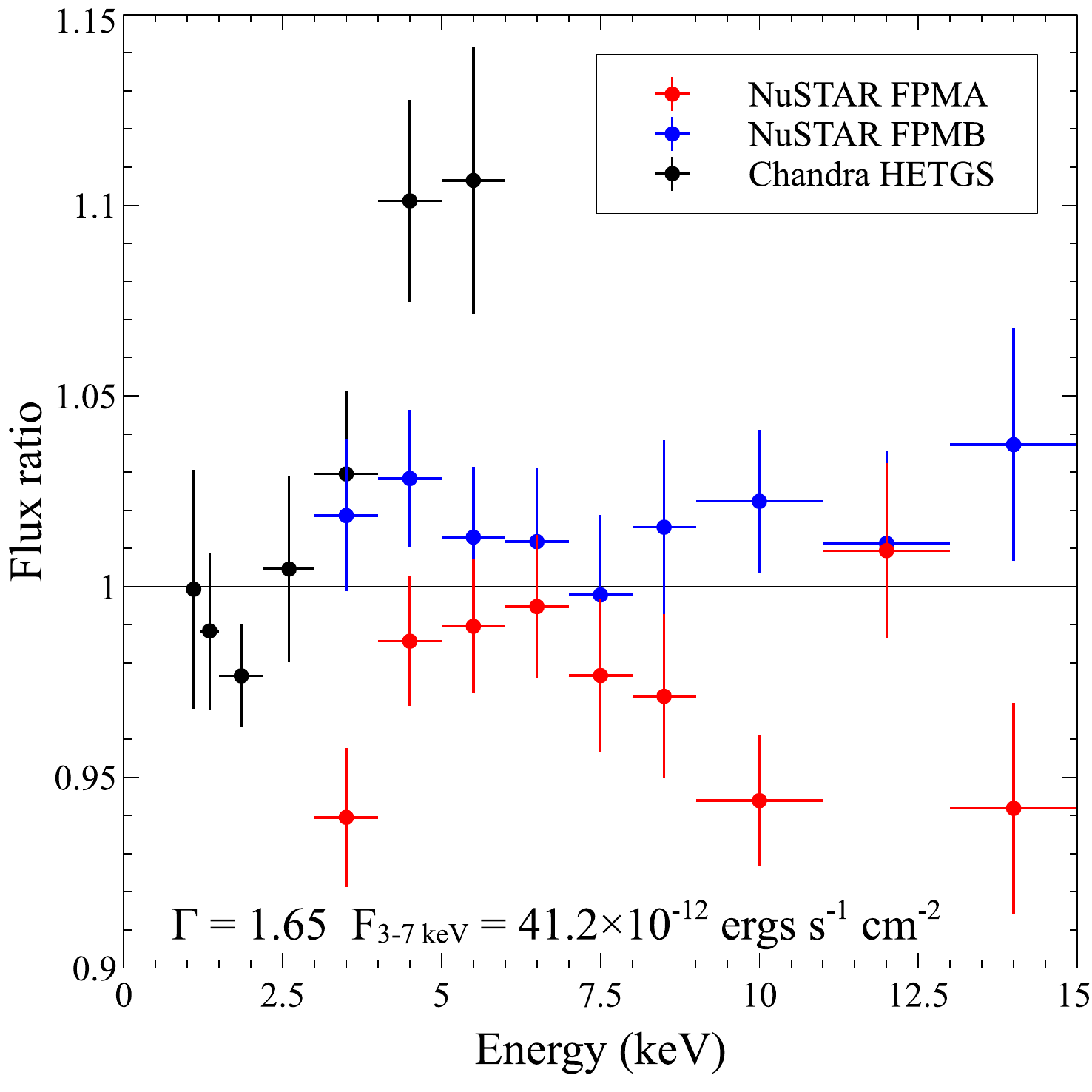}
\includegraphics[width=0.30\textwidth]{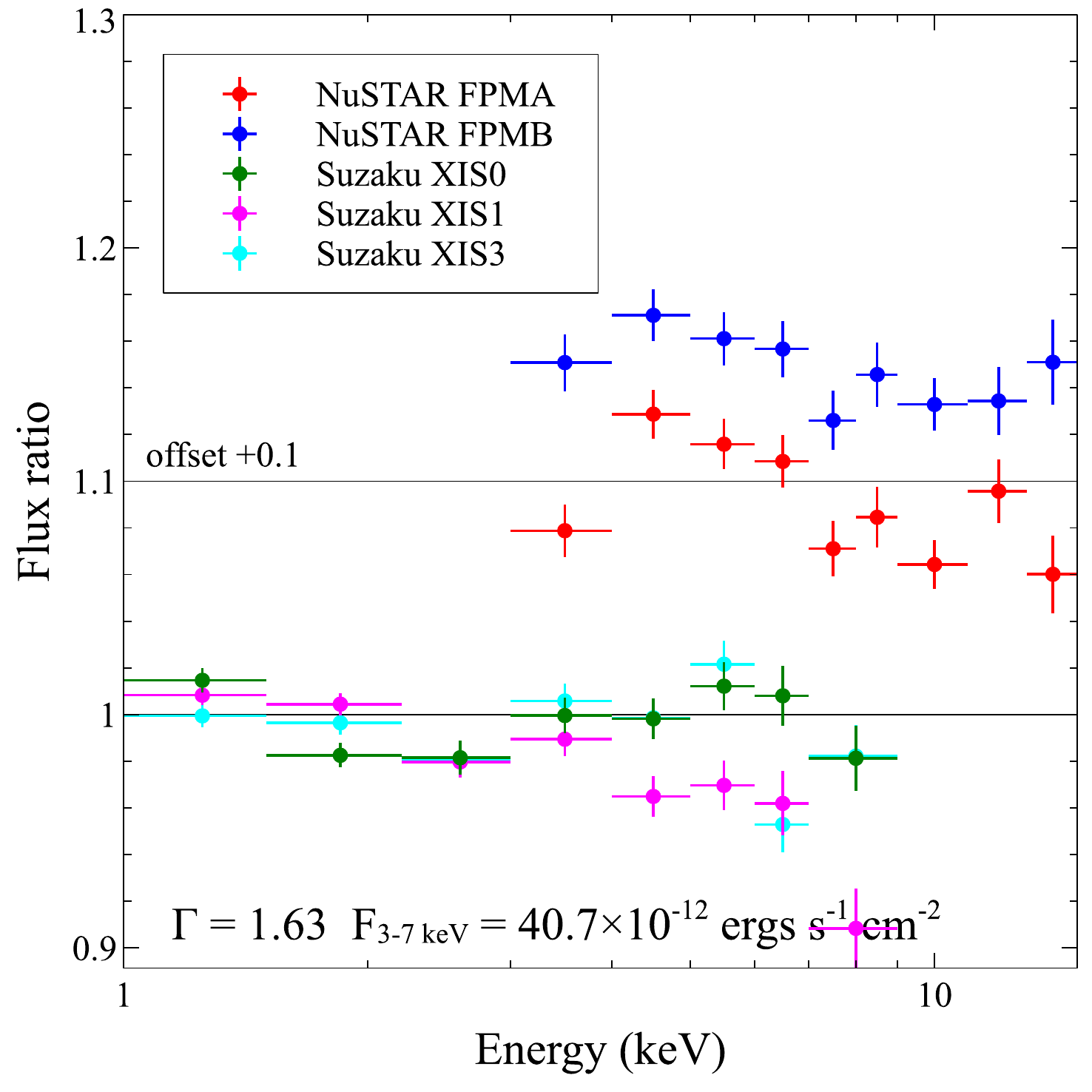}
\includegraphics[width=0.30\textwidth]{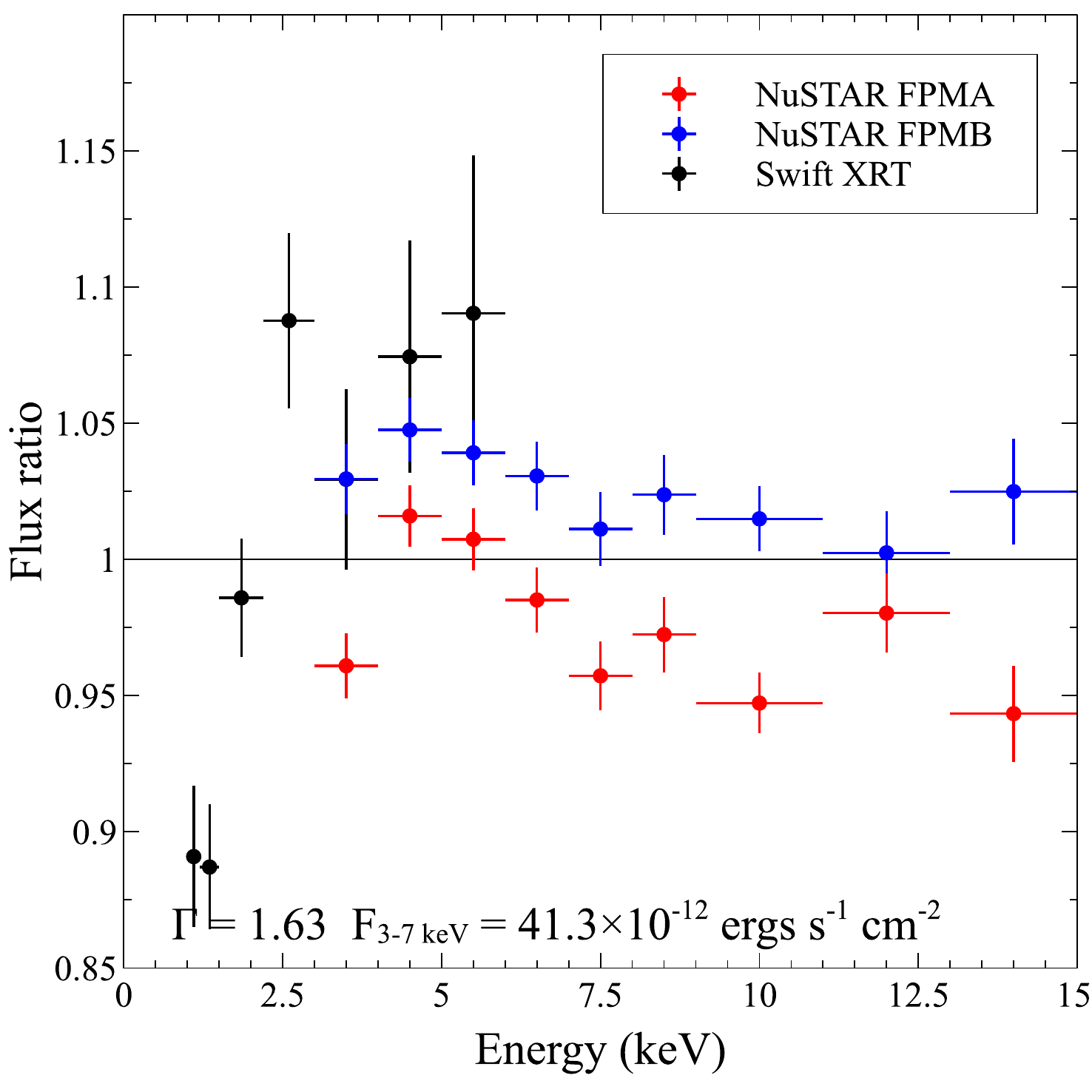}
\includegraphics[width=0.30\textwidth]{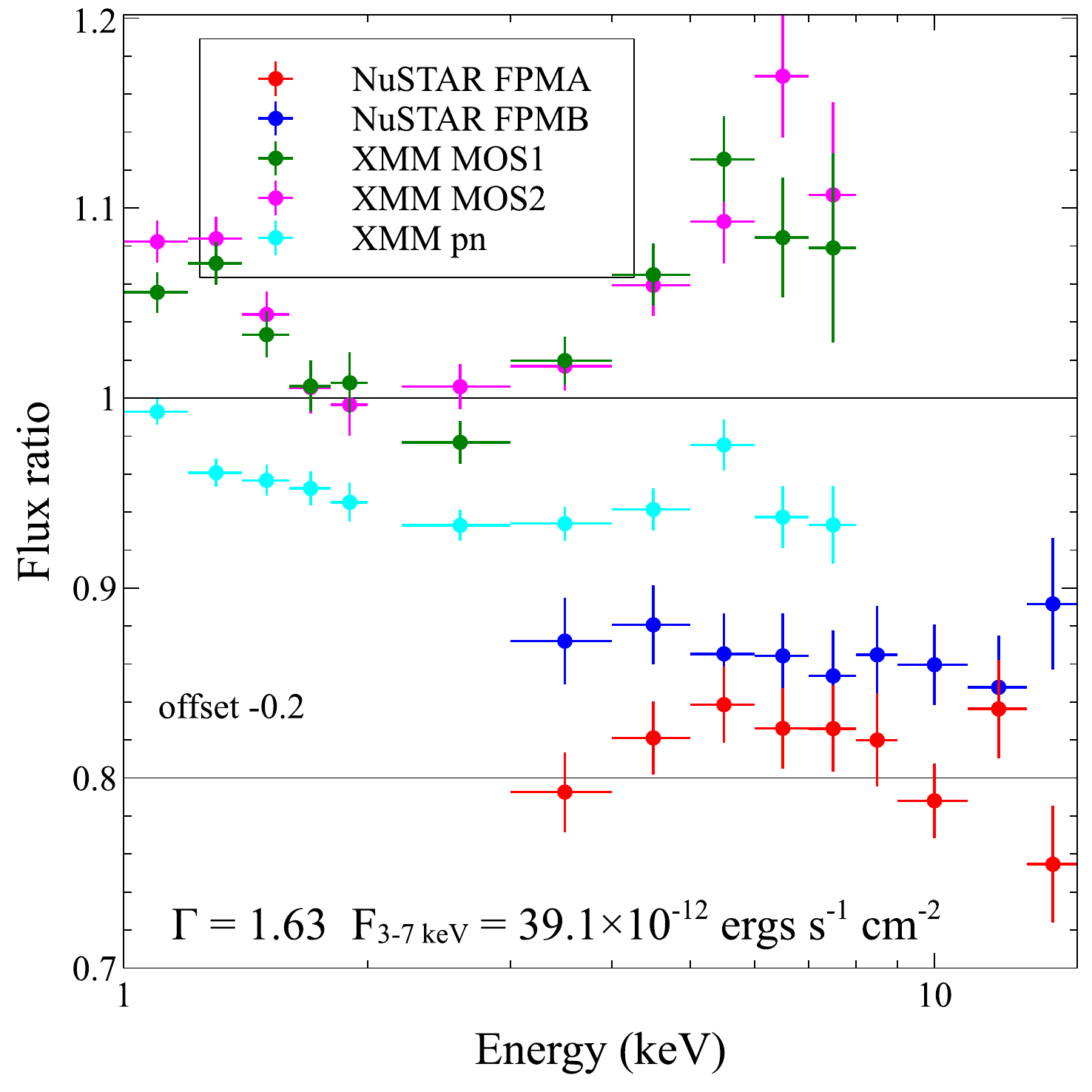}
\includegraphics[width=0.30\textwidth]{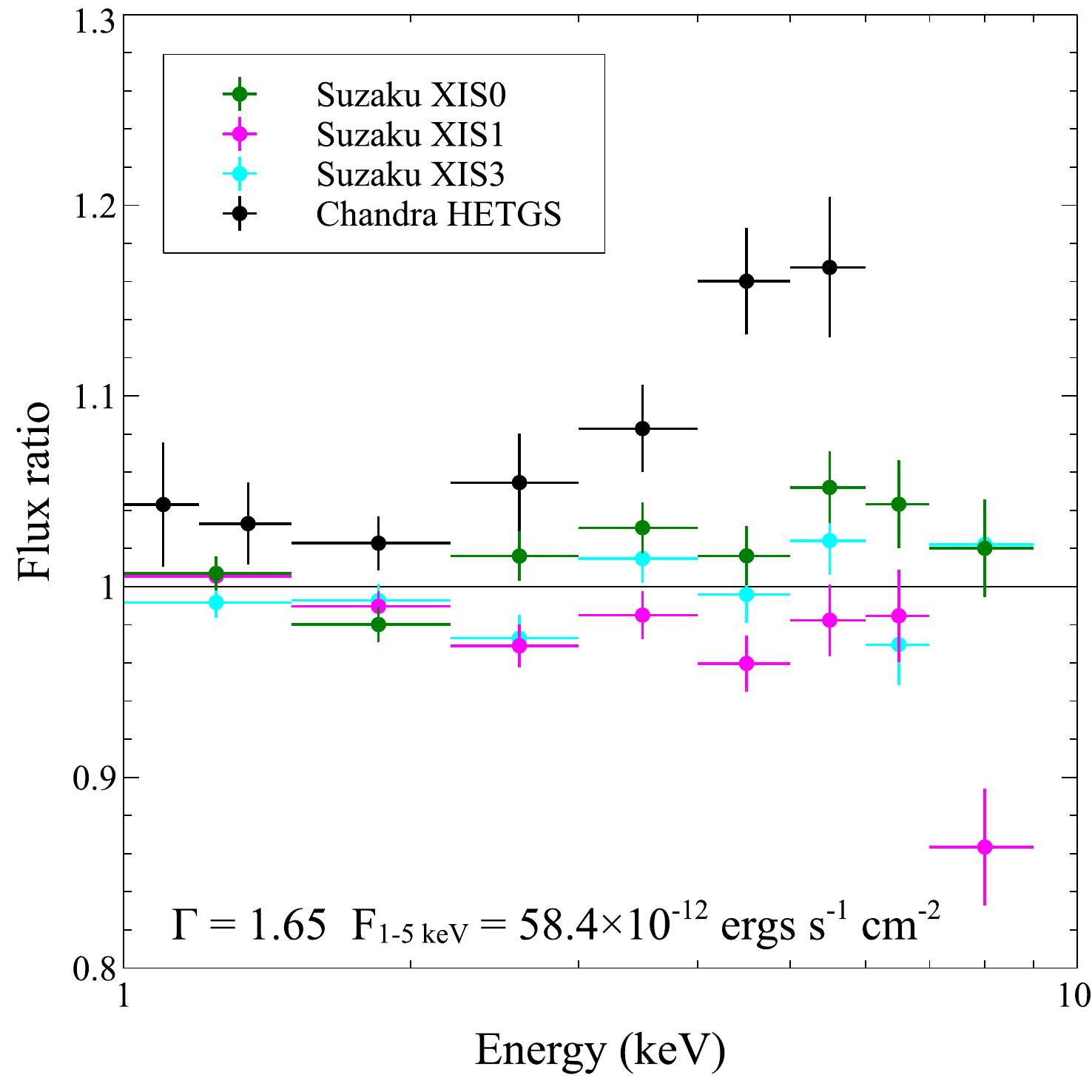}
\includegraphics[width=0.30\textwidth]{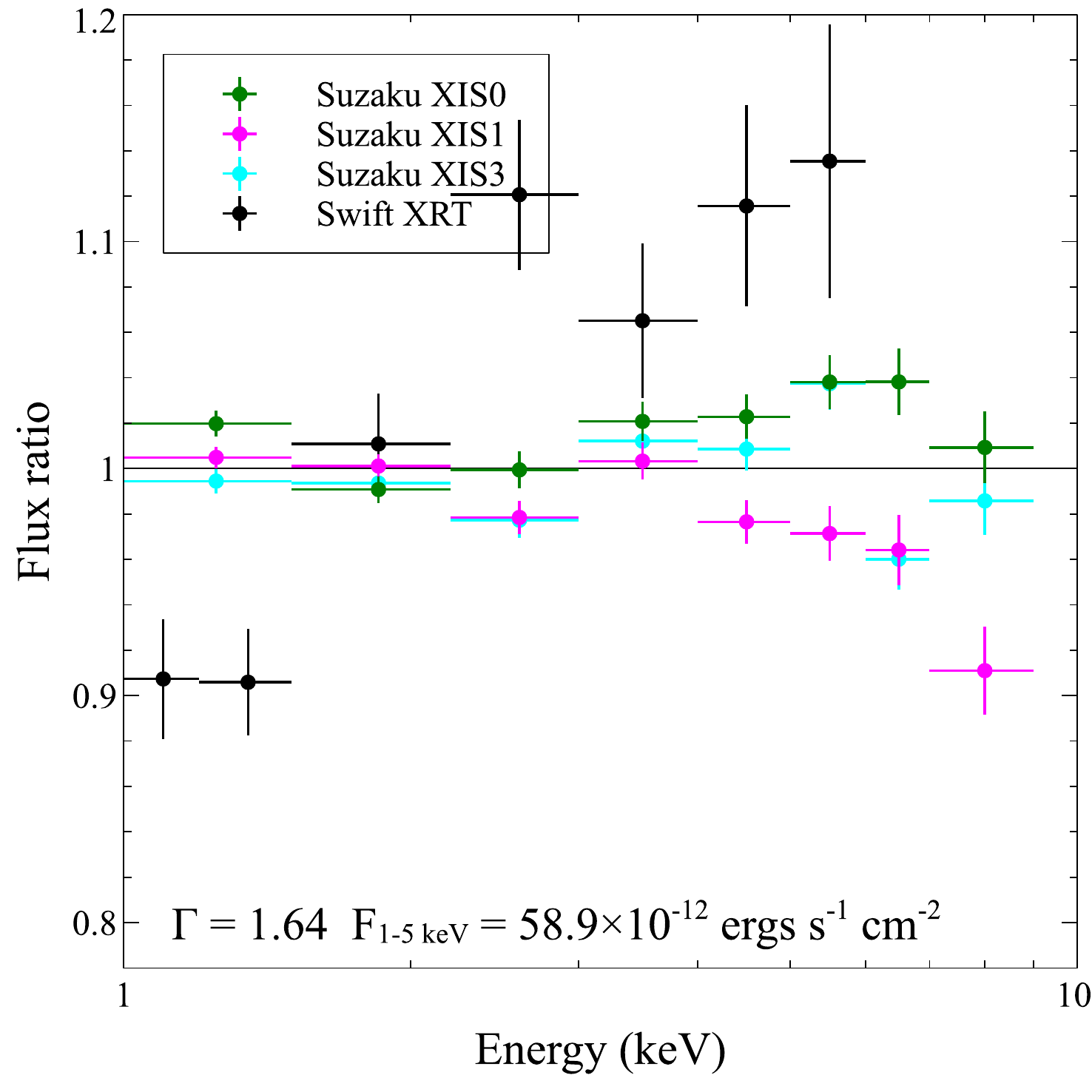}
\includegraphics[width=0.30\textwidth]{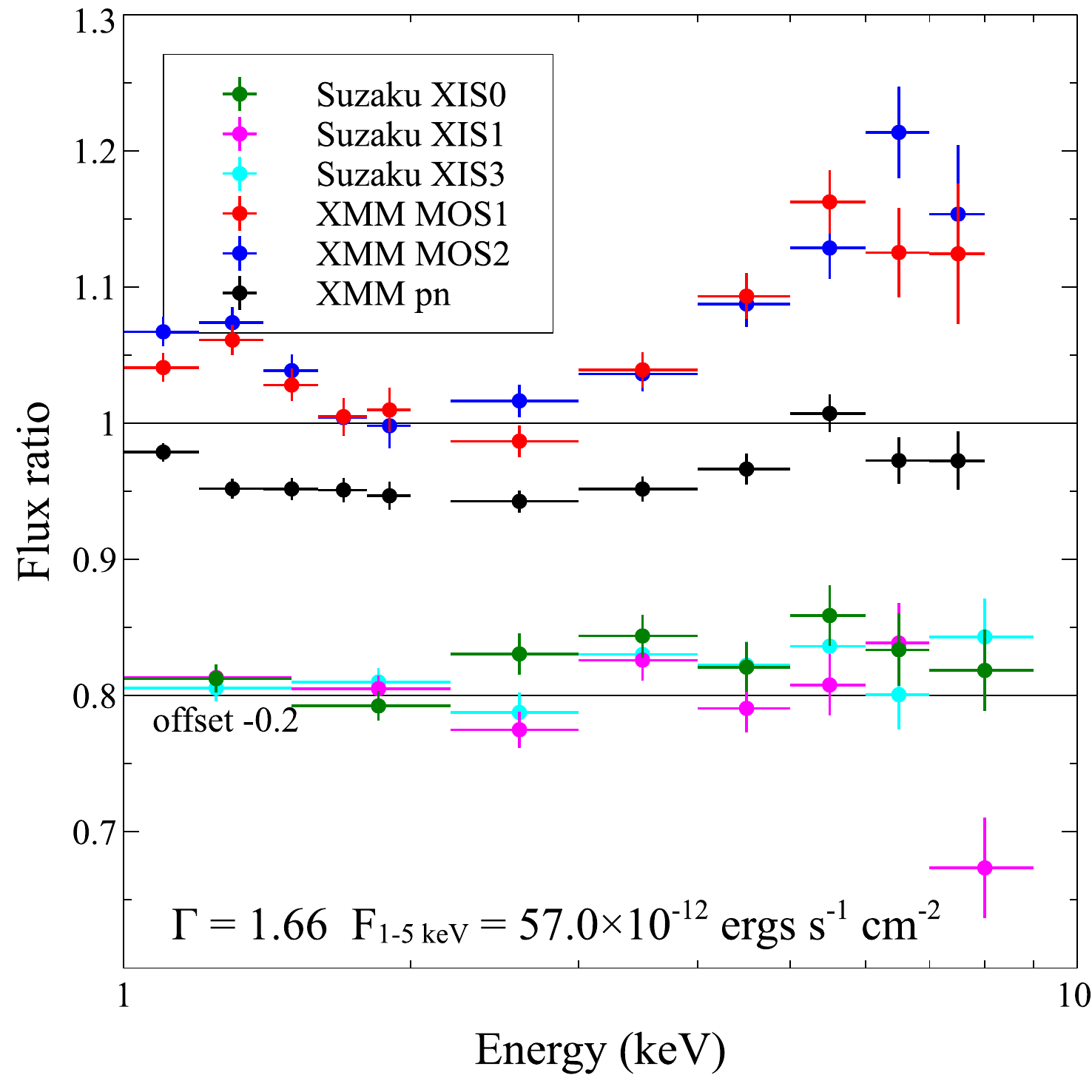}
\includegraphics[width=0.30\textwidth]{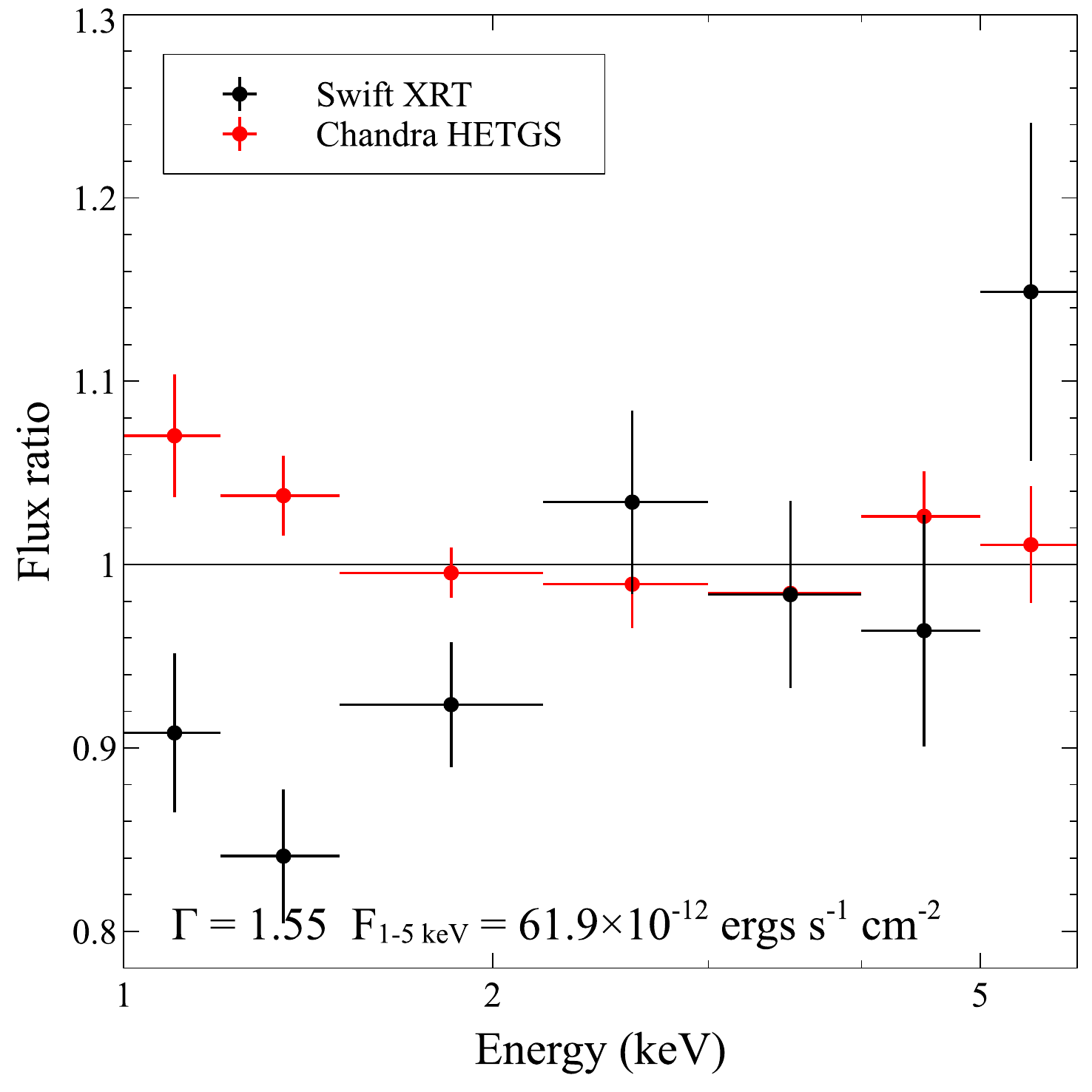}
\includegraphics[width=0.30\textwidth]{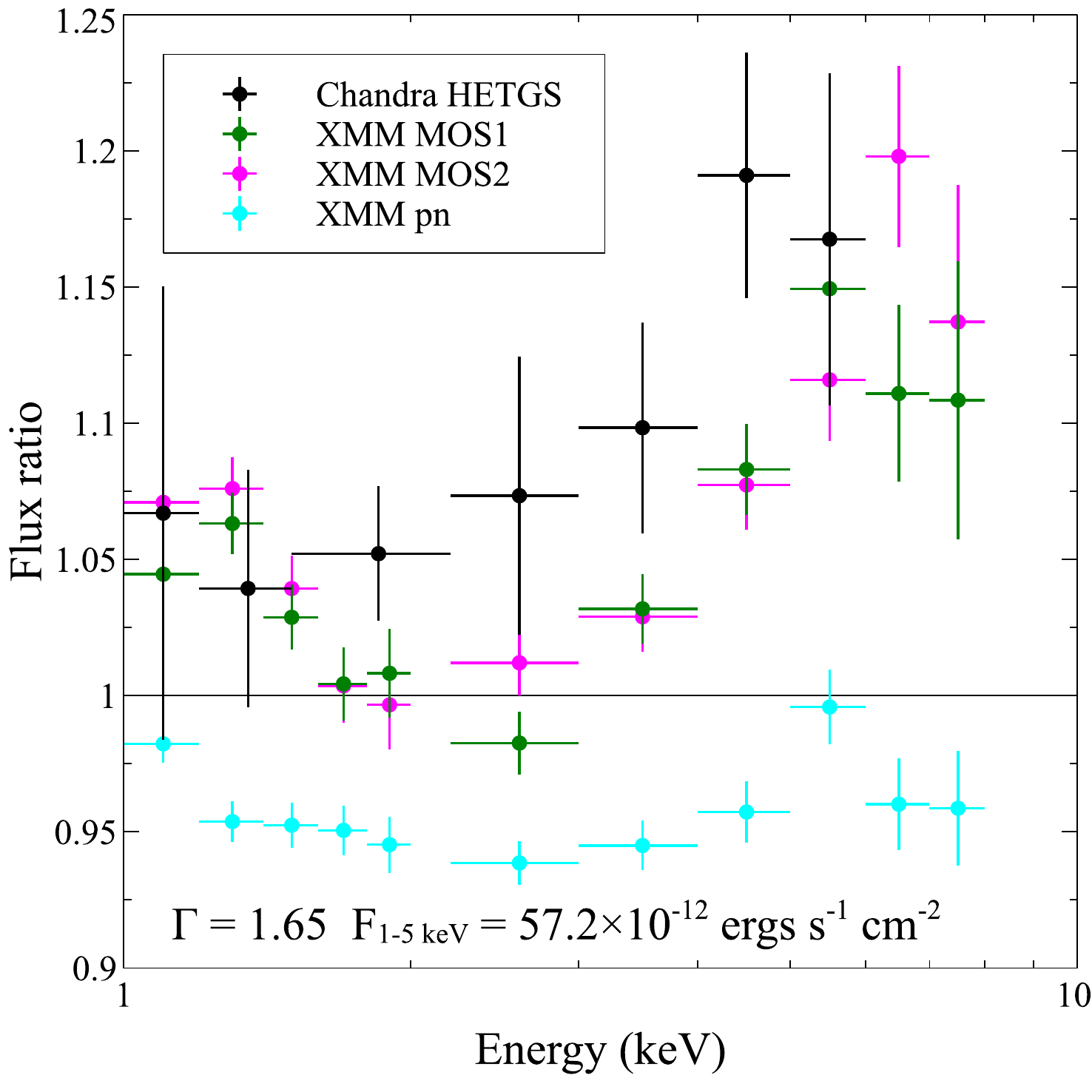}
\includegraphics[width=0.30\textwidth]{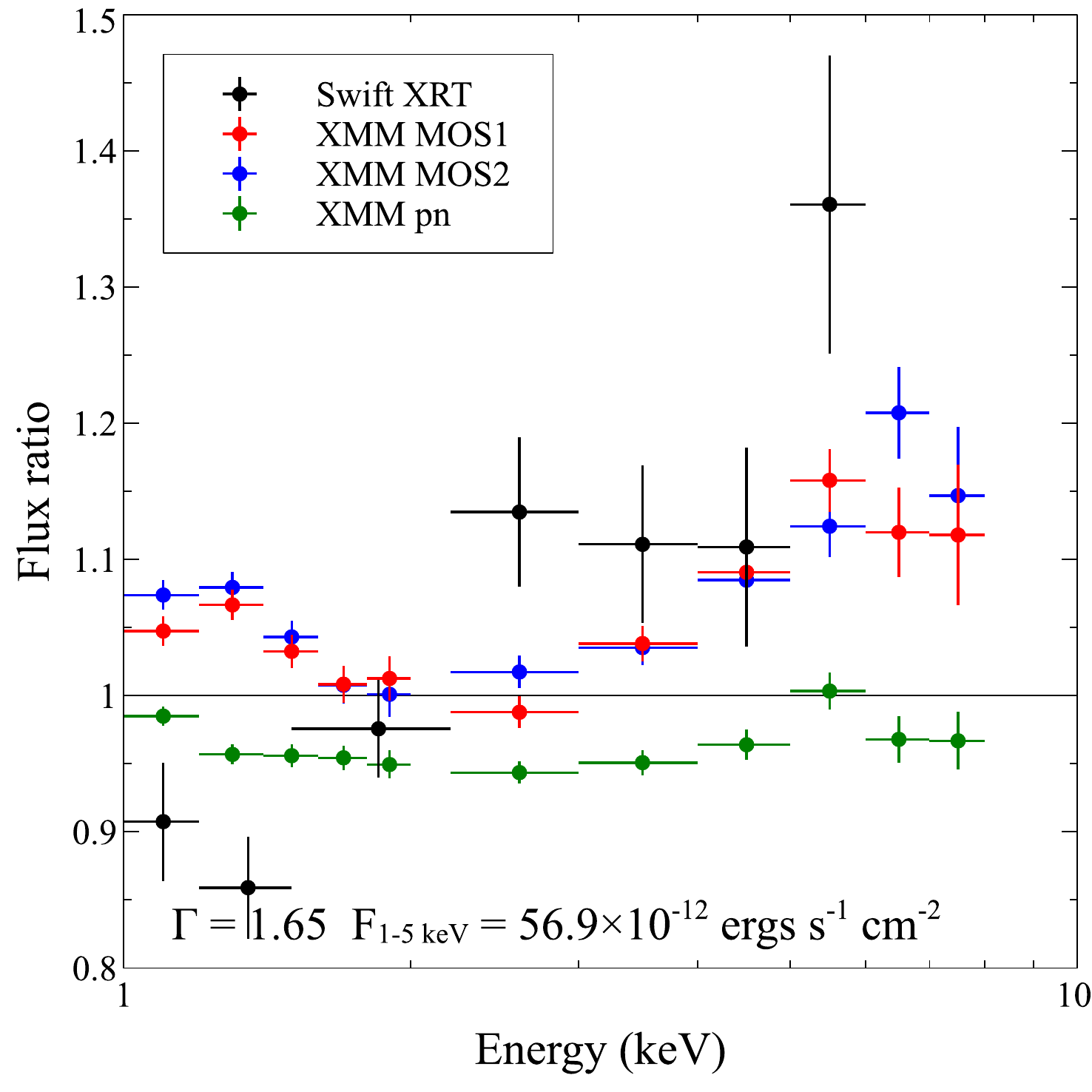}
\includegraphics[width=0.30\textwidth]{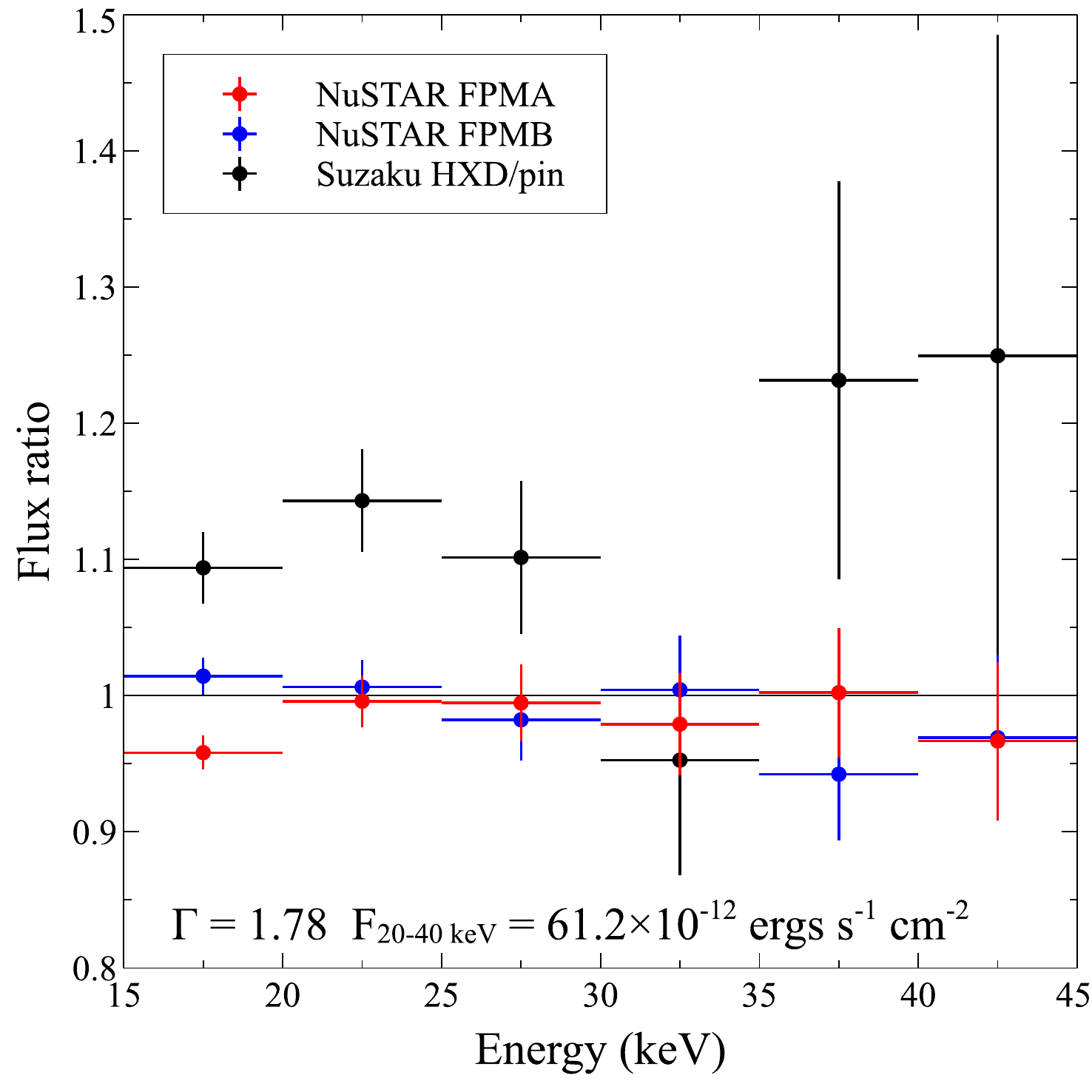}
\end{center}
\caption{Flux ratios for 3C273. The offset lines are for viewing purposes only.}
\label{qcratios}
\end{figure*}

\begin{figure*}
\begin{center}
\includegraphics[width=0.30\textwidth]{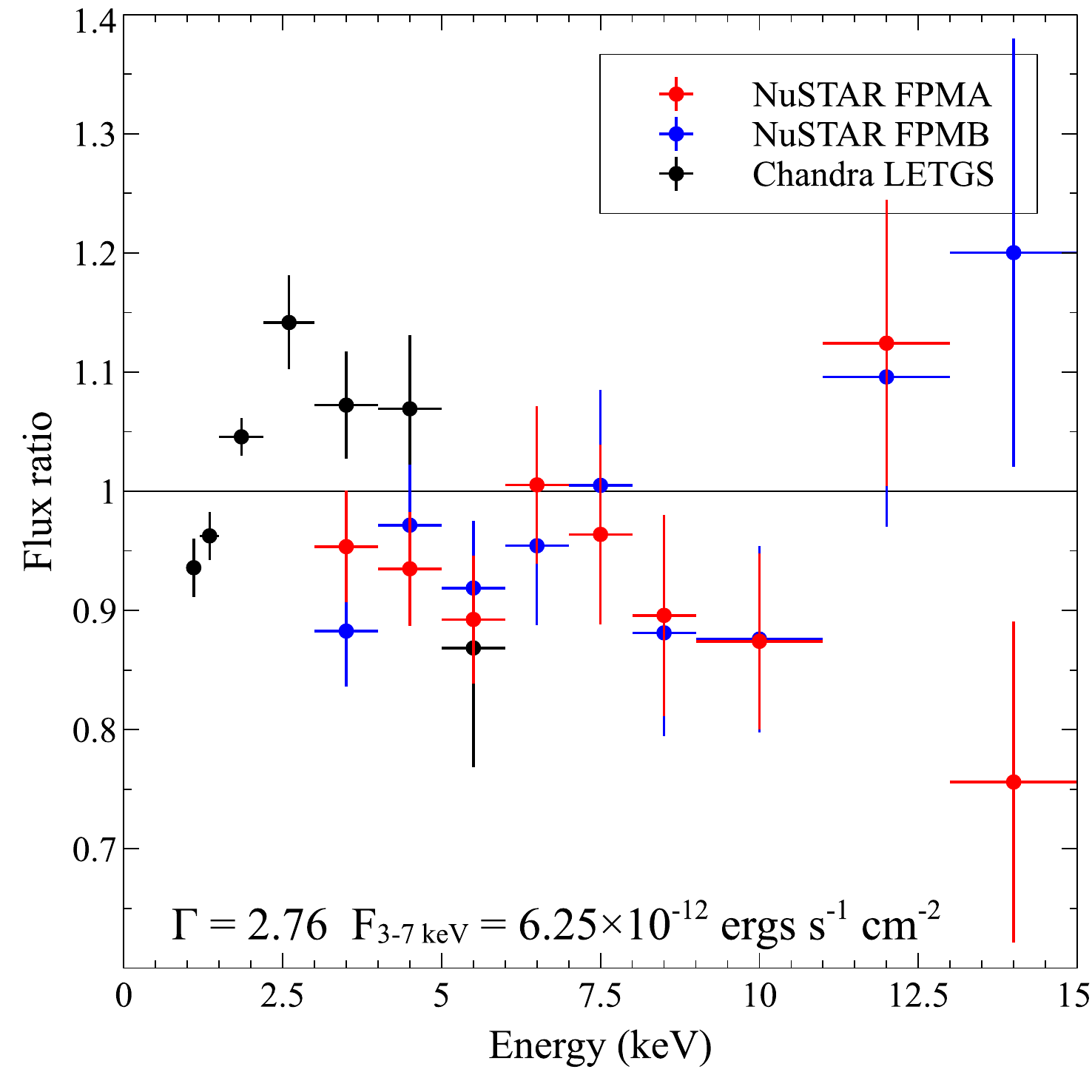}
\includegraphics[width=0.30\textwidth]{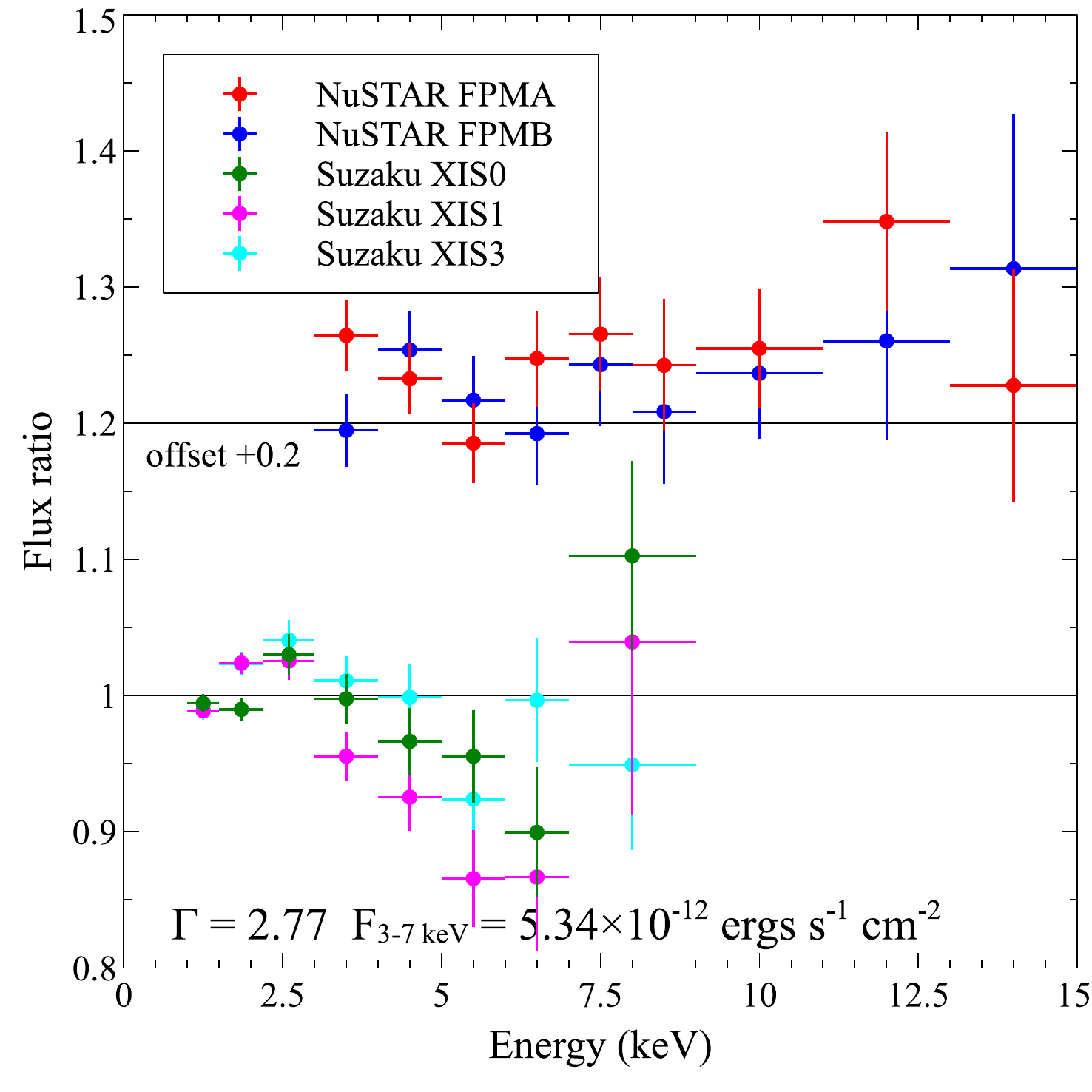}
\includegraphics[width=0.30\textwidth]{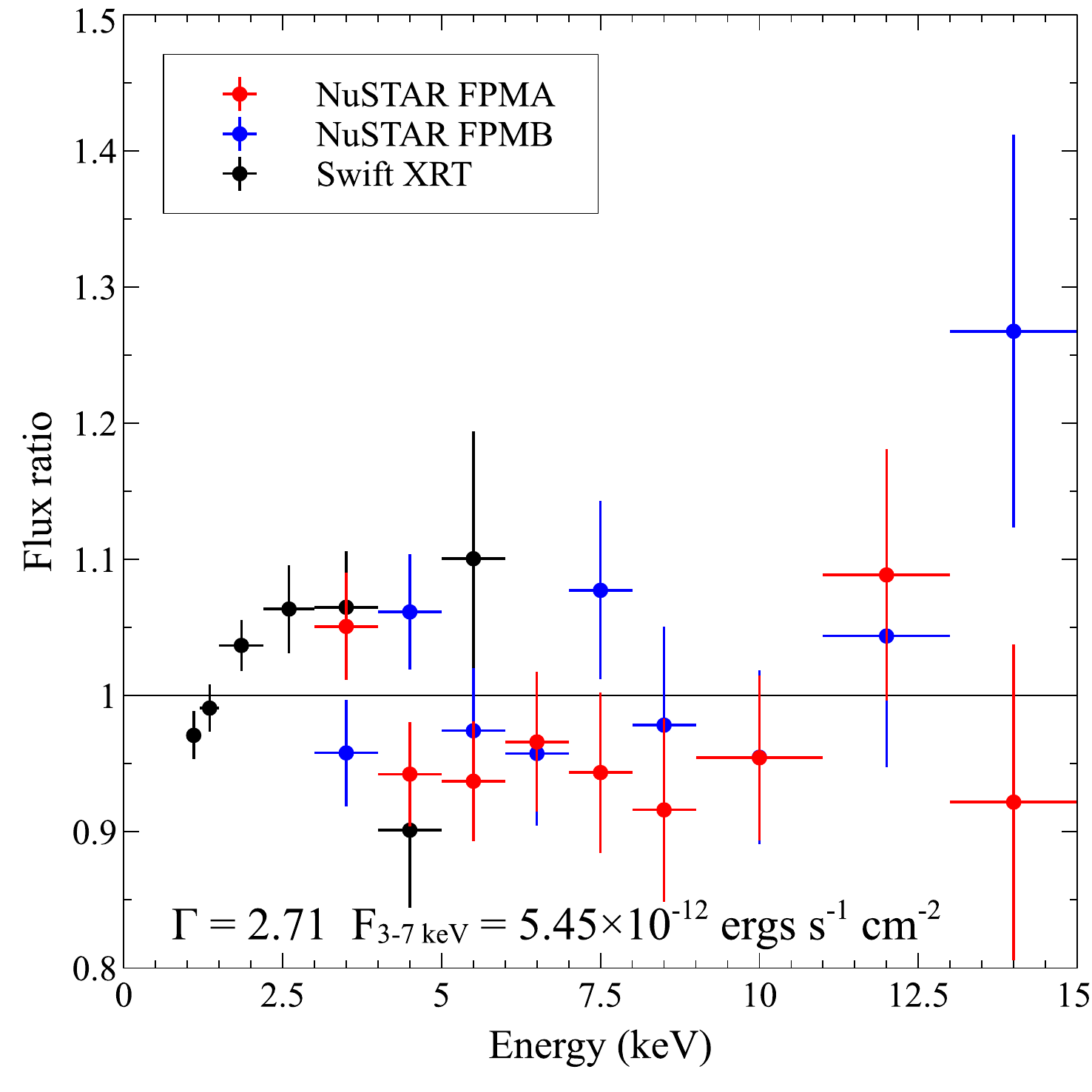}
\includegraphics[width=0.30\textwidth]{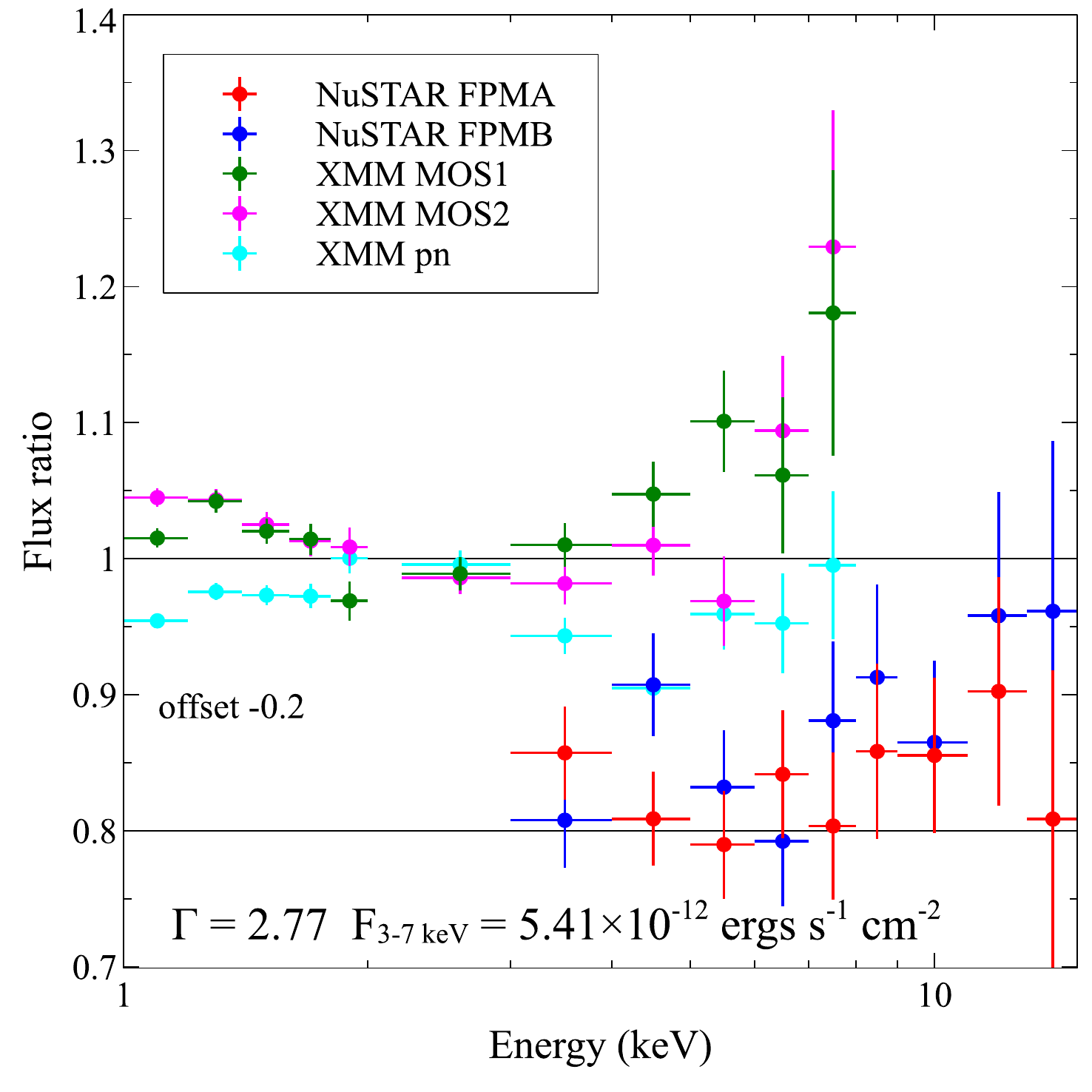}
\includegraphics[width=0.30\textwidth]{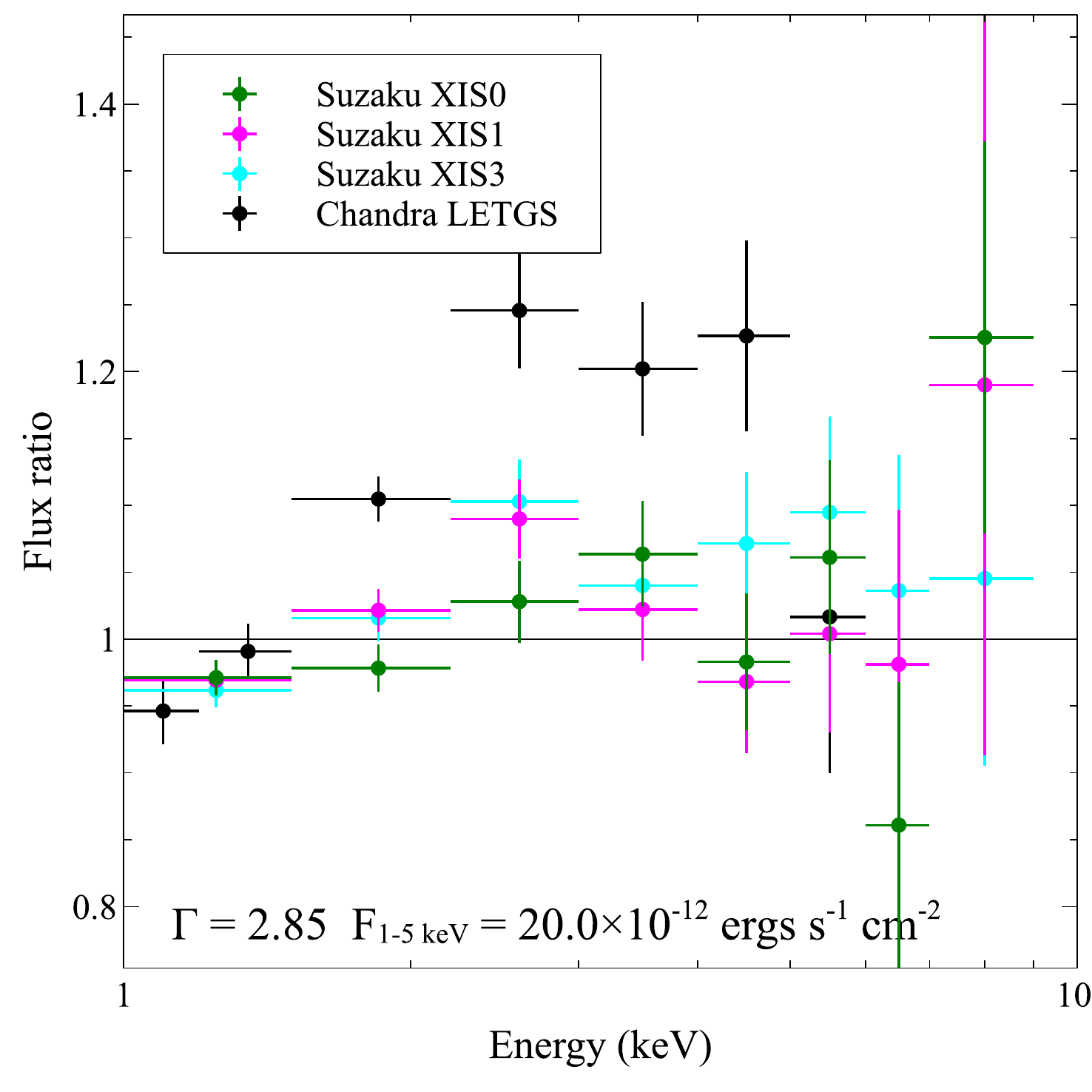}
\includegraphics[width=0.30\textwidth]{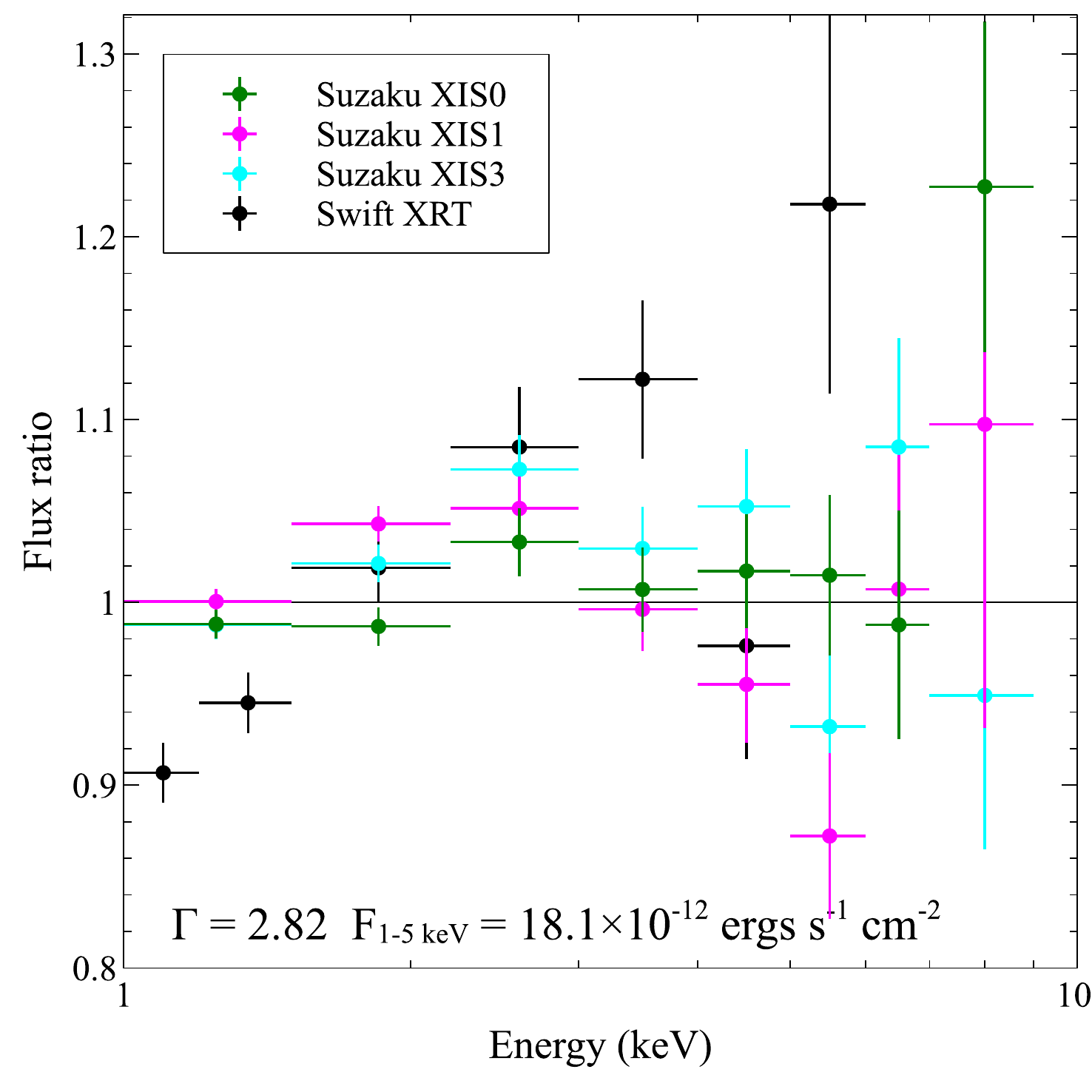}
\includegraphics[width=0.30\textwidth]{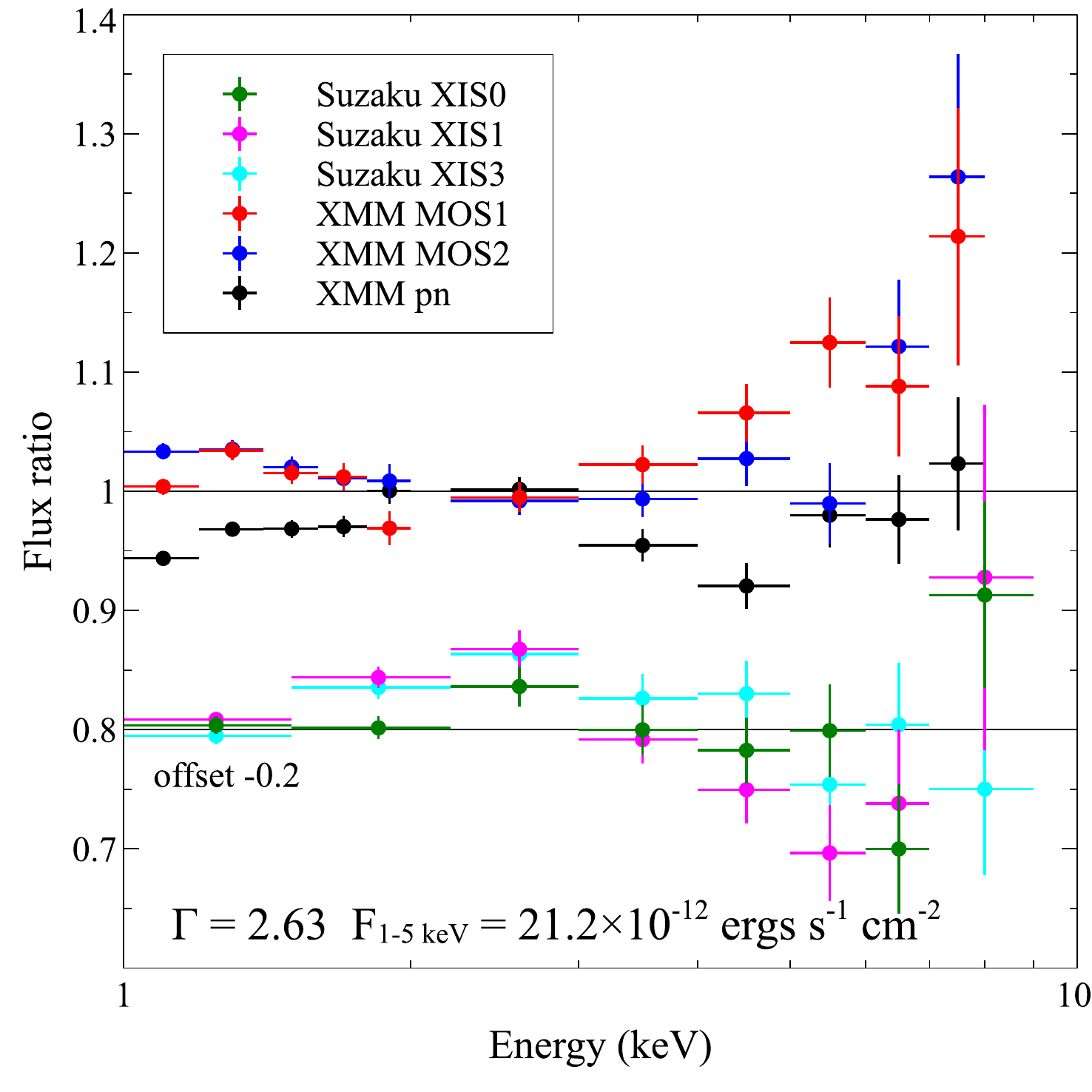}
\includegraphics[width=0.30\textwidth]{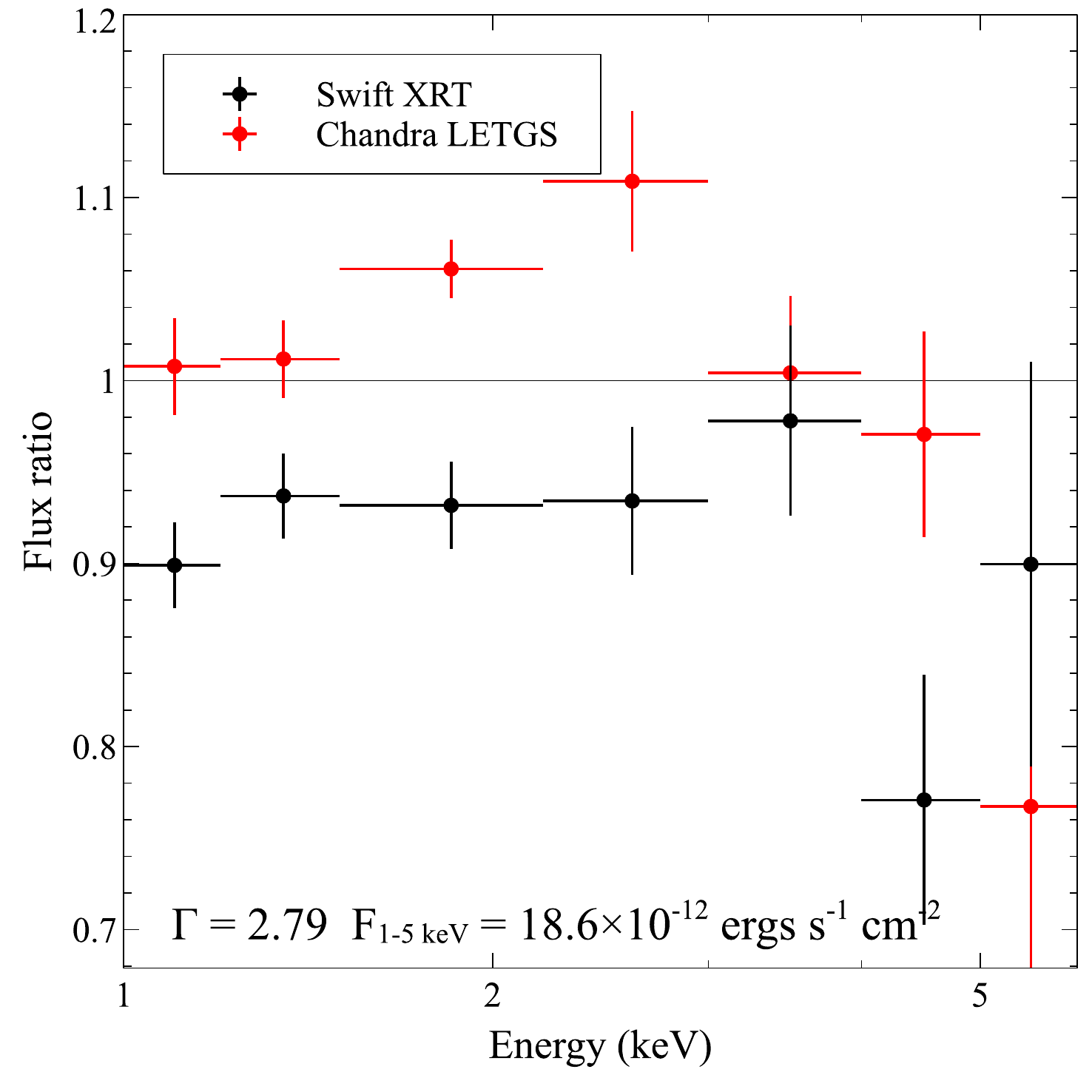}
\includegraphics[width=0.30\textwidth]{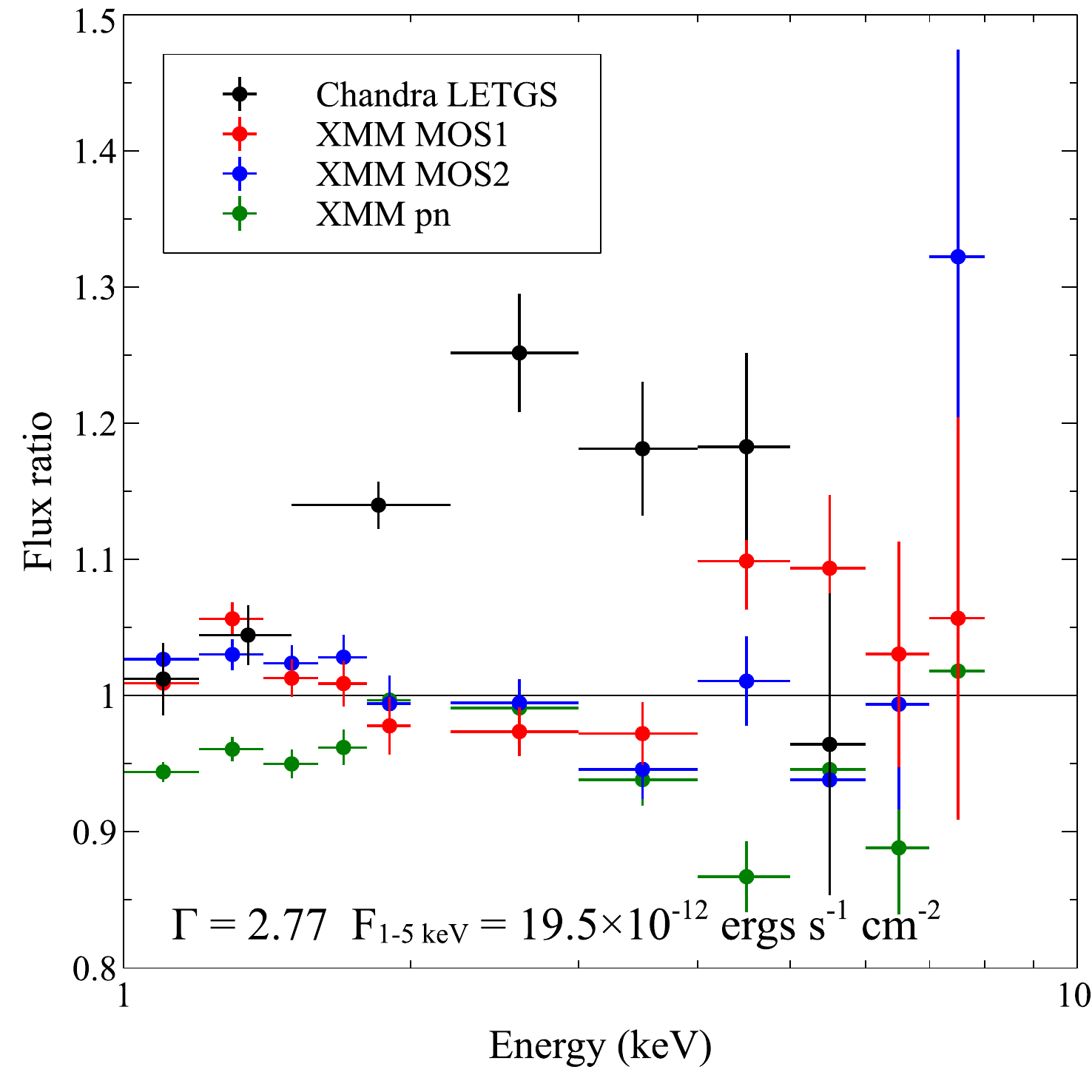}
\includegraphics[width=0.30\textwidth]{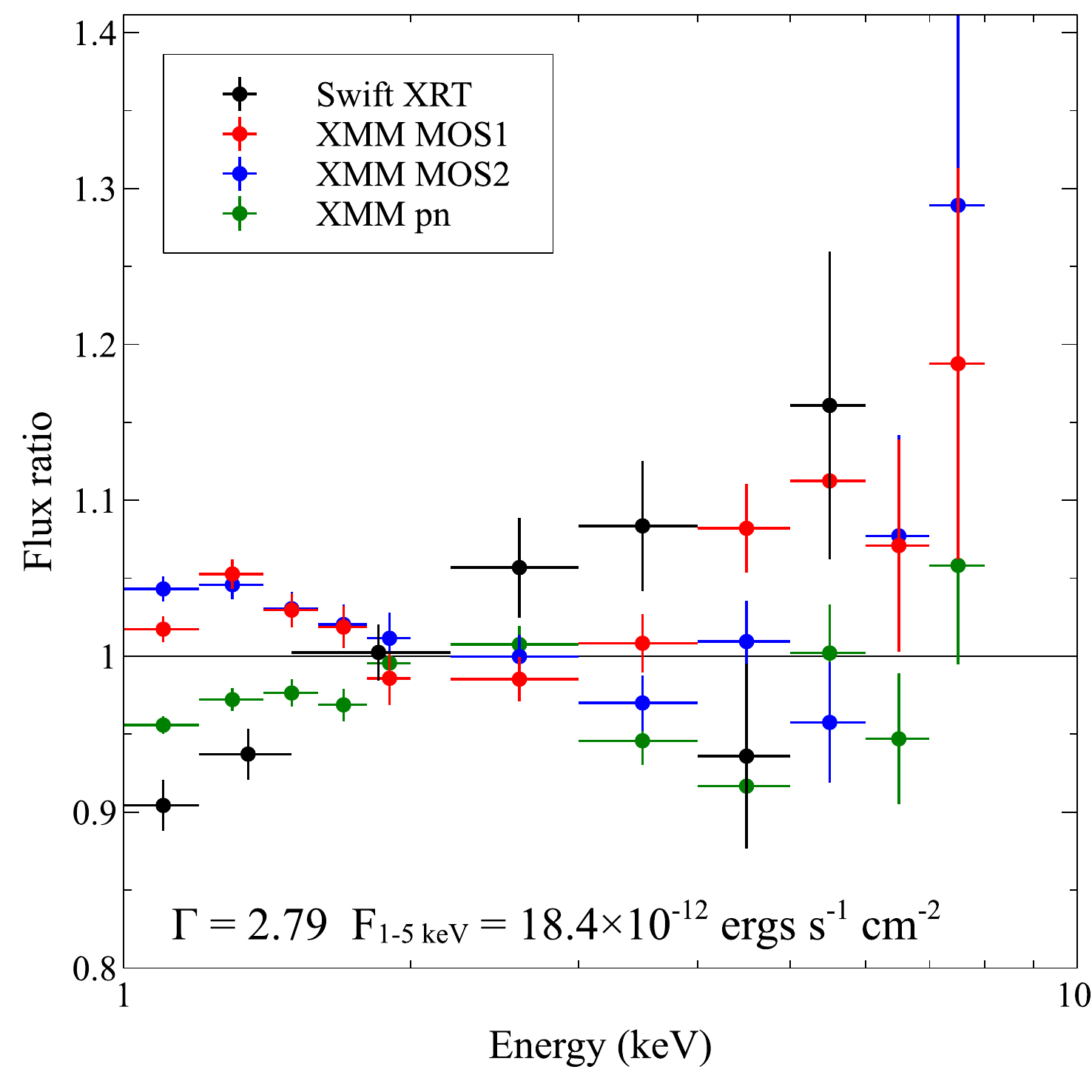}
\end{center}
\caption{Flux ratios for PKS2155-304. The offset lines are for viewing purposes only.}
\label{pksratios}
\end{figure*}

As can be seen from Figures \ref{qcratios} and \ref{pksratios} these cross-normalization constants may change with a different choice of overlapping energy band. Caution must be urged, though, not to read too much into features such as the convex trend observed in \suzaku\ XIS instruments in \pksb, which is not repeated for \qcs. For a feature to be called systematic, it should be observed in several independent observations and for different targets, and the non-repeated differences between instruments should instead be regarded as an indication of the magnitude of statistical fluctuations that might occur.

A trend that does repeat and has been observed in other cross-calibration papers, is the harder spectrum derived from LETGS and higher overall flux peaking at around $\sim$3--4\,keV with respect to \suzaku/XIS and \xmm/MOS \citep[see Figure 9,][]{Ishida2011}. Likewise, for both sources \swift/XRT shows a harder spectrum, and the XRT flux is typically less than other instruments below $\sim$3\,keV and higher above. The shallower slope is similar to \chandra, which has been confirmed also for G21.5-0.9 \citep{Tsujimoto2011}, where a new analysis with updated response files finds agreement between the two instruments\footnote{http://www.swift.ac.uk/analysis/xrt/files/SWIFT-XRT-CALDB-09\_v17.pdf}$^,$\footnote{http://www.swift.ac.uk/analysis/xrt/files/SWIFT-XRT-CALDB-09\_v18.pdf}. For \chandra\ the flux below 3\,keV is in agreement with \xmm/MOS and \suzaku/XIS, but rises to a peak of 10--15\% above 4\,keV with respect to the other instruments.

For these two sources, \xmm\ shows a harder spectrum above 3\,keV with respect to \suzaku, but this is not always the case as shown in \citet{Ishida2011} where several observations of \pksb\ were investigated. As already stated, pn measures a flux that is 5\% below that of MOS for energies less than 5\,keV. Above 5\,keV the difference in flux between pn and MOS increases to 10--15\%, which is the result of a softer pn slope.

There is good agreement between the measured slope of \suzaku/HXD and \nustar, but with a 10\% difference in flux. The statistical level of the overlapping energy bands of \nustar\ with the soft instruments is not high enough to reveal any systematic trends. By design, the average \nustar\ flux was set to be roughly between the other observatories.

\begin{table}
\centering
\caption{Cross-calibration \textit{cflux} $\times$ \textit{tbabs} $\times$ \textit{pow}}
\begin{tabular}{l|c|c|c}
\hline
Instrument & $\Gamma$ & Flux 3--7 keV & $\chi^2_{\mathrm{red}}$ \\
& & $10^{-12}$ erg cm$^{-2}$ s$^{-1}$ & \\
\hline
\multicolumn{3}{c}{3C\,273}\\
\hline
\hline
\nustar\ FPMA & 1.54 $\pm $0.06 & 40.1 $\pm $0.62 & 1.079 \\
\nustar\ FPMB & 1.65 $\pm $0.06 & 42.0 $\pm $0.67 & 0.673 \\
\chandra\ HETGS & 1.61 $\pm $0.07 & 44.8 $\pm $0.85 & 0.479 \\
\hline
\nustar\ FPMA & 1.59 $\pm $0.04 & 41.0 $\pm $0.38 & 1.052 \\
\nustar\ FPMB & 1.63 $\pm $0.04 & 43.1 $\pm $0.40 & 0.970 \\
\suzaku\ XIS0 & 1.61 $\pm $0.03 & 40.9 $\pm $0.31 & 1.040 \\
\suzaku\ XIS1 & 1.68 $\pm $0.03 & 39.6 $\pm $0.31 & 0.929 \\
\suzaku\ XIS3 & 1.68 $\pm $0.03 & 40.6 $\pm $0.30 & 1.089 \\
\hline
\nustar\ FPMA & 1.59 $\pm $0.04 & 41.0 $\pm $0.42 & 1.069 \\
\nustar\ FPMB & 1.63 $\pm $0.04 & 42.8 $\pm $0.21 & 0.936 \\
\swift\ XRT & 1.40 $\pm $0.15 & 45.4 $\pm $1.77 & 0.692 \\
\hline
\nustar\ FPMA & 1.55 $\pm $0.07 & 39.7 $\pm $0.68 & 1.085 \\
\nustar\ FPMB & 1.64 $\pm $0.07 & 41.9 $\pm $0.35 & 0.742 \\
\xmm\ MOS1 & 1.45 $\pm $0.06 & 42.0 $\pm $0.61 & 1.022 \\
\xmm\ MOS2 & 1.43 $\pm $0.06 & 42.1 $\pm $0.59 & 0.993 \\
\xmm\ pn & 1.65 $\pm $0.05 & 36.0 $\pm $0.46 & 1.037 \\
\hline
Instrument & $\Gamma$ & Flux 1--5 keV & $\chi^2_{\mathrm{red}}$ \\
& & $10^{-12}$ erg cm$^{-2}$ s$^{-1}$ & \\
\hline
\suzaku\ XIS0 & 1.64 $\pm $0.01 & 58.8 $\pm $0.47 & 0.993 \\
\suzaku\ XIS1 & 1.69 $\pm $0.01 & 57.3 $\pm $0.41 & 1.065 \\
\suzaku\ XIS3 & 1.65 $\pm $0.01 & 57.9 $\pm $0.45 & 0.999 \\
\chandra\ HETGS & 1.58 $\pm $0.01 & 63.5 $\pm $0.57 & 0.530 \\
\hline
\suzaku\ XIS0 & 1.64 $\pm $0.01 & 59.5 $\pm $0.30 & 0.973 \\
\suzaku\ XIS1 & 1.66 $\pm $0.01 & 58.5 $\pm $0.31 & 1.046 \\
\suzaku\ XIS3 & 1.63 $\pm $0.01 & 58.7 $\pm $0.29 & 0.977 \\
\swift\ XRT & 1.48 $\pm $0.04 & 61.2 $\pm $1.39 & 1.056 \\
\hline
\suzaku\ XIS0 & 1.64 $\pm $0.02 & 57.9 $\pm $0.56 & 0.968 \\
\suzaku\ XIS1 & 1.68 $\pm $0.00 & 57.0 $\pm $0.52 & 1.136 \\
\suzaku\ XIS3 & 1.65 $\pm $0.02 & 57.5 $\pm $0.51 & 1.004 \\
\xmm\ MOS1 & 1.66 $\pm $0.01 & 58.8 $\pm $0.47 & 1.149 \\
\xmm\ MOS2 & 1.67 $\pm $0.01 & 59.3 $\pm $0.46 & 1.066 \\
\xmm\ pn & 1.69 $\pm $0.01 & 53.9 $\pm $0.34 & 1.061 \\
\hline
\swift\ XRT & 1.46 $\pm $0.07 & 58.8 $\pm $2.13 & 1.524 \\
\chandra\ HETGS & 1.58 $\pm $0.01 & 63.5 $\pm $0.57 & 0.530 \\
\hline
\xmm\ MOS1 & 1.66 $\pm $0.01 & 58.8 $\pm $0.47 & 1.149 \\
\xmm\ MOS2 & 1.67 $\pm $0.01 & 59.3 $\pm $0.46 & 1.066 \\
\xmm\ pn & 1.69 $\pm $0.01 & 53.9 $\pm $0.34 & 1.061 \\
\chandra\ HETGS & 1.58 $\pm $0.01 & 63.4 $\pm $0.60 & 0.487 \\
\hline
\swift\ XRT & 1.46 $\pm $0.07 & 58.8 $\pm $2.13 & 1.548 \\
\xmm\ MOS1 & 1.66 $\pm $0.01 & 58.8 $\pm $0.47 & 1.149 \\
\xmm\ MOS2 & 1.67 $\pm $0.01 & 59.3 $\pm $0.46 & 1.066 \\
\xmm\ pn & 1.69 $\pm $0.01 & 53.9 $\pm $0.34 & 1.061 \\
\hline
Instrument & $\Gamma$ & Flux 20--40 keV & $\chi^2_{\mathrm{red}}$ \\
& & $10^{-12}$ erg cm$^{-2}$ s$^{-1}$ & \\
\hline
\nustar\ FPMA & 1.76 $\pm $0.06 & 60.3 $\pm $1.50 & 1.023 \\
\nustar\ FPMB & 1.85 $\pm $0.06 & 61.4 $\pm $1.59 & 0.978 \\
\suzaku\ HXD & 1.77 $\pm $0.10 & 68.0 $\pm $3.27 & 0.860 \\
\hline
\end{tabular}
\label{crossnorm2}

\label{qcstable}
\end{table}

\begin{table}
\centering
\caption{Cross-calibration \textit{cflux} $\times$ \textit{tbabs} $\times$ \textit{pow}}
\begin{tabular}{l|c|c|c}
\hline
Instrument & $\Gamma$ & Flux 3--7 keV & $\chi^2_{\mathrm{red}}$ \\
& & $10^{-12}$ erg cm$^{-2}$ s$^{-1}$ & \\ 
\hline
\multicolumn{3}{c}{PKS2155-304}\\
\hline
\hline
\nustar\ FPMA & 2.76 $\pm $0.20 & 5.88 $\pm $0.27 & 0.768 \\
\nustar\ FPMB & 2.65 $\pm $0.20 & 5.79 $\pm $0.14 & 1.089 \\
\chandra\ ACIS LETGS & 3.08 $\pm $0.21 & 6.42 $\pm $0.27 & 0.456 \\
\hline
\nustar\ FPMA & 2.85 $\pm $0.09 & 5.55 $\pm $0.18 & 0.974 \\
\nustar\ FPMB & 2.76 $\pm $0.10 & 5.44 $\pm $0.13 & 1.067 \\
\suzaku\ XIS0 & 2.98 $\pm $0.11 & 4.99 $\pm $0.12 & 1.116 \\
\suzaku\ XIS1 & 3.07 $\pm $0.12 & 4.80 $\pm $0.12 & 0.924 \\
\suzaku\ XIS3 & 2.95 $\pm $0.10 & 5.19 $\pm $0.12 & 0.966 \\
\hline
\nustar\ FPMA & 2.89 $\pm $0.15 & 5.40 $\pm $0.19 & 0.909 \\
\nustar\ FPMB & 2.72 $\pm $0.16 & 5.43 $\pm $0.20 & 1.211 \\
\swift\ XRT & 2.94 $\pm $0.26 & 5.45 $\pm $0.28 & 1.261 \\
\hline
\nustar\ FPMA & 2.83 $\pm $0.13 & 5.57 $\pm $0.17 & 0.950 \\
\nustar\ FPMB & 2.78 $\pm $0.13 & 5.64 $\pm $0.18 & 1.240 \\
\xmm\ MOS1 & 2.64 $\pm $0.09 & 5.66 $\pm $0.11 & 1.056 \\
\xmm\ MOS2 & 2.68 $\pm $0.09 & 5.41 $\pm $0.11 & 0.895 \\
\xmm\ pn & 2.77 $\pm $0.07 & 5.06 $\pm $0.09 & 1.047 \\
\hline
Instrument & $\Gamma$ & Flux 1--5 keV & $\chi^2_{\mathrm{red}}$ \\
& & $10^{-12}$ erg cm$^{-2}$ s$^{-1}$ & \\
\hline
\suzaku\ XIS0 & 2.91 $\pm $0.04 & 19.1 $\pm $0.36 & 1.192 \\
\suzaku\ XIS1 & 2.82 $\pm $0.04 & 20.0 $\pm $0.32 & 1.095 \\
\suzaku\ XIS3 & 2.79 $\pm $0.04 & 20.0 $\pm $0.34 & 0.907 \\
\chandra\ ACIS LETGS & 2.61 $\pm $0.03 & 22.1 $\pm $0.31 & 0.587 \\
\hline
\suzaku\ XIS0 & 2.81 $\pm $0.02 & 18.0 $\pm $0.18 & 1.187 \\
\suzaku\ XIS1 & 2.80 $\pm $0.02 & 18.4 $\pm $0.17 & 1.058 \\
\suzaku\ XIS3 & 2.76 $\pm $0.02 & 18.4 $\pm $0.16 & 0.953 \\
\swift\ XRT & 2.64 $\pm $0.04 & 18.2 $\pm $0.31 & 1.099 \\
\hline
\swift\ XRT & 2.62 $\pm $0.06 & 19.6 $\pm $0.49 & 1.145 \\
\chandra\ ACIS LETGS & 2.61 $\pm $0.03 & 22.1 $\pm $0.31 & 0.587 \\
\hline
\suzaku\ XIS0 & 2.80 $\pm $0.02 & 18.6 $\pm $0.17 & 1.215 \\
\suzaku\ XIS1 & 2.78 $\pm $0.02 & 19.0 $\pm $0.16 & 1.128 \\
\suzaku\ IS3 & 2.74 $\pm $0.02 & 19.0 $\pm $0.15 & 0.988 \\
\xmm\ MOS1 & 2.79 $\pm $0.01 & 18.8 $\pm $0.06 & 1.086 \\
\xmm\ MOS2 & 2.82 $\pm $0.01 & 18.8 $\pm $0.12 & 0.947 \\
\xmm\ pn & 2.76 $\pm $0.01 & 18.0 $\pm $0.08 & 1.077 \\
\hline
\xmm\ MOS1 & 2.80 $\pm $0.02 & 19.7 $\pm $0.19 & 1.099 \\
\xmm\ MOS2 & 2.82 $\pm $0.02 & 19.6 $\pm $0.18 & 1.012 \\
\xmm\ pn & 2.77 $\pm $0.01 & 18.7 $\pm $0.15 & 1.020 \\
\chandra\ ACIS LETGS & 2.61 $\pm $0.03 & 22.1 $\pm $0.31 & 0.587 \\
\hline
\swift\ XRT & 2.64 $\pm $0.04 & 18.2 $\pm $0.31 & 1.099 \\
\xmm\ MOS1 & 2.81 $\pm $0.01 & 18.8 $\pm $0.14 & 1.080 \\
\xmm\ MOS2 & 2.84 $\pm $0.01 & 18.7 $\pm $0.13 & 1.051 \\
\xmm\ pn & 2.78 $\pm $0.01 & 17.9 $\pm $0.12 & 0.957 \\
\hline
\end{tabular}
\label{crossnorm1}

\label{pkstable}
\end{table}

\begin{sidewaystable}
\footnotesize

\caption{Cross normalization constants (1-5\,keV)}
\begin{tabular}{|l|c|c|c|c|c|c|c|c|c|}
\hline
Top/Bottom & LETGS & HETGS & XIS0 & XIS1 & XIS3 & XRT & MOS1 & MOS2 & pn \\
\hline
\hline
LETGS & 1 & - & $ 0.87(2) $ & $ 0.90(1) $ & $ 0.91(2) $ & $ 0.89(2) $ &
$ 0.89(1) $ & $ 0.89(1) $ & $ 0.85(1) $ \\
\hline
HETGS & - & 1 & $ 0.93(1) $ & $ 0.90(1) $ & $ 0.91(1) $ & $ 0.93(3) $ &
$ 0.93(1) $ & $ 0.94(1) $ & $ 0.85(0) $ \\
\hline
XIS0 & $ 1.15(2) $ & $ 1.08(1) $ & 1 & $ 0.96(3) - 0.97(1) $ &
$ 0.99(1) - 1.04(3) $ & $ 1.02(2) - 1.03(2) $ & $ 1.01(1) - 1.01(1) $ &
$ 1.01(1) - 1.02(1) $ & $ 0.93(1) - 0.97(1) $ \\
\hline
XIS1 & $ 1.11(2) $ & $ 1.08(1) $ & $ 1.03(1) - 1.04(3) $ & 1 &
$ 1.03(1) - 1.08(3) $ & $ 0.99(1) - 1.05(2) $ & $ 0.99(0) - 1.03(1) $ &
$ 0.99(1) - 1.04(1) $ & $ 0.95(1) - 0.95(0) $ \\
\hline
XIS3 & $ 1.10(2) $ & $ 1.08(1) $ & $ 0.96(3) - 1.01(1) $ &
$ 0.92(3) - 0.98(1) $ & 1 & $ 0.99(1) - 1.04(2) $ & $ 0.99(0) - 1.02(1) $ &
$ 0.99(1) - 1.03(1) $ & $ 0.94(1) - 0.95(0) $ \\
\hline
XRT & $ 1.12(3) $ & $ 1.08(4) $ & $ 0.97(2) - 0.98(1) $ &
$ 0.95(2) - 1.01(1) $ & $ 0.96(2) - 1.01(1) $ & 1 & $ 1.00(3) - 1.03(1) $ &
$ 1.01(3) - 1.03(1) $ & $ 0.92(3) - 0.98(1) $ \\
\hline
MOS1 & $ 1.12(1) $ & $ 1.08(1) $ & $ 0.99(1) - 0.99(0) $ &
$ 0.97(1) - 1.01(0) $ & $ 0.98(1) - 1.01(0) $ & $ 0.97(1) - 1.00(3) $ & 1 &
$ 0.96(2) - 1.00(2) $ & $ 0.86(1) - 0.89(2) $ \\
\hline
MOS2 & $ 1.12(1) $ & $ 1.07(1) $ & $ 0.98(1) - 0.99(1) $ &
$ 0.96(1) - 1.01(1) $ & $ 0.97(1) - 1.01(1) $ & $ 0.97(1) - 0.99(3) $ &
$ 1.00(2) - 1.05(3) $ & 1 & $ 0.86(1) - 0.93(2) $ \\
\hline
pn & $ 1.18(1) $ & $ 1.17(1) $ & $ 1.03(1) - 1.07(1) $ & $ 1.05(1) - 1.06(1) $ &
$ 1.06(1) - 1.07(1) $ & $ 1.02(1) - 1.09(4) $ & $ 1.12(3) - 1.17(2) $ &
$ 1.07(2) - 1.17(2) $ & 1 \\
\hline
\end{tabular}

\label{fluxtable}

Note - cross-normalization constants from Table \ref{qcstable} and \ref{pkstable}. Where a range is given the instrument observed both sources, and the range are directly the lower and higher ratio. 

\tiny

\caption{Cross normalization constants (3-7\,keV \& 20-40 keV)}
\begin{tabular}{|l|c|c|c|c|c|c|c|c|c|c|c|}
\hline
Top/Bottom & LETGS & HETGS & FPMA\&FPMB & XIS0 & XIS1 & XIS3 & HXD\footnote{\tiny Energy range: 20-40\,keV} & XRT & MOS1 & MOS2 & pn \\
\hline
\hline
FPMA \& FPMB & $ 1.09(3) $ & $ 1.10(7) $ & 1 & $ 0.91(4) - 0.97(1) $ &
$ 0.87(4) - 0.94(1) $ & $ 0.95(4) - 0.97(1) $ & $ 1.12(6) $ &
$ 1.01(7) - 1.08(4) $ & $ 1.01(5) - 1.03(2) $ & $ 0.97(4) - 1.03(2) $ &
$ 0.88(2) - 0.90(4) $ \\
\hline
\end{tabular}

\label{fluxtable_2}

Note - cross-normalization constants from Table \ref{qcstable} and \ref{pkstable}. Where a range is given the instrument observed both sources, and the range are directly the lower and higher ratio.
\end{sidewaystable}

\section{Conclusion}
We have calculated flux ratios between the five observatories \chandra, \nustar, \suzaku, \swift, and \xmm\ for restricted energy bands 1--5\,keV, 3--7\,keV for those involving \nustar, and 20--40\,keV for \nustar\ and \suzaku/HXD. We stress that the cross-normalization constants derived from the fluxes are valid only in the specified energy bands, and do not inform on the differences in spectral slopes between instruments. Results from multiple observatories should be judiciously evaluated against the information provided in this paper, and it should be understood that in the absence of an absolute calibration source there is no instrument that is more ``correct" than the other. 

\acknowledgements
KKM was supported under NASA Contract No. NNG08FD60C, and made use of data from the NuSTAR mission, a project led by the California Institute of Technology, managed by the Jet Propulsion Laboratory, and funded by the National Aeronautics and Space Administration.

HLM was supported by the National Aeronautics and Space Administration (NASA) through the Smithsonian Astrophysical Observatory (SAO)
contract SV3-73016 to MIT for support of the Chandra X-Ray Center (CXC), which is operated by SAO for and on behalf of NASA under contract NAS8-03060. 

APB and KLP acknowledge support from the UK Space Agency.

EDM acknowledges funding from NASA grant NNX09AE58G to MIT to support the Suzaku XIS.

We thank the IACHEC for organizing the campaigns and providing the forum for development and discussion of the cross-calibration results that have led to this paper, and thank the referee for the comments and suggestions.
\\
\\
{\it Facility:} \facility{Chandra}, \facility{NuSTAR}, \facility{Swift}, \facility{Suzaku}, and \facility{XMM-Newton}

\bibliography{bib}
\bibliographystyle{jwaabib}

\end{document}